%% file: _TOI712.tex
\documentclass[twocolumn]{aastex631}
\usepackage{ragged2e}

\usepackage{xspace}
\usepackage{apjfonts}
\usepackage{float}
\usepackage{hyperref}

\input{newcommands}

\input{affiliations}

\usepackage{amsmath}

\shorttitle{TOI-712 system}
\shortauthors{Vach et al.}

\graphicspath{{./}{figures/}}

\begin{document}

\title{TOI-712: a system of adolescent mini-Neptunes extending to the habitable zone}

\input{authors}

\begin{abstract}
As an all-sky survey, NASA's \textit{TESS} mission is able to detect the brightest and rarest types of transiting planetary systems, including young planets that enable study of the evolutionary processes that occur within the first billion years. Here, we report the discovery of a young, multi-planet system orbiting the bright K4.5V star, TOI-712 ($V = 10.838$, $M_\star = 0.733_{-0.025}^{+0.026}\,M_\odot$, $R_\star = 0.674\pm0.016\,R_\odot$, $T_{\rm eff} = 4622_{-60}^{+61}$\,K). From the \textit{TESS} light curve, we measure a rotation period of 12.48 days, and derive an age between about $500$ Myr and 1.1 Gyr. The photometric observations reveal three transiting mini-Neptunes ($R_b = \bradiuse\,R_\oplus$, $R_c = 	\cradiuse\,R_\oplus$, $R_d = \dradiuse\,R_\oplus $), with orbital periods of $P_b = 9.531$ days, $P_c = 51.699$ days, and $P_d = 84.839$ days. After modeling the three-planet system, an additional Earth-sized candidate is identified, TOI-712.05 ($P = 4.32$ days, $R_P = 0.81 \pm 0.11\,R_\oplus$). We calculate that the habitable zone falls between 0.339 and 0.844\,au (82.7 and 325.3 days), placing TOI-712 d near its inner edge. Among planetary systems harboring temperate planets, TOI-712 ($T = 9.9$) stands out as a relatively young star bright enough to motivate further characterization. 

\end{abstract}


\section{Introduction} \label{sec:intro}
Planetary systems undergo many evolutionary processes within the first billion years, but the mechanisms that shape their physical properties and orbital architectures are not yet fully understood. Early interactions with the protoplanetary disk, with other planets, and with the host star can all influence the properties of the mature systems we observe today, but it can be difficult to constrain early processes by observing mature systems. Observing young ($<1$\,Gyr) planetary systems gives us a chance to more directly observe evolutionary processes (e.g., tidal circularization, atmospheric escape, cooling) and constrain their timescales.

Most stars form in groups, the densest of which persist for hundreds of millions of years in loosely bound stellar clusters and associations \citep[e.g.,][]{briceno2007}. Young stars rotate rapidly, which enhances magnetic field strength and leads to active surface regions \citep{hartmann1987}. These features produce large stellar activity signals and make it difficult to distinguish planetary signals from those of stellar origin \citep[e.g.,][]{Saar1997}. Early investigations of young planetary systems were largely composed of radial velocity (RV) surveys targeting known moving groups, stellar clusters, and associations \citep{Paulson2004, Sato2007, Lovis, Quinn2012, Quinn2014} but were limited to detecting giant planets due to the difficulty in detecting low-amplitude signals in the presence of increased stellar activity.

While these RV surveys have contributed to our understanding of planetary formation and evolution, transiting planets provide additional constraints on the physical evolution of planets by measuring planet radii and paving the way for subsequent atmospheric observations. Moreover, precise, space-based photometry is capable of identifying small transiting planets even in the presence of the large photometric rotational modulation that is characteristic of young stars. \Kepler\, \citep{Kepler}, and especially the ecliptic survey of \Ktwo\, \citep{k2} made possible the detection and characterization of small planets orbiting young stars. \Ktwo\, discoveries include planets in open clusters like the Hyades \citep{Mann2016, Livingston2018,Mann2018,Vanderburg2018} and Praesepe \citep{Mann2016,Obermeier2016,Pepper2017,Rizzuto2018,Livingston2019}, as well as associations like Upper Scorpius \citep{David2016,Mann2016b} and the Taurus-Auriga star forming region \citep{David2019a}. These discoveries led to the observation that young planets tend to have larger radii than their mature counterparts. While there may be a bias against detecting small planets orbiting young stars, the large radii of young planets are broadly consistent with predictions for planets undergoing atmospheric mass loss, though the observed sizes imply more mass loss than the models predict. Further study of the evolution of the planet size distribution can constrain the timescales of the process and therefore the mechanisms driving it. This result demonstrates the promise of young systems, but the K2 sample is relatively small and most of the host stars are too faint or active for mass measurements with radial velocities. To discover new benchmark young systems amenable to detailed characterization, and to build a large enough sample of young planets to derive strong constraints from their ensemble properties, many more young stars must be surveyed.

Fortunately, ESA's \textit{Gaia} mission \citep{Gaia2016,GaiaDR22018}, which has already released extremely precise astrometric and photometric measurements for more than a billion stars, has unveiled previously unknown young stellar groups and identified new members of existing groups \citep[e.g.,][]{Kounkel2019,Gagne2021}. As an all-sky survey, NASA's \tess\ mission \citep{Ricker2016} is searching the brightest of these stars for transiting planets. This has led to the discoveries of many more planets in known young groups \citep[e.g.,][]{Newton2019,Plavchan2020,Rizzuto2020,Mann2020,Tofflemire2021}. Moreover, because Sun-like stars spin down over time \citep[e.g.,][]{Barnes2007,Mamajek2008}, the \tess\ photometry used for transit searches can also identify young stars via their rapid rotation. In some cases, this leads to a revised young age for entire groups of stars \citep[the Pisces-Eridanus stream;][]{Meingast2019,Curtis2019b}. In other cases, \tess\ rotation periods reveal a young age for planet-hosting field stars \citep{Carleo2021,Zhou2021}.
Because \tess\ observes nearly the whole sky, it has begun to find the brightest, nearest examples of young planetary systems, which are amenable to detailed characterization through radial velocity masses and atmospheric measurement.

Though the size distribution of young planets has provided evidence for the physical evolution of small planets, many of them reside in systems with only one transiting planet, so their orbital architectures are not well constrained. In the same way that comparing the radii of young planets to their mature counterparts can illuminate the physical evolution of small planets, comparing the orbital architectures of young planetary systems to their mature counterparts can illuminate their orbital evolution.
\Kepler\, revealed that most short-period multi-planet systems are flat, with well-aligned, relatively circular orbits, and a slight preference to reside just outside of low-order mean motion resonances \citep{Lissauer2011,Fabrycky2014,Hadden2014}. However, to explain the observed population of systems with only one transiting planet, it has been suggested that there exists a sample of planets that is either characterized by high mutual misalignment or single planet systems \citep[e.g.,][]{Lissauer2011, fang2012, Ballard2016}. On the other hand, recent analysis of the mutual inclinations of \Kepler\ systems shows that they can be modeled by a single continuous distribution \citep{millholland2021}. Planetary system architectures at young ages can help us understand the processes that have shaped the observed characteristics of the more mature \Kepler\, sample. 
Multi-planet systems also offer an opportunity to constrain the system properties (e.g., masses, eccentricities, inclinations) via dynamical simulations, and to perform comparative studies with planets that have experienced the same stellar environment. For young planets, this may be particularly important in the context of constraining the timescale of thermal mass loss.

In this paper, we report the discovery of three mini-Neptunes transiting the adolescent, K4.5V, field star TOI-712, with periods of 9.53 days (TOI-712 b), 51.69 days (TOI-712 c), and 84.39 days (TOI-712 d). 
Additionally, we find a small candidate planet with a period of $\sim$4.3 days, but additional data is needed to confirm its planetary nature. In \rfsecl{obs}, we present our observational data obtained from \tess\ and ground-based facilities. 
In \rfsecl{stellar_params}, we describe the stellar characterization, including constraints on the age of TOI-712. 
In \rfsecl{exofast}, we provide our global analysis of the TOI-712 system, and report the derived properties of the planets. 
In \rfsecl{dynamics}, we present a dynamical stability analysis and explore the evolution of the TOI-712 system, including the potential survivability of a habitable zone (HZ) planet, given the location of TOI-712 d in the Venus zone at the inner edge of the HZ. We discuss our results in \rfsecl{discussion}.


\section{Observations}
\label{sec:obs}

\subsection{\textit{TESS} Photometry}\label{sec:tess}
NASA's Transiting Exoplanet Survey Satellite (\textit{TESS}) is an all-sky photometric survey, launched 18 April 2018. Its primary mission is to detect small planets around nearby, bright stars, ideal for follow-up observations for further characterization. Using 4 cameras, each $24^\circ \times 24^\circ$, \tess\ observes the sky in $24^\circ \times 96^\circ$ strips. In the two-year prime mission, these strips generally spanned from one of the ecliptic poles to near the ecliptic plane. Each observing sector lasts 27.4 days before stepping $\mysim27^\circ$\ to the next anti-solar sector. After a year (13 sectors) observing the southern ecliptic hemisphere, the spacecraft oriented to observe the north for a year\footnote{Some deviation from this strategy was necessary in the northern hemisphere to avoid regions of high scattered light due to the Earth and Moon, and some new pointing strategies are being employed in the extended mission to observe regions of the ecliptic plane}. Overlap of the fields of view of successive sectors at high ecliptic latitudes results in longer observation time spans, extending to nearly continuous year-long coverage at the poles. 

TIC 150151262 (TOI-712) was pre-selected for 2-minute cadence observations by the \tess\ mission on the basis of the star's size, brightness, and position on the sky near the southern ecliptic pole, the combination of which was expected to enable detection of small transiting planets. TOI-712 was observed on Camera 4 during the prime and extended missions between 2018 September 20 UTC and 2021 June 24 (sectors 3, 4, 6, 7, 9, 10, 11, 13, 27, 29, 30, 31, 32, 33, 34, 36, 37, and 39). During the extended mission, TOI-712 was selected for 20-second cadence observations in sector 27 as part of two Guest Investigator programs: G03068 (PI D. Kipping) and G03278 (PI A. Mayo).

The raw photometric data from \textit{TESS} was processed by the \tess\, Science Processing Operations Center (SPOC) located at NASA Ames Research Center. The SPOC pipeline \citep{Jenkins2010,Jenkins2016} extracts the light curve using Simple Aperture Photometry (SAP) and also corrects the light curve for instrument systematics using the Presearch Data Conditioning (PDC) algorithm \citep[see][]{Stumpe2012}. The resulting light curves are available through the Mikulski Archive for Space Telescopes (MAST) archive. After filtering for stellar variability and residual noise, the light curve was searched for transit signals using the SPOC Transiting Planet Search \citep[TPS,][]{jenkins2002, jenkins2010tps, jenkins2020tps}, which revealed two periodic transit events with periods of 9.53 days and 51.69 days. 
An initial limb-darkened transit model was fitted to the data \citep{li2019} and a suite of diagnostic tests were conducted to make or break confidence in the planetary interpretation of the signatures \citep{twicken2018}. The two candidates passed all the diagnostic tests and the sources of the transits were localized to within 1.9500 $\pm$ 4.1489 arc sec and 2.0800 $\pm$ 4.5217 arc sec, respectively. The TESS Science Office reviewed the DV Data Validation reports and issued an alert in May 2019 \citep{Guerrero2021}.
These were released as TESS Objects of Interest (TOIs), TOI-712.01 and TOI-712.02. An additional candidate with a period of 53.7 days was released as TOI-712.03, but only one transit event looked believable, and a system with 51- and 53-day planets would not be dynamically stable.

\begin{figure*}[h!]
    \centering
    \includegraphics[width=\linewidth]{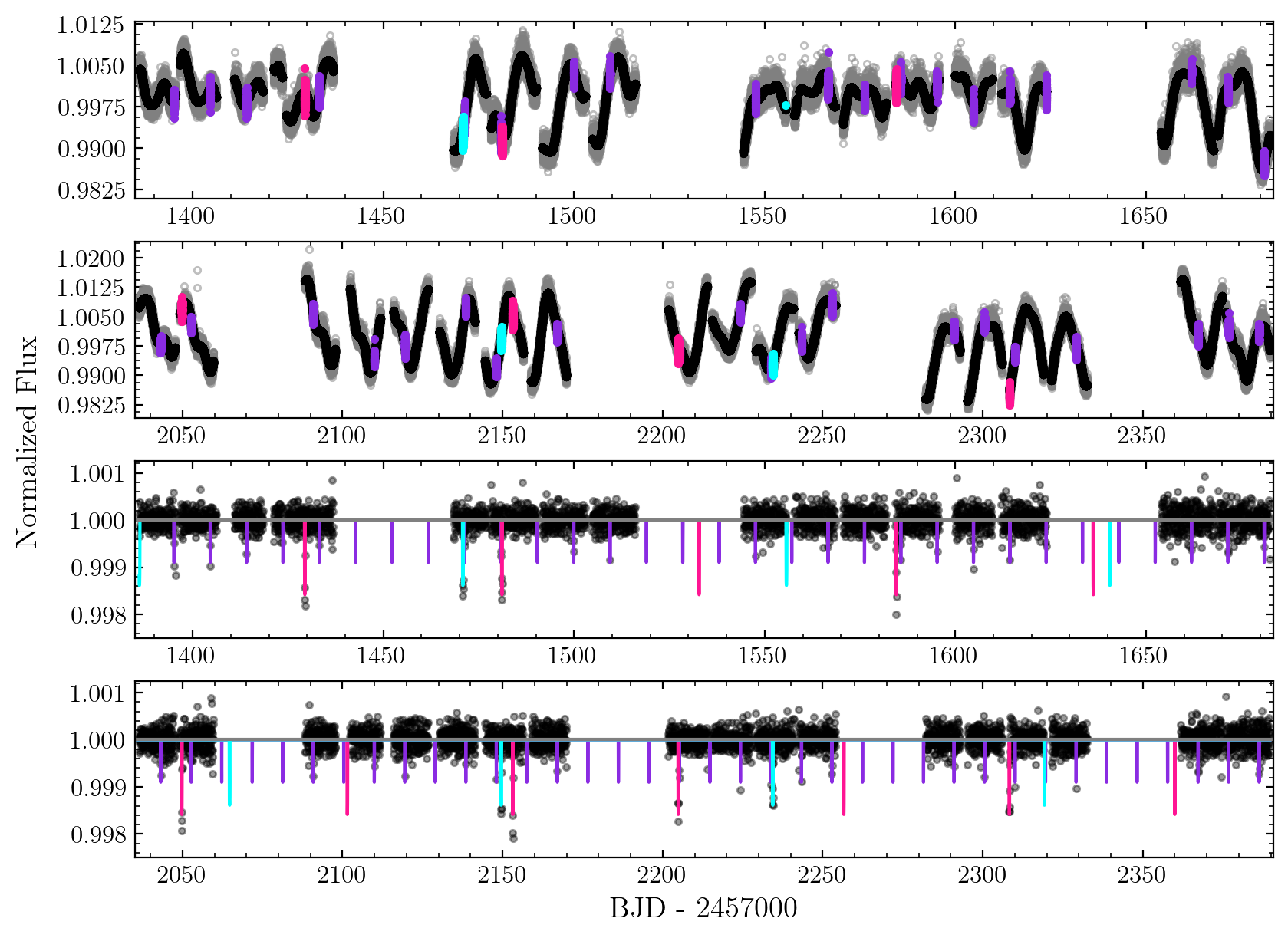}\\
    \includegraphics[width=0.98\linewidth]{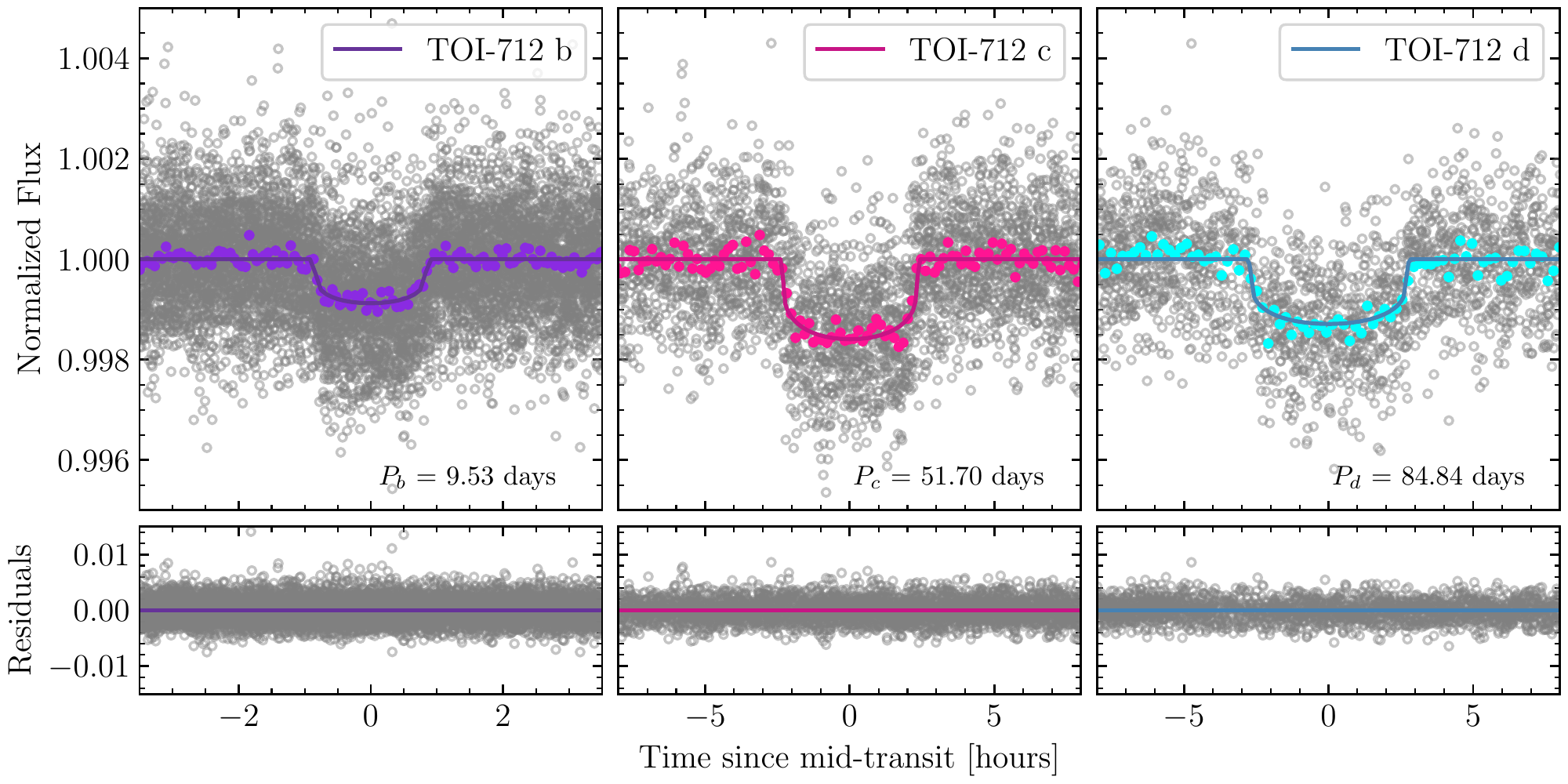}
    \caption{The upper two panels show the raw \tess\ light curves from year 1 and year 3. \tess\ two-minute cadences are plotted in gray and the in-transit points for each planet are colored according to the legend. The spline model is shown in black. The middle two panels show the flattened light curves after removal of the spline, and the best-fit {\tt EXOFASTv2} model is plotted for each planet. The lower panels show the phase folded transits and residuals for each planet. Colored circles indicate binned data, and the colored lines are the best-fit {\tt EXOFASTv2} models.}
    \label{fig:pdcsap}
\end{figure*}

We note that the algorithms applied by SPOC are successful for the vast majority of stars, but the stellar variability of young stars often requires a boutique analysis to properly remove stellar signal and identify real transits \citep[see, e.g.,][]{Rizzuto2017, hedges2021}. Given the apparent presence of this additional transit-like event, we chose to re-extract the light curve while correcting for spacecraft systematics, following \citet{Vanderburg:2019}. We started with the SPOC simple aperture photometry (SAP) light curve and fit a model to the time series consisting of a basis spline (with breakpoints spaced every 1.0 days to adequately model the stellar variability) and systematics parameters relating to the means and standard deviations of the spacecraft quaterion time series within each exposure (see \citealt{Vanderburg:2019}), the fast timescale (band 3) cotrending basis vectors from the SPOC PDC analysis, and a high-pass-filtered time series of the flux in the SPOC background aperture. We fit for up to quadratic relationships in each systematics parameter. We solved for the best-fit coefficients of our free parameters using a matrix inversion technique, iteratively excluding 3$\sigma$ outliers from the fit until it reached convergence. We then subtracted the best-fit systematics vectors from the SAP light curve, and corrected for diluting flux in the optimal aperture using the SPOC \texttt{crowdsap} parameter, to yield a final, corrected light curve. While there are 24 stars within two \tess\ pixels, none of them are brighter than $T=17$, and as a result the aperture typically includes only about $2\%$ dilution. The uncertainty in this correction, propagated to the derived parameters, is negligible compared to other sources of uncertainty. We performed the same extraction for the sector 27 twenty-second data, and then binned to 2 minutes. While the overall scatter in our extracted light curve is similar to the PDCSAP scatter, the systematics during periods of increased pointing jitter are reduced. This helped us identify additional transits of the third candidate, for which we determined a period of 84.839 days. We note that SPOC later identified another transit of this candidate in the extended mission. It was assigned a period of 678.7 days (eight times too long) and released as TOI-712.04, though we now know that both TOI-712.03 (53 days) and TOI-712.04 (679 days) correspond to the 84.8-day candidate. The three candidates have transit depths close to 1 mmag, implying sizes between $2$\ and $3\,\re$, and durations ranging from 1.7 to 5.7 hours, as expected for the range of orbital periods. Details of the transit fitting are discussed in \rfsecl{exofast}. Our extracted light curve is shown in Figure \ref{fig:pdcsap}.

\subsection{Ground-based Time-series Photometry}
\label{sec:sg1}

We conducted ground-based photometric follow-up observations for \thisstar as part of the {\em TESS} Follow-up Observing Program\footnote{https://tess.mit.edu/followup} \citep[TFOP;][]{collins:2019}. We used the {\tt TESS Transit Finder}, which is a customized version of the {\tt Tapir} software package \citep{Jensen:2013}, to schedule our transit observations. The observations are described below and summarized in \rftabl{SG1}.
\begin{figure}
    \centering
    \includegraphics[width=\linewidth]{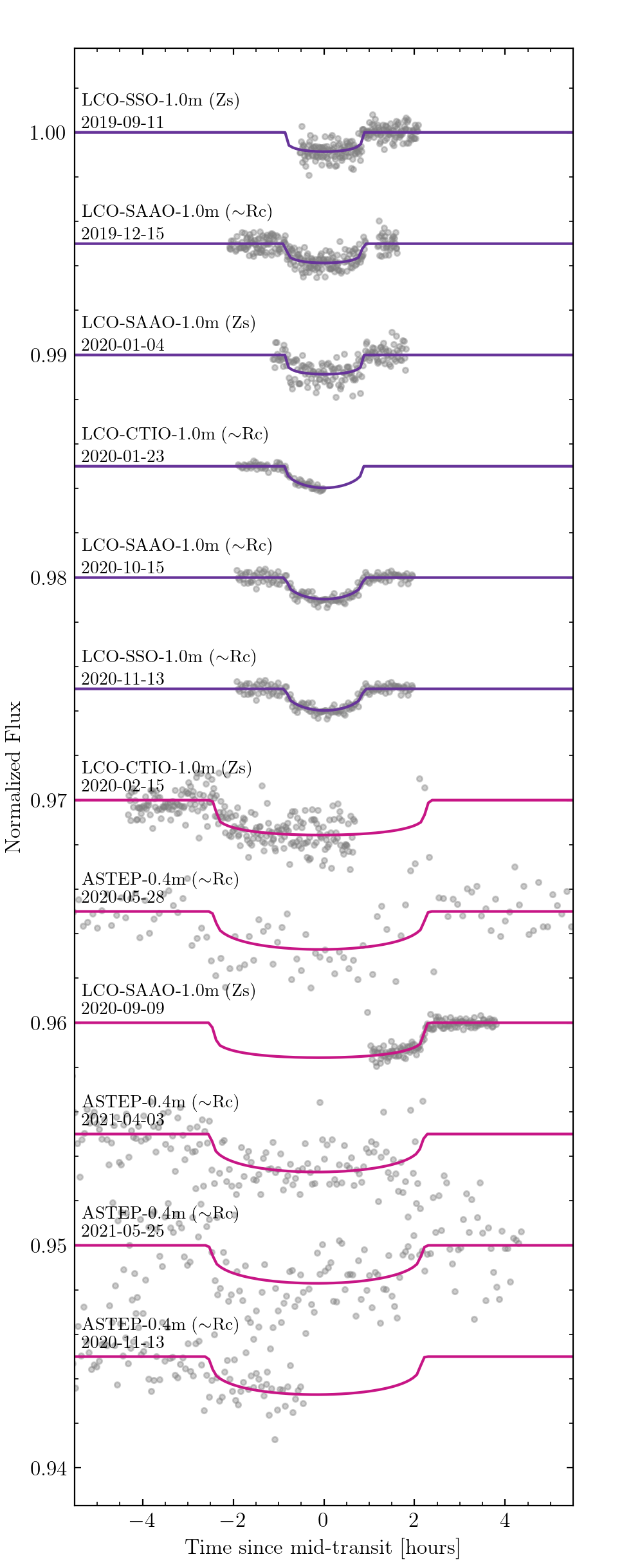}
    \caption{Twelve ground-based follow-up transits with \texttt{EXOFASTv2} best-fit models of \thisstarb\ (purple lines) and \thisstarc (pink lines). Label indicate the instrument, the filter, and the date of observation (see \rftabl{SG1}).}
    \label{fig:lcotoi712b}
\end{figure}

\subsubsection{LCOGT}
We observed seven \thisstar b and three \thisstar c full or partial transits from the Las Cumbres Observatory Global Telescope \citep[LCOGT;][]{Brown:2013} 1.0\,m network. The $4096\times4096$ LCOGT SINISTRO cameras have an image scale of $0\farcs389$ per pixel, resulting in a $26\arcmin\times26\arcmin$ field of view. The images were calibrated by the standard LCOGT {\tt BANZAI} pipeline \citep{McCully:2018}, and photometric data were extracted using {\tt AstroImageJ} \citep{Collins:2017}.

We include all observations in the global fit (see \rfsecl{exofast}) with the exception of the egress observation from 2019-12-25, which suffered from poor precision. It was nevertheless an important observation that confirmed that the transit occurrs at the location of the target star and is not a nearby eclipsing binary unresolved by \tess.

\input{toi712.lco.tex}

\subsubsection{ASTEP}
Four transits of TOI-712 c were observed in 2020 and 2021 with the Antarctica Search for Transiting ExoPlanets (ASTEP) program on the East Antarctic plateau \citep{guillot2015, Mekarnia:2016}. This includes two full transits observed on 2020-05-28 and 2021-05-25. The 0.4\,m telescope is equipped with an FLI Proline science camera with a KAF-16801E, $4096\times 4096$ front-illuminated CCD. The camera has an image scale of $0\farcs93$\,pixel$^{-1}$ resulting in a 1$^{o}\times 1^{o}$ corrected field of view. The focal instrument dichroic plate splits the beam into a blue wavelength channel for guiding, and a non-filtered red science channel roughly matching an Rc transmission curve. The data are processed on-site using an automated IDL-based pipeline described in \citet{Abe:2013}.

\input{rvtable}

\subsubsection{Hazelwood}
We observed an ingress of \thisstar c from Hazelwood Observatory near Churchill, Victoria, Australia. The 0.32\,m telescope is equipped with a $2184\times1472$ SBIG STT3200 camera. The image scale is 0$\farcs$55 pixel$^{-1}$, resulting in a $20\arcmin\times14\arcmin$ field of view. The images were calibrated and the photometric data were extracted using {\tt AstroImageJ}.

This event was the same as the one observed by LCOGT on 2019-12-25. It shows evidence for systematics in transit, and we therefore do not include it in our fit, but as the first observation of this candidate, we note that it represented a valuable confirmation that the transit occurs on the target star and is not a nearby eclipsing binary.

\subsection{CHIRON Spectroscopy}

We obtained three spectra of \thisstar with the CHIRON high-resolution echelle spectrograph on the 1.5\,m SMARTS telescope at the Cerro Tololo Inter-American Observatory (CTIO), Chile \citep{Tokovinin2013}. Observations were taken on UT 2019-Sep-20, 2019-Oct-31, and 2021-Apr-10 in slicer mode, which employs a fiber-fed image slicer to deliver light to the spectrograph, achieving a resolving power of $R\mysim80,000$\ across the wavelength range $4100$--$8700$\,\AA. The CHIRON spectra can be used to characterize the stellar properties, search for massive outer companions, and rule out false positive scenarios that would induce large radial-velocity (RV) variations or exhibit line profile variations.

We extracted the RVs of \thisstar by fitting the line profiles of each spectrum, which were measured via least-squares deconvolution (LSD) of the observed spectra against synthetic templates \citep{Donati1997}. The RVs, which reveal no significant velocity variation beyond the uncertainties ($\mysim50$\,\ms) are listed in \rftabl{rvs}. We also find no evidence for a composite spectrum or line profile variations.

\subsection{High-resolution Imaging}
\label{sec:hri}

\begin{figure}
    \centering
    \includegraphics[width=\linewidth]{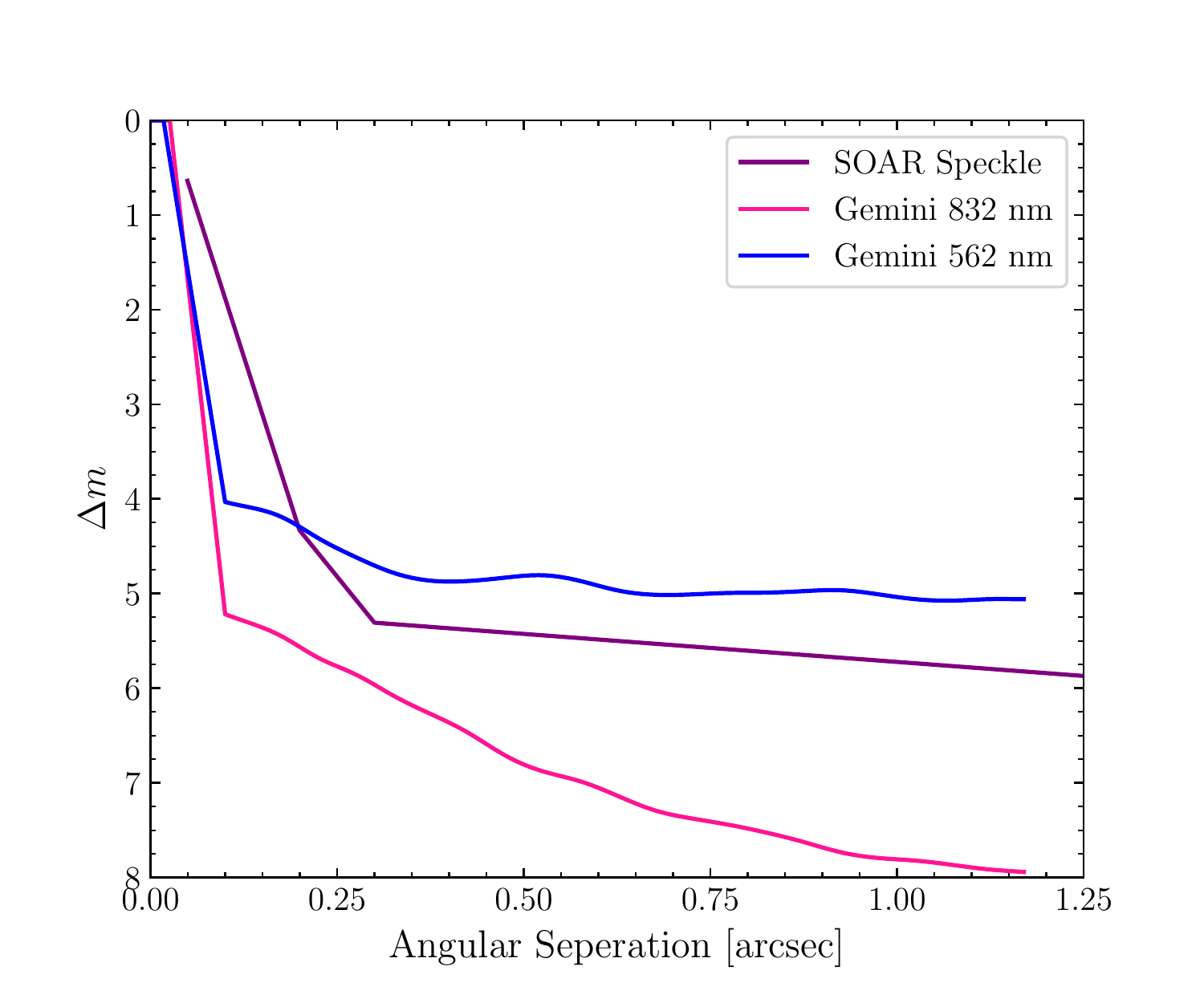}
    \caption{The $5\sigma$ contrast curves of \thisstar, observed by the SOAR speckle imager in the $I$-band (green), and the Zorro speckle imager at 562nm (blue) and 832nm (pink). Zorro at 832nm achieves contrast ratios better than 5 magnitudes at $0\farcs1$, 7 magnitudes at $0\farcs5$, and nearly 8 magnitudes at $1\farcs2$. An eclipsing binary would have to be within 7 magnitudes to produce an event as deep as \thisstarb.}
    \label{fig:speckle}
\end{figure}

We searched for close visual companions to \thisstar\ in $I$\ band using the HRCam speckle imager on the $4.1$\,m Southern Astrophysical Research (SOAR) telescope \citep{Tokovinin:2018, Ziegler:2018} on UT 2019-Oct-16, and simultaneously at $562$\,nm and $832$\,nm using the Zorro\footnote{\url{https://www.gemini.edu/sciops/instruments/alopeke-zorro/}} speckle imager on the 8\,m Gemini South Telescope on UT 2020-Jan-10. We achieved contrast ratios better than $5$\ magnitudes at a separation of $0\farcs1$\ and $>7$\ magnitudes beyond $\mysim0\farcs5$\ in the Zorro $832$\,nm data, which have a field of view of $2\farcs5\times2\farcs5$\ and are therefore sensitive to companions within about $1\farcs2$\ of \thisstar. Outside of $1\farcs5$, {\it Gaia} can exclude the presence of stellar sources bright enough to produce the observed transit signals \citep{Ziegler:2018}. Closer stars unresolved by \textit{Gaia} will often manifest as elevated astrometric noise \citep[see, e.g.,][]{Evans2018}, but \thisstar\ has ${\rm RUWE}=0.994$, providing no evidence for excess astrometric motion\footnote{See \citet{Lindegren2018} for a description of the Gaia astrometric solution. A description of the Renormalised Unit Weight Error (RUWE) can be found at \href{https://www.cosmos.esa.int/web/gaia/public-dpac-documents}{https://www.cosmos.esa.int/web/gaia/public-dpac-documents}}. The $5\sigma$ contrast curves from our speckle imaging are shown in \rffigl{speckle}. 

\section{Stellar Parameters}
\label{sec:stellar_params}
\subsection{Spectroscopic Classification}

We extract the effective temperature (\teff), surface gravity (\logg), and metallicity (\feh) of \thisstar\ from the CHIRON spectrum by matching against a library of observed spectra that have been previously classified by SPC \citep{Buchhave:2012}, with interpolation performed via a gradient-boosting regressor implemented in the \texttt{scikit-learn} python module. We find $\teff=4767 \pm 50$\ K, $\logg=4.63 \pm 0.10$, and $\feh=-0.15 \pm 0.08$. The projected rotational velocity, \vsini, is derived following \citet{Gray:2005} and \citet{Zhou:2018}, fitting broadening kernels to the instrumental, macroturbulent, and rotational line profiles. We estimate $\vsini=2.0\pm1.0$\,\kms\ and $\vmac=1.4\pm1.0$\,\kms, though the instrumental resolving power is only $3.7$\,\kms; for such slow rotation, these estimates may suffer systematic offsets. The projected rotational velocity is consistent with the equatorial rotation velocity calculated from the stellar radius and rotation period, but it is not precise enough to confidently state that the stellar spin and orbital axes are well aligned. 

\subsection{The Age of \thisstar}\label{sec:age}
\subsubsection{Gyrochronology}

\begin{figure*} [ht!]
    \centering
    \includegraphics[width=\linewidth]{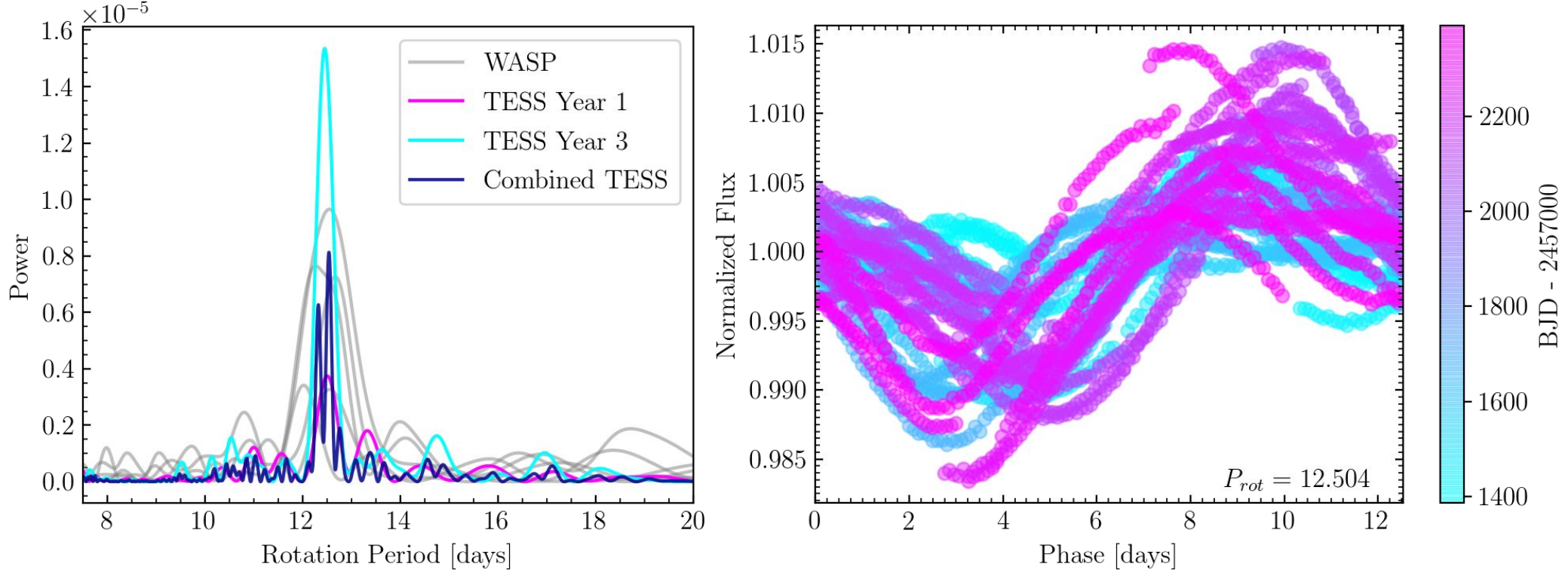}
    \caption{\textit{Left:} The \tess\, light curve shows a strong rotational modulation. The Lomb-Scargle periodogram of the combined TESS light curves is shown in pink, the periodogram for the extended mission is shown in teal, and the periodogram for the combined TESS light curves is shown in navy.. Each season of the \textit{WASP} observations is shown in grey. The rotation period of $\sim$12.5 days is consistently measured in the seventeen sectors observed by \tess\, and the five seasons observed by \textit{WASP}.
    \\ \textit{Right:} Normalized \tess\, light curve detrended using the methods described in \citet{Vanderburg:2019}, phase folded over the measured stellar rotation period extracted from the combined primary and extended \tess\, observations. The phased points are colored based on the time of observation, showing that while the phase curve morphology changes slightly over time, the measured period is consistent with data from both Year 1 (blue tint) and Year 3 (pink tint).}
    \label{fig:lmbscgl}
\end{figure*}

The \tess\, light curve of \thisstar\ shows rotational modulation with an amplitude of $\mysim2\%$, indicative of active regions on the stellar surface like those expected for a young star. We estimate the rotation period with multiple techniques, first using a Lomb-Scargle periodogram \citep{lomb1976,scargle1982}. We assign the rotation period to the period of maximum power, with the uncertainty taken to be the half width at half maximum. However, the analysis is slightly complicated by the long time span of observations. Visual inspection of the light curve (\rffigl{pdcsap}) reveals changes in the morphology of the rotational phase curve on a relatively short timescale. The first year of \tess\ observations includes two epochs of a more complex, double-peaked phase curve and two epochs of a single-peaked phase curve. The third year is primarily single-peaked. If the dominant spot latitudes also change between the prime and extended missions, we might expect a slight change in the measured rotation period due to differential rotation, which is both predicted \citep[e.g.,][]{kuker2011} and measured for some rapidly rotating stars \citep[e.g.,][]{collier2007}. A periodogram of the full time series would then be sensitive to the phasing of the signals even though we should not expect or require them to be in phase. Indeed, this is what we see in \rffigl{lmbscgl}, which shows that the periodogram of the combined \tess\ light curve only allows a few discrete periods, while individual years of data each exhibit broader peaks that are consistent with one another. We therefore measure the rotation period in each year separately.

From the \tess\, year 1 light curves, we measure a stellar rotation period of 12.505 $\pm$ 0.441 days, and the \tess\, year 3 light curves yields a rotation period of 12.452 $\pm$ 0.629 days. Additionally, \textit{WASP} photometry spanning five observing seasons from 2008 to 2012 measures a rotation period of 12.503 days. The periodograms from the \tess\, and \textit{WASP} observations are shown in \rffigl{lmbscgl}, and the binned \tess\, light curve, phase-folded to the measured rotation period, is shown in \rffigl{lmbscgl}.
As a check on the Lomb-Scargle periods, we also fit the full \tess\ light curve using a Gaussian process with a quasi-periodic kernel. This yields a period consistent with the Lomb-Scargle results, $P_{rot} = 12.538$ days. We adopt a weighted average of rotation periods from individual seasons, $P_{\rm rot} = 12.454$ days.

Finally, we caution that we are observing the rotation period at the latitude of the active regions. The unknown strength of differential rotation and current spot latitudes contribute additional (systematic) uncertainties if interpreting the observed rotation period as the equatorial rotation period. If differential rotation in \thisstar\ is similar to that of the Sun, then the true equatorial rotation period could be different by $10\%$\ or more.

We adopt the activity-rotation-age relation as described in \citet{Mamajek2008}, to estimate the age of \thisstar. In particular, we use the equation:

\begin{equation}
    P_{\textrm{rot}}\left(B - V, t\right) = a\left[\left(B-V\right) - c\right]^bt^n
\end{equation}

where $P_{\textrm{rot}}$ is the rotational period of the star, in days, $t$ is the stellar age in Myr, $B-V$\ is the color index, $a = 0.407 \pm 0.021$ and $b = 0.325 \pm 0.024$ are gyrochronolgy constants, $c = 0.495 \pm 0.010$ is the 'color singularity`, and $n = 0.566 \pm 0.008$ is the time-dependent power law.  Using the stellar rotation period above, and $B-V = 1.265 \pm 0.152$, we estimate a gyrochronological age of $492 \pm 103$ Myr. However, we caution that the formal uncertainty is quite likely underestimated. First, systematic errors may affect the determination of $P_{\rm rot}$, as described above. More importantly, \thisstar is outside of the color range ($0.5 < B - V < 0.9$) in which the \citet{Mamajek2008} relation is well calibrated, as their sample consisted of primarily Sun-like stars and \thisstar\ is relatively red. More recent investigations of the gyrochronology of K dwarfs, enabled by precise photometry of adolescent stars from \Ktwo\ and \tess, suggest that in contrast to the G dwarfs, K dwarfs experience an epoch of stalled spin-down. While it is true that K dwarfs are expected to have rotation periods similar to \thisstar\ at 500 Myr, they may also have the same rotation period at twice that age \citep{Curtis2019a}. As a result, from its rotation alone, we can only estimate the age of \thisstar\ to be between about $0.5$\ and $1.1$\ Gyr.

\subsubsection{Comparison with stellar populations}

\begin{figure}[!t]
    \centering
    \includegraphics[width=\linewidth]{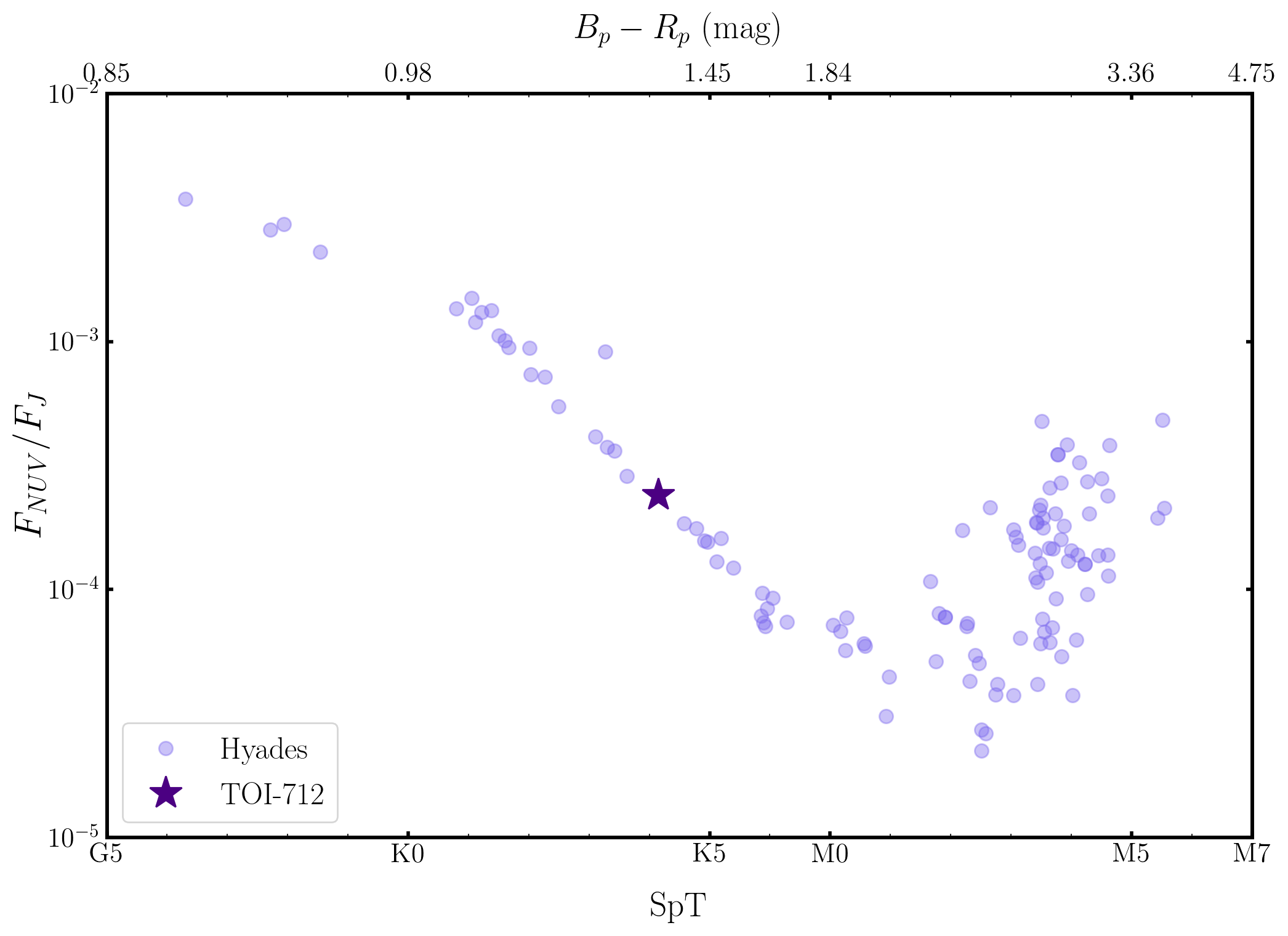}
    \caption{Ratio of NUV to J-band flux plotted against spectral type for Hyades members (light purple dots) according to Friend Finder. Younger stars are expected to exhibit enhanced NUV fluxes, while older stars should produce less high energy flux. The dark purple star shows the location of \thisstar, which suggests an age consistent with that of the Hyades ($\mysim625$\,Myr).}
    \label{fig:galex}
\end{figure}

Since recent work has identified many new members of known stellar groups, we explore whether \thisstar\ might belong to a known association. We use the online tool BANYAN $\Sigma$ \citep{banyan}, to assess possible membership in known groups, but the result suggests that it is not part of any previously known groups.

Similarly, there are likely many sparse groups that are currently unrecognized, but the large volume of public data (from Gaia, \tess, broadband photometric surveys, and spectroscopic surveys) makes possible a targeted search for comoving, coeval companions to a star of interest. We searched for evidence of a previously unknown comoving population of stars via Friend Finder\footnote{\href{https://github.com/adamkraus/Comove}{https://github.com/adamkraus/Comove}}. Within 25 pc of \thisstar, we find only two stars with similar kinematics, but this is likely no more than would be expected by chance. \thisstar has either escaped from its birth environment or formed in such a small environment that almost none of its stellar siblings remain within $<25$ parsecs. While no convincing kinematic matches were found, the age can also be constrained by the precise UV photometry provided by GALEX \citep[e.g.,][]{Shkolnik2011, Findeis2010}. Friend Finder illustrates that the near-UV (NUV) flux of \thisstar is consistent with the NUV sequence of the stars populating the Hyades cluster (see \rffigl{galex}), which has a well constrained age of $\sim$625 Myr  \citep[see, e.g.,][]{perryman1998,martin2018} . Older stars should exhibit lower levels of NUV flux. This provides supporting evidence that the age of \thisstar is consistent with the age of the Hyades, though if NUV flux and rotation evolve together, it is possible that NUV flux would plateau while spin-down stalls. If that were the case, then the NUV flux of Hyades K dwarfs could be similar to that of somewhat older stars.  \citet{Richey-Yowell:2019} do show that the NUV flux is inversely related to stellar age in K dwarfs, but that study included no stars between Hyades age and field stars. NUV sequences in slightly older clusters like NGC 752 \citep[1.3 Gyr;][]{agueros2018} and Ruprecht 147 \citep[2.7 Gyr; e.g.,][]{torres2020,curtis2020} might help resolve the age of \thisstar, but because of their distance, GALEX provides only non-informative upper limits for most of the K dwarf members.

Given the age ambiguity associated with the stalled spin-down of K dwarfs, we ultimately adopt an age of $830^{+250}_{-280}$\ Myr and bounded between 0.3 and 1.3 Gyr, as determined by the global fit described in \rfsecl{exofast}.

\subsubsection{The stellar flare rate of TOI-712}

\thisstar\ was observed with 20-second cadence in sector 27, and during that time a flare was observed at time BTJD 2054.672 with a maximum flux excursion of $\mysim10$\%. Though it lasted only three 20-second cadences, two independent flare-finding algorithms (\citealt{medina2020}; \citealt{stella2020,feinstein2020}) identify it as a high-probability flare. We calculate an energy in the TESS bandpass of $8.75\times10^{32}$\,ergs, and we derive a flare rate of approximately $10^{-3}$\ flares per day above an energy of $2.32\times10^{32}$\,ergs. While there are no well calibrated flare rates for adolescent K dwarfs, this flare rate falls slightly below the observed rates for younger K dwarfs \citep{feinstein2020}. This does not resolve the gyrochronology age degeneracy, but the presence of any significant flares further supports that this is an adolescent star. 

Finally, we note that this flare lasted only 1 minute, which implies that there may be flares in the two-minute data that last only one cadence and appear as outliers of a few percent or less. This should not be surprising, as many previous studies with \tess\ and \kep\ data have focused on active stars with flares lasting many minutes \citep[][]{gunther2020,medina2020,feinstein2020} or longer \citep[e.g.,][]{davenport2016}, but there is an indication that flare energies in the optical correlate with flare duration \citep{maehara2015}.
While it is beyond the scope of this work, \tess\ data could offer a means to calibrate flare rates of adolescent stars, which may be prone to short duration and low energy flares that appear as a single modest outlier in two-minute cadence.  Distinguishing single-cadence flares from instrumental effects is difficult, but a statistical approach may be feasible, particularly given the large number of TESS light curves of both young and old stars. 

\input{stellar}

~\\
\section{EXOFASTv2 Global Fit}
\label{sec:exofast}

To derive the properties of \thisstar and its planets, we fit the system with the exoplanet modeling software EXOFASTv2 \citep{Eastman:2019}. We used all available 2-minute photometric data from \tess\ and the ground-based light curves described in \rfsecl{sg1} in conjunction with broadband flux measurements (\rftabl{stellar}), the Gaia DR2 parallax, spectroscopic stellar parameters, and the MESA Isochrones and Stellar Tracks (MIST) stellar isochrones \citep{Paxton2011, Paxton2013, Paxton2015, Choi2016, Dotter2016}. In our Markov Chain Monte Carlo (MCMC) fit, we set priors on the stellar metallicity derived from the CHIRON spectra, the parallax measured by Gaia, and the upper limit on extinction from the dust maps of \citet{Schlegel1998} and \citet{Schlafly2011}. At each step in the MCMC, the quadratic limb-darkening coefficients are interpolated from the values tabulated by \citet{Claret2017}. Following \citet{Ford2006}, we require the Gelman-Rubin statistic of all parameters to be $<1.01$\ and the number of independent draws to exceed 1000 to indicate convergence.

The results of our global fit are reported in \rftabl{exofast}, and the best-fit models are plotted with the phase-folded light curves in \rffigl{pdcsap}. Notably, the radii (\bradius\,\re, \cradius\,\re, \dradius\,\re) suggest that all three planets are mini-Neptunes, likely to possess significant volatile envelopes. The eccentricities of the outer planets are relatively low, and while the eccentricity of \thisstarb\ is not inconsistent with zero, it is poorly constrained. The medians of the marginalized posteriors are reported in \rftabl{exofast}, but the eccentricity posterior for planet b is broad and non-Gaussian.  This is important to note, because while we present further analysis based on the broad {\tt EXOFASTv2} posteriors, our expectation based on \kep\ multi-planet systems is that all three planets will have eccentricities much closer to zero than the medians of the posteriors reported in \rftabl{exofast} \citep[see, e.g.,][]{moorhead2011,vaneylen2015,he2020}. While multi-planet systems can exhibit transit time variations due to mutual dynamical interactions, the {\tt EXOFASTv2} transit times are consistent with a linear ephemeris. 

\subsection{Validation of the \thisstar\ planets}

In the previous section, we presented a model of the planetary system orbiting \thisstar, but have not formally validated the signals as bona fide planets. In particular, we note that while we detect the transits of \thisstarb\ and c from the ground in small apertures that localize the transit signal to within a few arcseconds of \thisstar, there is no ground-based transit detection of \thisstard. Given the large \tess\ pixels, the signal could potentially be produced by a faint, nearby eclipsing binary (NEB). Gaia detects 25 stars within 2 pixels ($42\farcs0$), though the centroid analysis made available in the SPOC Data Validation report rules out NEBs more distant than $18\farcs0$. The brightest of the remaining stars is 8.1 magnitudes fainter in the \tess\ bandpass, and therefore contributes a flux fraction of only 0.0006 (0.6 mmag). The transit of \thisstard\ is deeper than 1 mmag, so none of the nearby stars can be responsible for the transit signal. The high resolution imaging and Gaia astrometry discussed in \rfsecl{hri} rule out nearly all stellar companions within 1 arcsecond, and the CHIRON RVs exclude many close, bound eclipsing binaries that would have accelerated \thisstar\ during the 600-day observing timespan.

Previous analysis has shown that the overwhelming majority of \kep\ candidates in multi-planet systems are real planets, and moreover that systems with more than 2 candidates are even more likely to be real \citep[e.g.,][]{latham2011,lissauer2012}. The same cannot be assumed for \tess, mainly because of the increased potential for NEBs due to its large pixels and, in some cases, crowded fields. However, our follow-up observations have excluded the presence of contaminating eclipsing binaries even more confidently than is the case for \kep\ candidates. As a result, we consider \thisstarb, c, and d to be bona fide planets.

\input{toi712_ef2}

\begin{figure}[t!]
    \centering
    \includegraphics[width=\linewidth]{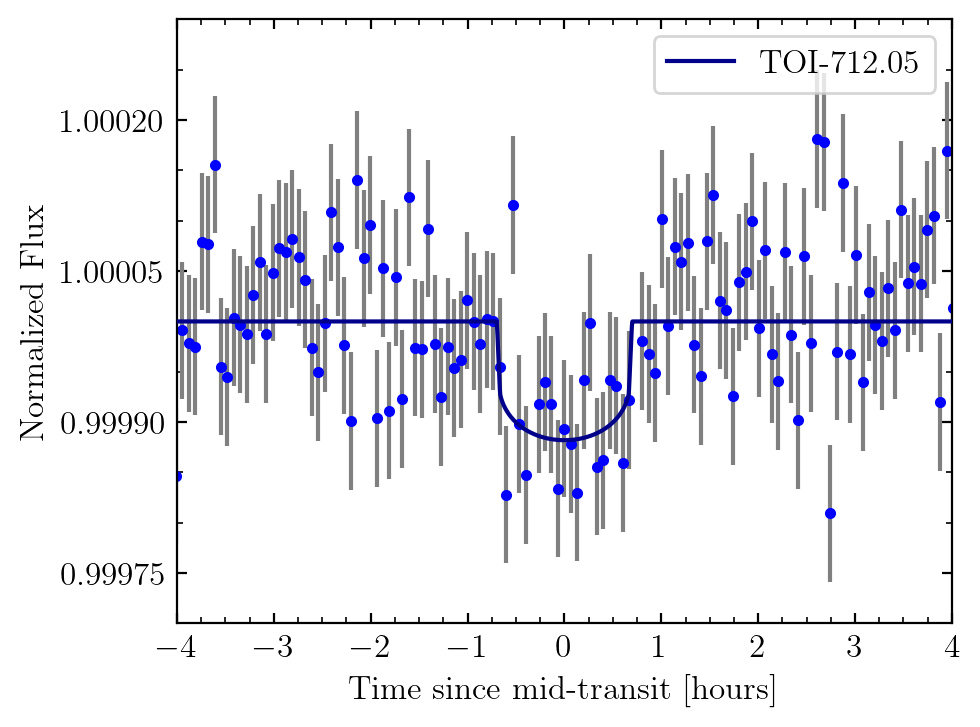}
    \caption{A TLS analysis of the three planet EXOFASTv2 Global Fit residuals reveals an additional transit signal, with a period of $\sim$4.32 days. The phase folded, binned light curve from \tess\, is shown with a normalized flux.}
    \label{fig:toi712.05}
\end{figure}

\subsection{TOI-712.05}

Upon subtracting the best-fit EXOFASTv2 three planet model from the \tess\ light curve, we perform a Transit Least Squares \citep[TLS,][]{tls} analysis on the residuals to search for the signature of any additional transiting planets. TLS reveals a $\sim$4.32-day candidate signal. We refer to this signal as TOI-712.05. A new EXOFASTv2 fit including all four candidates converged, with no change to the stellar parameters or properties of the other three candidates. The derived properties of this candidate are shown in Table \rftabl{toi712_05tab}, and the phase folded \tess\, light curve is shown in Figure \ref{fig:toi712.05}. If real, TOI-712.05 would be slightly smaller than the Earth, $R_{P} = 0.81\,R_\oplus$.

We present the results for TOI-712.05 separately because we have less confidence that it is real. Notably, the signal is not clearly present in the full frame images (FFIs). While this could be due to smearing of the short-duration transit in the 30- and 10-minute cadence data, it may also be due to a difference in the apertures used to produce the FFI and 2-minute light curves. If this were the case, it could suggest that the signal originates on a nearby star at the edge of the 2-minute aperture. To test this possibility, we ran two experiments. First, we extracted FFI light curves for each of the stars within $2.5\arcmin$\ and folded them at the ephemeris of TOI-712.05. None showed signs of an eclipse that would cause the observed signal. Because the sensitivity of the FFIs to a shallow, short duration event may be compromised we also extracted 2-minute light curves using different apertures. 

\input{toi712.05}

Planetary candidates that are released as TOIs also undergo ephemeris matching against all other signals detected by the pipeline, in order to test for a false positive introduced by a nearby transit-like signal. TOI-712.05 was not released as a TOI, so we performed this test using all signals detected by MIT's Quick Look Pipeline \citep[QLP;][]{qlp2020a,qlp2020b}, and we did not find any ephemeris matches.

It is also possible that SPOC introduced the transit signal while processing the light curve since \thisstar is relatively active. As a check, we ran TLS on the unprocessed SAP light curve to see if the transit is still present. After flattening, and removing the transits from the three planets, the phase folded SAP light curve does show the transit signal for \thisstar.05.

A final concern is that the signal could be introduced by stellar activity. If that were the case, we might expect that the signal would not appear consistent in time due to evolution of the stellar signal, which is apparent by eye in the unflattened \tess\ light curve. We tested this by comparing the signal in Year 1 to the signal in Year 3. To within the uncertainties, the two are consistent, as would be expected for a transiting object.

While TOI-712.05 passed the above tests, it remains a low signal-to-noise event, so we present it here as a candidate planet. Its planetary nature may be solidified as the signal strength increases in future \tess\ extended missions, or perhaps by additional analysis of existing data.

\section{Habitable Zone and Venus Zone}
\label{sec:hz}

\begin{figure}
    \begin{center}
    \includegraphics[width=\linewidth]{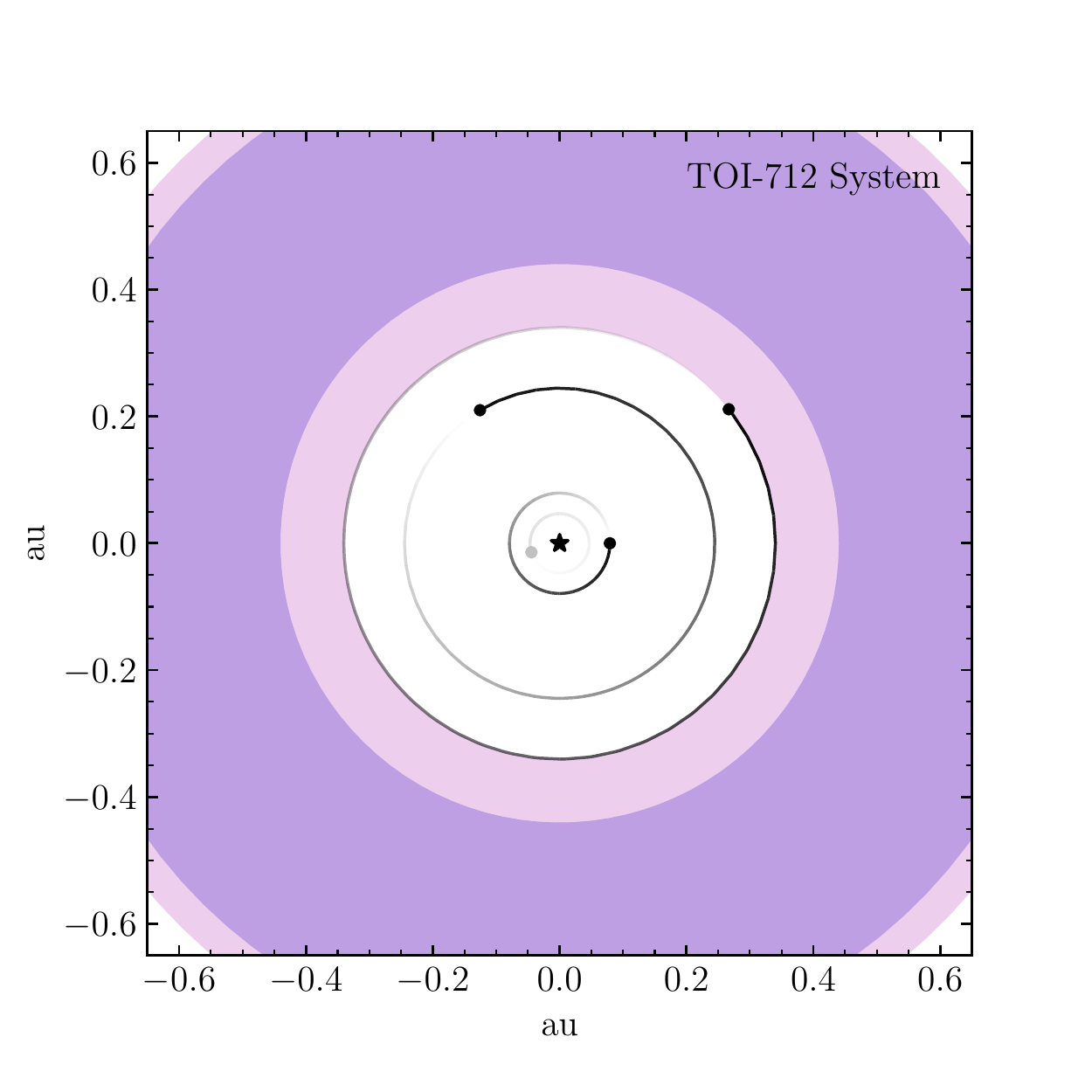}
    \end{center}
    \caption{A top-down view of the TOI-712 system, where the known planets are plotted in black and the candidate planet TOI-712.05 is plotted in gray. The conservative habitable zone (purple) spans 0.436 to 0.802 au, and the optimistic habitable zone (pink) extends from 0.344 to 0.846 au, coinciding with the orbit of \thisstard\ at its inner edge.}
    \label{fig:hz}
\end{figure}

As described above, the three detected planets in the system are all mini-Neptune in terms of size. While they are too large to expect Earth-like compositions, their locations with respect to the Habitable Zone (HZ) and the Venus Zone (VZ) can motivate further analysis and future observations. The HZ boundaries have been previously calculated from Earth-based climate models that determine the radiative balance conditions for retaining surface liquid water \citet{kasting1993a,kopparapu2013a,kopparapu2014}. The conservative HZ (CHZ) is defined by the runaway greenhouse limit at the inner boundary, and the maximum CO$_2$ greenhouse limit at the outer boundary \citep{kane2016c}. The optimistic HZ (OHZ) is an extension to the conservative HZ based on empirical evidence regarding retention of surface liquid water on Venus (inner boundary) and Mars (outer boundary) during their evolutionary histories \citep{kane2016c}. The location of the HZ boundaries are sensitively related to the stellar parameters \citep{kane2014a}, provided in Table~\ref{tab:exofast}. Adopting the methodology of \citet{kopparapu2014}, we calculate the CHZ boundaries for the TOI-712 system as 0.436--0.802~au, and the OHZ boundaries as 0.344--0.846~au. The HZ regions are represented in Figure~\ref{fig:hz}, which shows a top-down view of the TOI-712 system. Because the eccentricities are poorly constrained, we plot circular orbits, though note that modest eccentricities are not ruled out. \thisstard\ lies at the extreme inner edge of the OHZ.

\begin{figure*}
    \begin{center}
    \includegraphics[width=\textwidth]{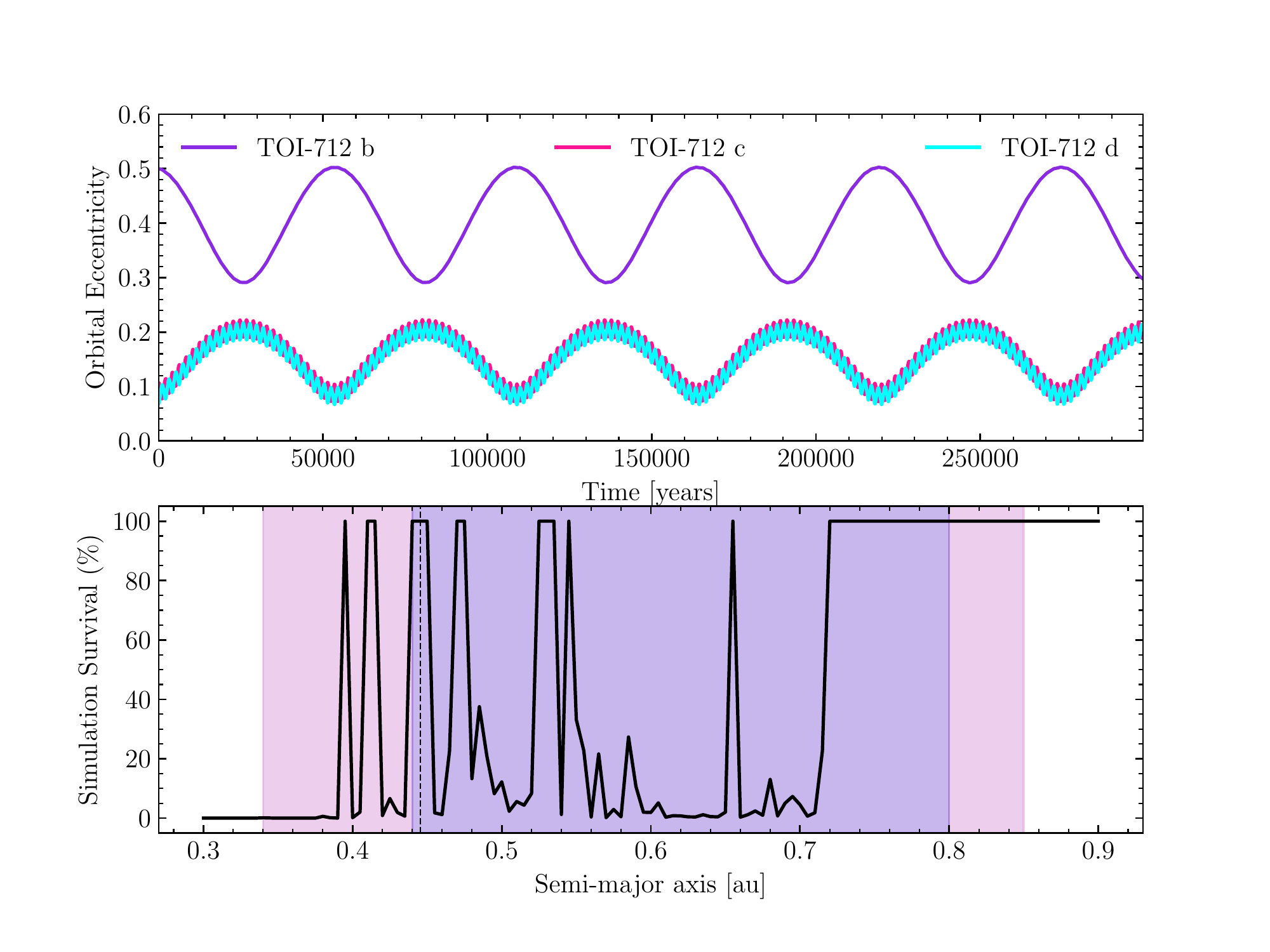} 
    \end{center}
    \caption{Results of the dynamical simulations, described in Section~\ref{sec:dynamics}. Top: The eccentricity variations of the known planets for a period of 300,000 years, showing the stability of the system and the high frequency eccentricity oscillations of planets c and d. Bottom: The dynamical simulation results for an Earth-mass planet located in the semi-major axis range 0.3--0.9~au, shown as percentage survival of the simulation as a function of semi-major axis (solid line). As for Figure~\ref{fig:hz}, the CHZ and OHZ are shown in light green and dark green, respectively. The vertical dashed line indicates an example stable resonance location (3:2).}
    \label{fig:stability}
\end{figure*}

Though none of the TOI-712 planets lie unambiguously within the HZ, they do lie within the VZ, defined as the region interior to the runaway greenhouse limit and exterior to the atmospheric mass loss limit \citep{kane2014e}. Such planets provide exceptionally useful targets for atmospheric characterization, both individually and in the context of describing the processes that occur within the VZ \citep{kempton2018,ostberg2019}. The TOI-712 planets are unlikely to have rocky surfaces, but their atmospheric response to the local radiation environment can provide useful information regarding the processes driving atmospheric erosion. 
Moreover, because the planetary system extends (at least) to the inner edge of the HZ, their presence can place dynamical constraints on the existence of exterior planets that may lie within the HZ.

\section{Dynamical analysis}
\label{sec:dynamics}

The orbital parameters of the TOI-712 planets (see Table~\ref{tab:exofast}) provide an opportunity to examine the orbital dynamics that exists within the system, and the dependence on their proximity to each other. Such dynamical interactions are particularly important for assessing eccentricity constraints for the known planets, and for placing dynamical limits on the presence of other potential planets in the system \citep{gladman1993,chambers1996,hadden2018b}. Sampling from the posteriors of the {\tt EXOFASTv2} fit, we performed N-body integrations using the Mercury Integrator Package \citep{chambers1999} and adopted a methodology similar to that described by \citet{kane2014b,kane2016d,kane2019c}. We used a hybrid symplectic/Bulirsch-Stoer integrator with a Jacobi coordinate system \citep{wisdom1991,wisdom2006b}, and adopted a time resolution of 0.2~days to adequately sample the orbit of the inner planet. Each of the simulations were executed for a duration of $10^7$ simulation years.

Shown in the top panel of Figure~\ref{fig:stability} is a 300,000~year segment of the results from a simulation initialized with eccentricities near the median of the posteriors. While the true eccentricities may be closer to zero, the system remains stable even for relatively high eccentricities, and this simulation illustrates some of the mutual interactions experienced by the planets. The outer planets (c and d), which lie near, but not within, the 8:5 mean motion resonance (MMR), exhibit high frequency angular momentum exchanges that cause frequent fluctuations in their eccentricities. In turn, planets c and d interact with the innermost known planet (b), resulting in high amplitude oscillations in their respective eccentricities with a period of $\sim$50,000~years. These complex dynamical interactions raise the question of whether the known planets may exclude potential terrestrial planets from residing within the HZ. To test for this, we conducted an exhaustive suite of $10^7$~year dynamical simulations using the same methodology as described above---with initial conditions drawn randomly from the {\tt EXOFASTv2} posteriors---but with the insertion of a hypothetical Earth-mass planet located in the semi-major axis range of 0.3--0.9~au, in steps of 0.05~au. The simulations were allowed to run until either the simulation completed, or planet-planet interactions caused the Earth-mass planet to be swallowed by the star or ejected from the system. The bottom panel of Figure~\ref{fig:stability} shows the results from this simulation in terms of the percentage of the simulations that survived the full $10^7$ years as a function of the semi-major axis for the test locations of the Earth-mass planet, indicated by the solid line. Orbits beyond $\mysim0.7$\ au are robustly stable, and some planets closer than 0.7 au survive near MMRs. Since there are three interior planets, the locations of MMR are numerous and complicated, but we highlight one example of a resonance stable location corresponding to the 3:2 MMR of an HZ planet with planet d.

\begin{figure}[t!]
    \centering
    \includegraphics[width=1.05\linewidth]{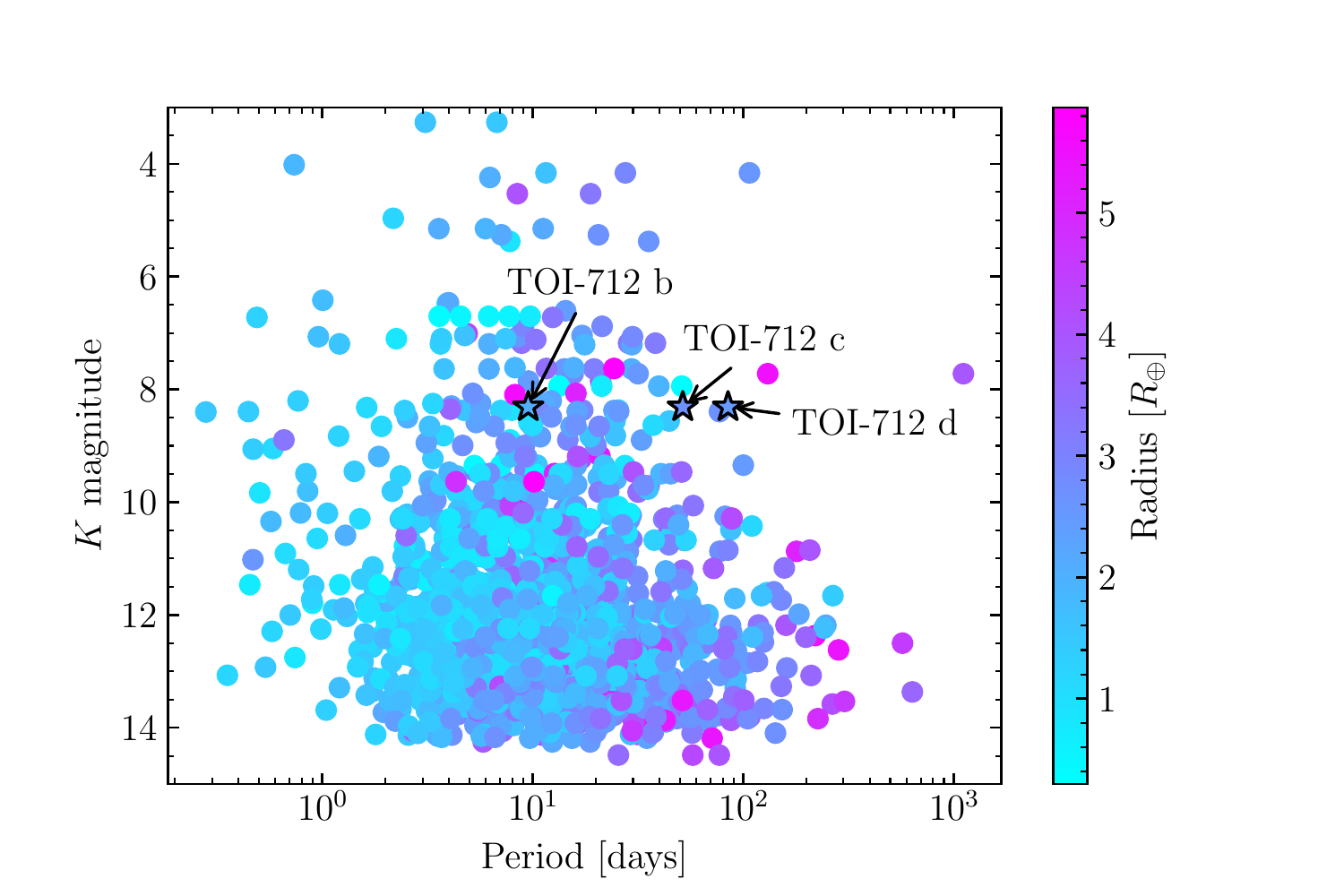}
    \caption{Known validated small planets ($R_P \leq 6.0R_\oplus$) in multi-planet systems (as of 29 October 2021, via NASA's Exoplanet Archive). The host stars' $K$-magnitudes are plotted against the orbital periods. The points are colored based on the planet radius. \thisstar d is one of the longest period, small planets orbiting a bright star. The brightness of the host allows for the planetary radii to be well constrained through photometric observations with relatively few observed transits, and also enables future characterization with other facilities.}
    \label{fig:brightness}
\end{figure}

\section{Discussion}
\label{sec:discussion}
 
\subsection{\thisstar among known planets}
\label{sec:knownplanets}

Multi-planet systems provide a laboratory for comparative studies of exoplanets in an environment controlled for stellar properties and age, and the ability to leverage dynamical interactions to constrain planetary masses and orbital architectures. This avenue for characterization is especially important when other methods like radial velocity monitoring are difficult---for example, due to stellar activity, faint host stars, or small or long-period planets that induce low-amplitude orbital motion. Because short-period planets are more likely to transit, relatively few systems include long-period planets, and most that do were discovered by \kep\ orbiting faint stars. As a result, \thisstar\ stands out as one of the brightest long-period planetary systems currently known (see \rffigl{brightness}). While many mini-Neptunes transit bright stars, most reside close to their star with high equilibrium temperatures, so \thisstar\ represents an opportunity to study temperate mini-Neptunes, which are likely to have experienced different formation environments and evolutionary histories than their short-period counterparts. 

\begin{figure}[t!]
    \centering
    \includegraphics[width=1.05\linewidth]{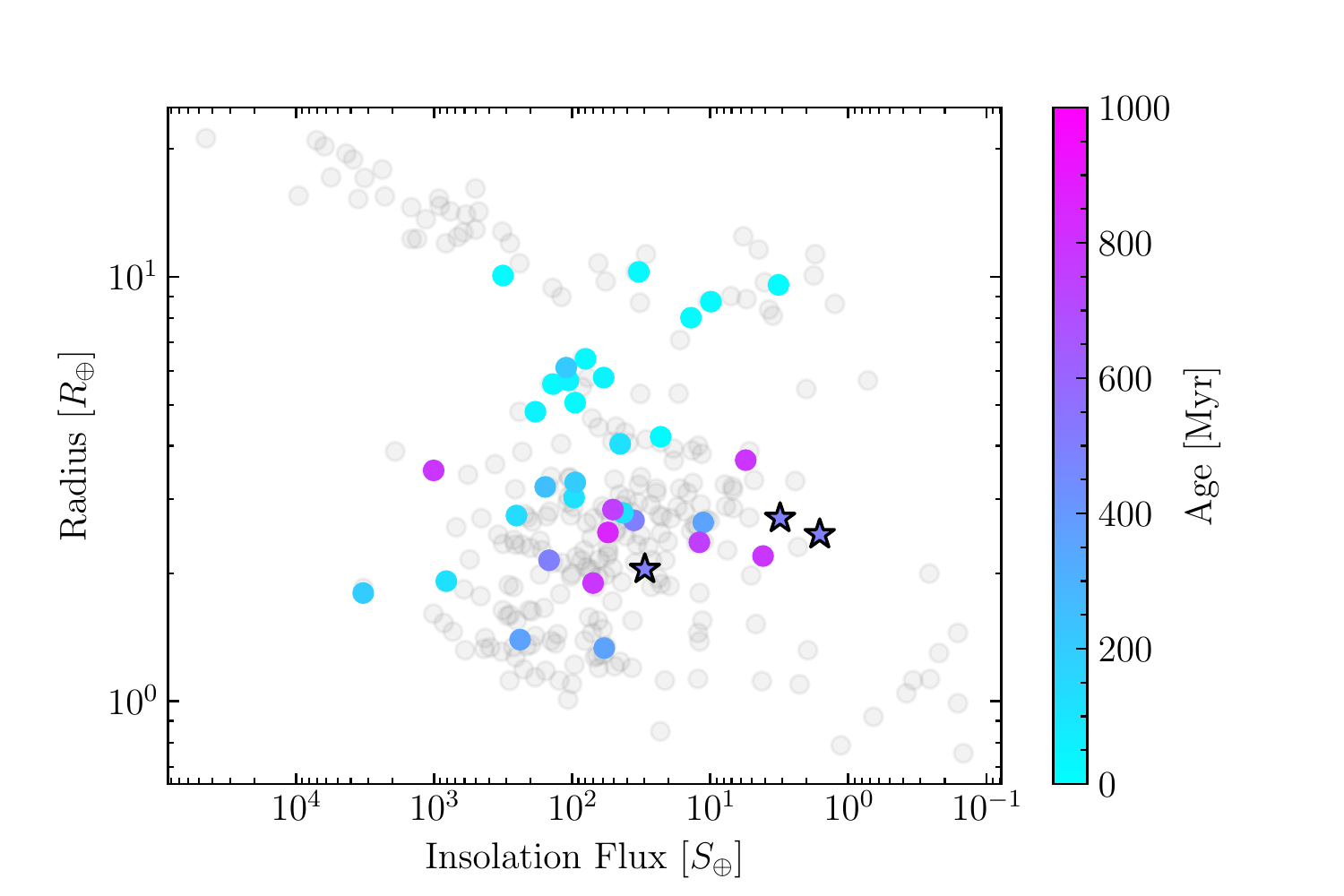}
    \caption{Young planets ($\leq$1 Gyr) shown with insolation flux received from host star, in terms of Earth's insolation flux. The planetary radii for the \thisstar system are smaller than most of the youngest planets, though there are no young planets with comparable insolation fluxes.}
    \label{fig:isoage}
\end{figure}

Previous investigations of young planets have suggested that they tend have larger radii than their older counterparts \citep[e.g.,][]{Mann2017,david:2021}, which is qualitatively consistent with atmospheric mass loss, but the details of the process and the timescale over which it occurs are not precisely known. Moreover, most known young planets have relatively short orbital periods for which photoevaporation is likely an important process \citep[e.g.,][]{owen:2013, jin:2018}, so whether a mechanism like core-powered mass loss \citep{ginzburg2018} drives significant atmospheric evolution in temperate planets is an open question. We illustrate in \rffigl{isoage} that the radii of the \thisstar\ planets are smaller than the youngest planets, but there are no young planets with insolation fluxes similar to the outer \thisstar\ planets.

For \thisstar, it is unclear if the planets are small because they have experienced significant atmospheric loss,
or because they simply formed smaller than the young planets discovered to date. The first possibility was discussed in \rfsecl{age}; if they are older than 1 Gyr, their small sizes may not be a powerful constraint. On the other hand, the outer planets should not experience significant mass loss from photoevaporation, and \citet{david:2021} show evidence for evolution in the radius gap as late as 2 to 3 Gyr, so even if the planets are $\mysim 1$\ Gyr old, core-powered mass loss may operate on long enough timescales that they can provide constraints on the process, particularly at long periods where fewer planets are known. Even at short periods, where the radius valley is well characterized, \thisstarb\ is of note. At $R_p = \bradius$\ \rearth, it sits near the edge of the radius valley for its size and incident flux. If the timescale for core-powered mass loss is as long as 2 Gyr or more, it could be losing atmosphere now and ultimately be destined to become a rocky world. Future advances in understanding K dwarf gyrochronology and activity may provide a more precise age and a more definitive conclusion about the evolution of these planets.

If \thisstar\ \textit{is} as young as gyrochronology suggests, then the small planet sizes are more likely to reflect the formation process rather than (or in addition to) the mass loss history. In particular, \thisstarc\ and d orbit at greater distances from their host star than other known young planets, and may differ in composition. Under the assumption that mini Neptunes form in situ, local conditions in the protoplanetary disk during formation would differ for the long-period \thisstar\ planets, the outer of which resides at the edge of the habitable zone. More data, such as transmission spectroscopy, could shed light on the compositional differences between hot and temperate planets. While the brightest examples of such planets are statistically likely to be old, adolescent examples like \thisstarc\ and d, for which core-powered mass loss may not have had time to strip significant atmosphere, could provide complementary information about formation by probing primordial compositions. 

\subsection{Future characterization of \thisstar}

As discussed above, young planets---including those orbiting \thisstar---open avenues for observational constraints on the dynamical and physical evolution of planetary systems. While young stars can be challenging to observe, future observations of the \thisstar\ system remain promising. 

Our modest precision RVs suggest some variation beyond the internal uncertainty estimates, which is consistent with expectations for an adolescent star with rotating surface features. Given the rotation ($\vsini=2$\,\kms) and the amplitude of photometric modulation ($\mysim 1$\%), we estimate there should be tens of meters per second RV variation from stellar surface features \citep[see, e.g.,][]{Saar1997}. Since the predicted RV amplitudes induced by the planets are only a few meters per second (see \rftabl{exofast}), attempts to measure the planet masses with stabilized spectrographs would be difficult without significant advances in stellar activity modeling. 

Dynamical interactions between the planets, if detected via transit timing variations or transit duration variations, could help determine their masses and eccentricities. There is currently no evidence for transit timing variations, but \thisstar\ will be observed by \tess\ in future extended missions, providing longer baselines over which to measure TTVs. Moreover, the signal-to-noise ratio (SNR) of the TOI-712.05 transit should increase if it is a real companion, and we can therefore more confidently confirm or refute its existence.  Similarly, additional transiting planets in the system may become apparent with the inclusion of more data. In particular, we showed in \rfsecl{hz} that there are dynamically stable orbits for small planets in the habitable zone of \thisstar. We also know that systems of small planets tend to be packed, so it would not be surprising to find an additional planet between \thisstarb\ and c ($10\lesssim P \lesssim 51$\ days). While planets larger than \thisstarb\ ($2\,\re$) would have likely been detected already, smaller planets may be revealed in future \tess\ data. Better knowledge of the system architecture and improved measurements of orbital properties will make dynamical modeling more powerful, which also feeds back into improved planetary properties.

\begin{figure}[t!]
    \centering
    \includegraphics[width = 1.05\linewidth]{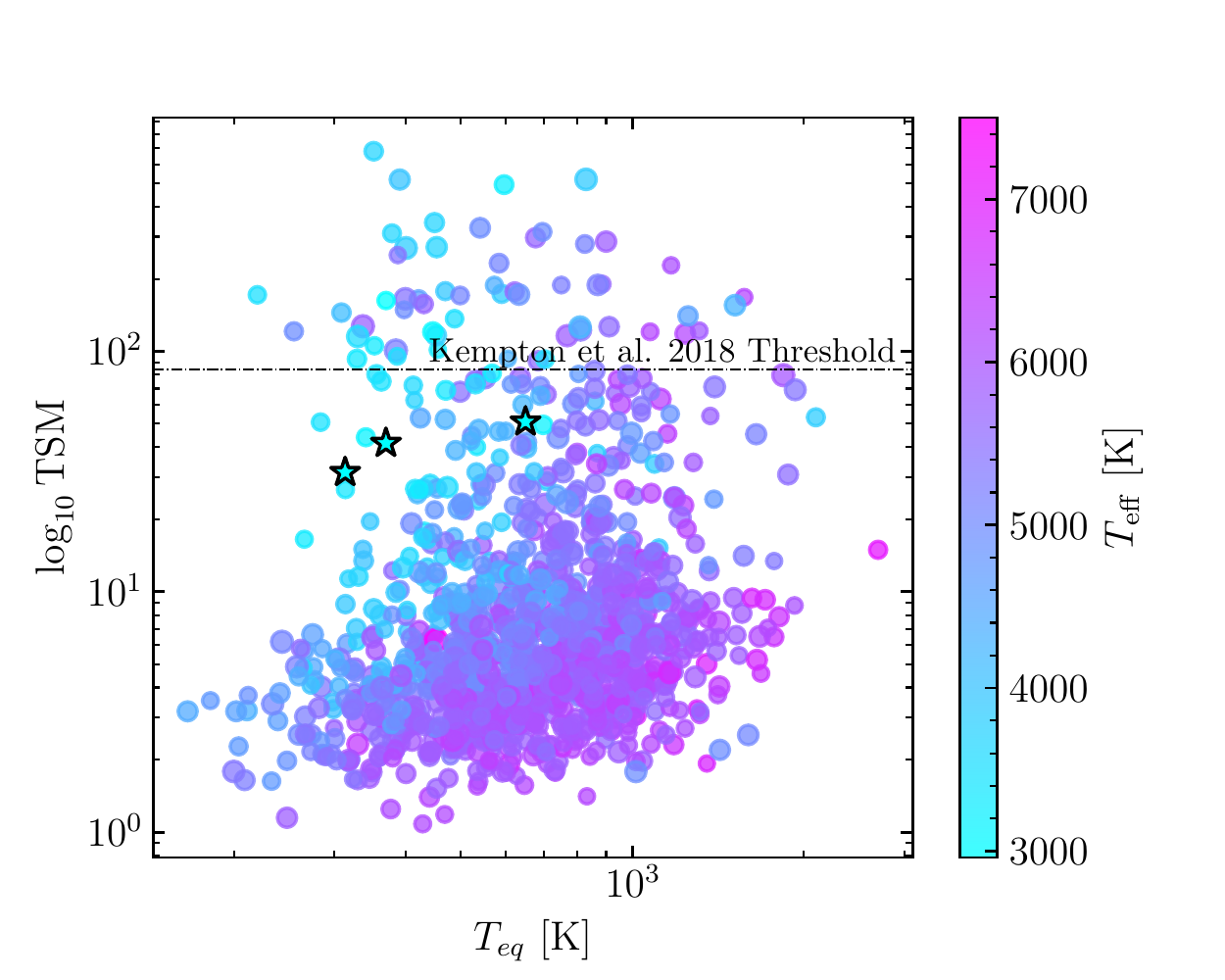}
    \caption{Calculated Transmission Spectroscopy Metric (TSM) for transiting, long period, mini-Neptunes from NASA's Exoplanet Archive (as of 29 October 2021). The log of the TSM is plotted as a function of equilibrium temperature, $T_{eq}$. Each planet is colored based on the effective temperature of the host star, $T_{\rm eff}$. The dashed lines denote the suggested JWST TSM cuttoff for mini-Neptunes as described in \citet{kempton2018}. We show the calculated TSM for planets ranging between $2.0 \leq R_P \leq 4.0\,R_\oplus$. \thisstarb, \thisstarc, and \thisstard are shown by the starred points, with a black border. There are very few temperate planets with a high TSM, so while \thisstarc and \thisstard fall below the suggested threshold, they are promising targets for studying atmospheric evolution in the absence of strong irradiation.}
    \label{fig:tsm}
\end{figure}

As discussed in \rfsecl{knownplanets}, the atmospheric composition of the outer planets could inform our understanding of the formation of temperate mini-Neptunes. We note that \thisstar\ is located $1.4^\circ$ from the Southern Ecliptic Pole, in the JWST continuous viewing zone, and it stands out as one of the brightest stars hosting long-period transiting planets (\rffigl{brightness}). The star is relatively small, and as a result, transmission spectroscopy of the planets may be possible. Following \citet{kempton2018}, we calculate the Transmission Spectroscopy Metric (TSM) of the planets to assess the prospect for transmission spectroscopy with JWST. We find ${\rm TSM}_b = 53.22$, ${\rm TSM}_c = 41.35$, and ${\rm TSM}_d = 31.88$. Based on expected yields from \tess, \citet{kempton2018} suggested that a survey with JWST should adopt a TSM threshold of 92 for planets of this size. However, this threshold is purely based on the SNR and ignores the outsized scientific value of some individual planets that happen to have lower expected SNRs. \thisstarc\ and d are examples of planets that would not meet the cutoff for such a survey, but may still be worth observing. \rffigl{tsm} shows the TSM values of transiting mini-Neptunes ($2.0 \leq R_P \leq 4.0 R_\oplus$) as a function of equilibrium temperature and stellar effective temperature. It is clear that although the planets lie below the TSM threshold for the predicted \tess\ yield, they stand out among the real sample as two of the best temperate mini-Neptunes. When also considering the age of the system, these planets become more scientifically valuable and may form the basis for a comparative study of temperate planets as a function of age.

\section{Conclusion}
Young planetary systems offer a glimpse into planetary evolution, most of which occurs within the first billion years after formation. In this paper, we presented the discovery of three planets transiting the adolescent K dwarf \thisstar. We determine the stellar age from the \tess\ rotation period, but because adolescent K dwarfs experience stalled spin-down, can only determine its age to be in the range of $500$\,Myr to $1.1$\,Gyr. There is no evidence that it is a part of a moving group, so we conclude that \thisstar is an adolescent field star.

Our global analysis determined precise planetary radii ($R_b = \bradiuse\,R_\oplus$, $R_c = \cradiuse\,R_\oplus$, $R_d = \dradiuse\,R_\oplus$), indicating that all three planets likely possess significant volatile envelopes. We identify an additional 4.32-day rocky planet candidate (\thisstar.05; $R_P = 0.81 R_\oplus$), but do not claim it as a validated planet due to its low SNR. \thisstard orbits at the inner edge of the optimistic habitable zone, making it the first such planet orbiting a young star. As a result of their cool temperatures and relatively bright host star, \thisstarc\ and d represent promising targets to study the evolution of temperate small planets, for which thermal mass loss cannot shape planetary atmospheres. 

Dynamical simulations of the system reveal it to be stable and find that small planets in the conservative habitable zone could also survive. Future \tess\ observations may be sensitive to such objects, if they exist, and can confirm or refute the small inner rocky planet candidate.

\facilities{\tess, LCOGT, ASTEP, Hazelwood}

\software{\texttt{AstroImageJ} \citep{Collins:2017}, \texttt{astropy} \citep{astropy}, \texttt{batman} \citep{batman} \texttt{EXOFASTv2} \citep{Eastman:2019}, \texttt{exoplanet} \citep{exoplanet}, \texttt{Matplotlib} \citep{matplotlib}, \texttt{Numpy} \citep{numpy}, \texttt{TAPIR} \citep{Jensen:2013}}

\section*{Acknowledgements}
%
%

We thank Adina Feinstein for helpful discussion regarding stellar flares in \tess\ data. 

SV and SNQ acknowledge support from the TESS Guest Investigator Program G03268 and NASA grant number 80NSSC21K1056.
MNG acknowledges support from the European Space Agency (ESA) as an ESA Research Fellow.

This work makes use of observations from the LCOGT network. Part of the LCOGT telescope time was granted by NOIRLab through the Mid-Scale Innovations Program (MSIP). MSIP is funded by NSF.

Some of the observations in the paper made use of the High-Resolution Imaging instrument Zorro obtained under Gemini LLP Proposal Number: GN/S-2021A-LP-105. Zorro was funded by the NASA Exoplanet Exploration Program and built at the NASA Ames Research Center by Steve B. Howell, Nic Scott, Elliott P. Horch, and Emmett Quigley. Zorro was mounted on the Gemini North (and/or South) telescope of the international Gemini Observatory, a program of NSF’s OIR Lab, which is managed by the Association of Universities for Research in Astronomy (AURA) under a cooperative agreement with the National Science Foundation. on behalf of the Gemini partnership: the National Science Foundation (United States), National Research Council (Canada), Agencia Nacional de Investigación y Desarrollo (Chile), Ministerio de Ciencia, Tecnología e Innovación (Argentina), Ministério da Ciência, Tecnologia, Inovaçõese Comunicações (Brazil), and Korea Astronomy and Space Science Institute (Republic of Korea).

This research received funding from the European Research Council (ERC) under the European Union's Horizon 2020 research and innovation programme (grant agreement n$^\circ$ 803193/BEBOP), and from the Science and Technology Facilities Council (STFC; grant n$^\circ$ ST/S00193X/1). 

This work makes use of observations from the ASTEP telescope. ASTEP benefited from the support of the French and Italian polar agencies IPEV and PNRA in the framework of the Concordia station program.

Funding for
the TESS mission is provided by NASA’s Science Mission directorate. We acknowledge the use of TESS public data from pipelines
at the TESS Science Office and at the TESS Science Processing Operations Center. Resources supporting this work were provided by
the NASA High-End Computing (HEC) Program through the NASA
Advanced Supercomputing (NAS) Division at Ames Research Center for the production of the SPOC data products. This research has
made use of the Exoplanet Follow-up Observation Program website,
which is operated by the California Institute of Technology, under
contract with the National Aeronautics and Space Administration
under the Exoplanet Exploration Program. This paper includes data
collected by the TESS mission, which are publicly available from
the Mikulski Archive for Space Telescopes (MAST). 

This work has made use of data from the European Space Agency (ESA) mission
Gaia (https://www.cosmos.esa.int/gaia), processed by the
Gaia Data Processing and Analysis Consortium (DPAC, https:
//www.cosmos.esa.int/web/gaia/dpac/consortium). Funding for the DPAC has been provided by national institutions, in particular the institutions participating in the Gaia Multilateral Agreement.
\bibliography{refs}{}
\bibliographystyle{aasjournal}

\end{document}

%% file: newcommands.tex
\newcommand{\bjdtdb}{\ensuremath{\rm {BJD_{TDB}}}}
\newcommand{\feh}{\ensuremath{\left[{\rm Fe}/{\rm H}\right]}}

\newcommand{\teff}{\ensuremath{T_{\rm eff}}\xspace}

\newcommand{\logg}{\ensuremath{\log g}}
\newcommand\vsini{\ifmmode{v\sin{i_\star}}\else $v\sin{i_\star}$\fi}
\newcommand{\vmac}{\ensuremath{v_{\rm mac}}}
\newcommand\sini{\ifmmode{\sin{i_\star}}\else $\sin{i_\star}$\fi}

\newcommand{\msun}{\ensuremath{\,M_\Sun}}
\newcommand{\rsun}{\ensuremath{\,R_\Sun}}
\newcommand{\lsun}{\ensuremath{\,L_\Sun}}

\newcommand{\rearth}{\ensuremath{\,R_{\rm \Earth}}\xspace}

\newcommand{\re}{\ensuremath{\,R_{\rm \Earth}}\xspace}
\newcommand{\me}{\ensuremath{\,M_{\rm \Earth}}\xspace}

\newcommand{\fave}{\langle F \rangle}
\newcommand{\fluxcgs}{10$^9$ erg s$^{-1}$ cm$^{-2}$}

\newcommand{\Kepler}{{\it Kepler}}
\newcommand{\kep}{{\it Kepler}}
\newcommand{\Ktwo}{{\it K2}}
\newcommand{\tess}{{\it TESS}}

\newcommand\mysim{\mathord{\sim}}
\newcommand{\kms}{\,km\,s$^{-1}$}
\newcommand{\ms}{\,m\,s$^{-1}$}

\newcommand{\thisstar}{TOI-712\xspace}

\newcommand{\thisstarb}{TOI-712\,b\xspace}
\newcommand{\thisstarc}{TOI-712\,c\xspace}
\newcommand{\thisstard}{TOI-712\,d\xspace}

\newcommand{\bradius}{\ensuremath{2.049}}
\newcommand{\bradiuse}{\ensuremath{2.049^{+0.12}_{-0.080}}}
\newcommand{\cradius}{\ensuremath{2.701}}
\newcommand{\cradiuse}{\ensuremath{2.701^{+0.092}_{-0.082}}}
\newcommand{\dradius}{\ensuremath{2.474}}
\newcommand{\dradiuse}{\ensuremath{2.474^{+0.090}_{-0.082}}}

\newcommand{\rffigl}[1]{Figure~\ref{fig:#1}}

\newcommand{\rfsecl}[1]{\mbox{Section \ref{sec:#1}}}

\newcommand{\rftabl}[1]{Table~\ref{tab:#1}}

%% file: affiliations.tex
\newcommand{\cfa}{Center for Astrophysics \textbar{} Harvard \& Smithsonian, 60 Garden Street, Cambridge, MA 02138, USA}

\newcommand{\utaustin}{Department of Astronomy, The University of Texas at Austin, Austin, TX 78712, USA}
\newcommand{\MIT}{Department of Physics and Kavli Institute for Astrophysics and Space Research, Massachusetts Institute of Technology, Cambridge, MA 02139, USA}
\newcommand{\eaps}{Department of Earth, Atmospheric, and Planetary Sciences, Massachusetts Institute of Technology, 77 Massachusetts Avenue, Cambridge, MA 02139, USA}
\newcommand{\mitaero}{Department of Aeronautics and Astronautics, Massachusetts Institute of Technology, 125 Massachusetts Avenue, Cambridge, MA 02139, USA}

\newcommand{\riverside}{Department of Earth Sciences, University of California,
Riverside, CA 92521, USA}
\newcommand{\usq}{Centre for Astrophysics, University of Southern Queensland, West Street, Toowoomba, QLD 4350, Australia}
\newcommand{\ames}{NASA Ames Research Center, Moffett Field, CA, 94035, USA}

\newcommand{\princeton}{Department of Astrophysical Sciences, Princeton University, 4 Ivy Lane, Princeton, NJ 08544, USA}

\newcommand{\gsfc}{Exoplanets and Stellar Astrophysics Laboratory, Code 667, NASA Goddard Space Flight Center, Greenbelt, MD 20771, USA}

\newcommand{\oca}{Universit\'e C\^ote d'Azur, Observatoire de la C\^ote d'Azur, CNRS, Laboratoire Lagrange, Bd de l'Observatoire, CS 34229, 06304 Nice cedex 4, France}

\newcommand{\esafellow}{\altaffiliation{ESA Research Fellow}}

%% file: authors.tex

\author[0000-0002-0786-7307]{Sydney Vach}
\affiliation{\cfa}

\author[0000-0002-8964-8377]{Samuel N. Quinn} 
\affiliation{\cfa}

\author[0000-0001-7246-5438]{Andrew Vanderburg} 
\affiliation{\MIT}

\author[0000-0002-7084-0529]{Stephen R. Kane} 
\affiliation{\riverside}

\author[0000-0001-6588-9574]{Karen A.\ Collins} 
\affiliation{\cfa}

\author[0000-0001-9811-568X]{Adam L. Kraus} 
\affiliation{\utaustin}

\author[0000-0002-4891-3517]{George Zhou} 
\affiliation{\usq}

\author[0000-0001-8726-3134]{Amber A. Medina}
\affiliation{\utaustin}



\author[0000-0001-8227-1020]{Richard P. Schwarz}
\affiliation{Patashnick Voorheesville Observatory, Voorheesville, NY 12186, USA}

\author[0000-0003-2781-3207]{Kevin I.\ Collins}
\affiliation{George Mason University, 4400 University Drive, Fairfax, VA, 22030 USA}

\author[0000-0003-2239-0567]{Dennis M.\ Conti}
\affiliation{American Association of Variable Star Observers, 49 Bay State Road, Cambridge, MA 02138, USA}

\author[0000-0003-2163-1437]{Chris Stockdale}
\affiliation{Hazelwood Observatory, Australia}

\author[0000-0001-8879-7138]{Bob Massey}
\affil{Villa '39 Observatory, Landers, CA 92285, USA}
 
\author{Olga Suarez}
\affiliation{\oca}

\author[0000-0002-7188-8428]{Tristan Guillot}
\affiliation{\oca}

\author[0000-0001-5000-7292]{Djamel Mekarnia}
\affiliation{\oca}

\author[0000-0002-0856-4527]{Lyu Abe}
\affiliation{\oca}

\author[0000-0002-3937-630X]{Georgina Dransfield}
\affiliation{School of Physics \& Astronomy, University of Birmingham, Edgbaston, Birmingham, B15 2TT, UK}

\author[0000-0001-7866-8738]{Nicolas Crouzet}
\affiliation{European Space Agency (ESA), European Space Research and Technology Centre (ESTEC), Keplerlaan 1, 2201 AZ Noordwijk, The Netherlands}

\author[0000-0002-5510-8751]{Amaury H. M. J. Triaud}
\affiliation{School of Physics \& Astronomy, University of Birmingham, Edgbaston, Birmingham, B15 2TT, UK}

\author[0000-0003-3914-3546]{François-Xavier Schmider}
\affiliation{\oca}

\author{Abelkrim Agabi}
\affiliation{\oca}


\author{Marco Buttu}
\affiliation{ INAF Osservatorio Astronomico di Cagliari, Via della Scienza 5 - 09047 Selargius CA, Italy}

\author[0000-0001-9800-6248]{Elise Furlan}
\affiliation{NASA Exoplanet Science Institute, Caltech/IPAC, Mail Code 100-22, 1200 E. California Blvd., Pasadena, CA 91125, USA}
 
\author[0000-0003-2519-6161]{Crystal~L.~Gnilka}
\affiliation{NASA Exoplanet Science Institute, Caltech/IPAC, Mail Code 100-22, 1200 E. California Blvd., Pasadena, CA 91125, USA}
\affil{\ames}

\author[0000-0002-2532-2853]{Steve B.\ Howell} 
\affiliation{\ames}

\author[0000-0002-0619-7639]{Carl Ziegler}
\affiliation{Department of Physics, Engineering and Astronomy, Stephen F. Austin State University, 1936 North St, Nacogdoches, TX 75962, USA}

\author[0000-0001-7124-4094]{C\'{e}sar Brice\~{n}o}
\affiliation{Cerro Tololo Inter-American Observatory/NSF’s NOIRLab, Casilla 603, La Serena, Chile}

\author{Nicholas Law}
\affiliation{Department of Physics and Astronomy, The University of North Carolina at Chapel Hill, Chapel Hill, NC 27599-3255, USA}

\author[0000-0003-3654-1602]{Andrew W. Mann}
\affiliation{Department of Physics and Astronomy, The University of North Carolina at Chapel Hill, Chapel Hill, NC 27599-3255, USA}


\author{Alexander~Rudat}
\affiliation{\MIT}

\author{Knicole~D.~Colon}
\affiliation{\gsfc}

\author[0000-0003-4724-745X]{Mark E. Rose}
\affiliation{\ames}

\author{Michelle Kunimoto}
\affiliation{\MIT}

\author[0000-0002-3164-9086]{Maximilian N. Günther}
\esafellow
\affiliation{European Space Agency (ESA), European Space Research and Technology Centre (ESTEC), Keplerlaan 1, 2201 AZ Noordwijk, The Netherlands}

\author[0000-0002-9003-484X]{David Charbonneau}
\affiliation{\cfa}

\author[0000-0002-5741-3047]{David R. Ciardi}
\affiliation{NASA Exoplanet Science Institute, Caltech/IPAC, Mail Code 100-22, 1200 E. California Blvd., Pasadena, CA 91125, USA}



\author[0000-0003-2058-6662]{George R. Ricker}
\affiliation{\MIT}

\author[0000-0001-6763-6562]{Roland K. Vanderspek}
\affiliation{\MIT}

\author[0000-0001-9911-7388]{David W. Latham}
\affiliation{\cfa}

\author{Sara Seager}
\affiliation{\MIT}
\affiliation{\eaps}
\affiliation{\mitaero}

\author[0000-0002-4265-047X]{Joshua N. Winn}
\affiliation{\princeton}

\author[0000-0002-4715-9460]{Jon M. Jenkins}
\affiliation{\ames}

%% file: toi712.lco.tex
\begin{deluxetable}{llcl}
\label{tab:SG1}
\tablewidth{\textwidth}
\tablecaption{Ground-based transit observations of \thisstar}
\tablehead{\colhead{Date (UT)~~~} &
           \colhead{Telescope~~~} &
           \colhead{Filter~~~}&
           \colhead{Coverage~~~}
           }
\startdata
\multicolumn{4}{l}{\textbf{\thisstarb}} \\
2019-09-11 & LCO-SSO-1.0m & Zs & Egress\\
2019-12-15 & LCO-SAAO-1.0m & Zs & Full\\
2020-01-04 & LCO-SAAO-1.0m & Zs & Full\\
2020-01-23 & LCO-CTIO-1.0m & B & Ingress\\
2020-10-15 & LCO-SAAO-1.0m & B & Full\\
2020-11-13 & LCO-SSO-1.0m & Zs & Full\\
2020-12-12 & LCO-CTIO-1.0m & ip & Full \\\hline
\multicolumn{4}{l}{\textbf{\thisstarc}} \\
2019-12-25$^\dagger$ & LCO-SSO-1.0m & Zs & Egress\\
2019-12-25$^\dagger$ & Hazelwood-0.32m & Rc & Ingress\\
2020-02-15 & LCO-CTIO-1.0m & Zs & Ingress\\
2020-05-28 & ASTEP-0.4m & $\sim$Rc & Full\\
2020-09-09 & LCO-SAAO-1.0m & Zs & Egress\\
2021-04-03 & ASTEP-0.4m & $\sim$Rc & Ingress\\
2021-05-25 & ASTEP-0.4m & $\sim$Rc & Full\\
2021-09-05 & ASTEP-0.4m & $\sim$Rc & Ingress\\
\enddata
\footnotesize{ 
\justifying
\vspace{6pt}
\noindent \textbf{Note.} \noindent $^\dagger$\ indicates an observation that was not used in the global fit. \\[1ex]
The LCOGT filter designations Zs and ip indicate Pan-STARRS $z$-short band and Sloan $i'$ band, respectively. The LCOGT 1\,m telescope nodes used are at Siding Spring Observatory (SSO), Cerro Tololo Inter-American Observatory (CTIO), and South Africa Astronomical Observatory (SAAO). See text regarding the ASTEP observation band.}
\vspace{-16pt}
\end{deluxetable}

%% file: rvtable.tex
\begin{deluxetable}{lcc}
\label{tab:rvs}
\tablewidth{\textwidth}
\tablecaption{CHIRON radial velocities of \thisstar}
\tablehead{\colhead{~~~~~~~${\rm BJD_{TDB}}$~~~~~~~} &
           \colhead{~~~~~~~~~~~~RV~~~~~~~~~~~~} &
           \colhead{~~~~~~~~~~~~$\sigma_{\rm RV}$~~~~~~~~~~~~} \\[-1ex]
           \colhead{} &
           \colhead{(\kms)} &
           \colhead{(\kms)}
           }
\startdata
2458746.897310 & 3.822 & 0.064 \\
2458787.812670 & 3.967 & 0.066 \\
2459314.531560 & 3.953 & 0.036 \\
\enddata
\footnotesize{ 
\justifying
\vspace{6pt}
\noindent \textbf{Note.} The CHIRON radial velocities show no significant variation that might have been indicative of a bound stellar companion.}
\vspace{-16pt}
\end{deluxetable}

%% file: stellar.tex
\begin{table}
\scriptsize
\centering
\caption{Literature Properties for \thisstar}
\begin{tabular*}{\columnwidth}{l @{\extracolsep{\fill}} lcr}
  \hline
  \hline
 & \multicolumn{3}{c}{TIC 150151262} \\
Other & \multicolumn{3}{c}{TYC 8905-00810-1} \\
identifiers	  & \multicolumn{3}{c}{2MASS J06114467-6549335} \\
 & \multicolumn{3}{c}{WISE J061144.66-654933.1} \\
\hline
\hline
Parameter & Description & Value & Source\\
\hline 
$\alpha_{J2000}$	&R.A. & 06:11:44.6729007687& 1	\\
$\delta_{J2000}$	&decl. & -65:49:33.499607825& 1	\\
\\
$T$	& \tess\ $T$ mag	& 9.9059 $\pm$ 0.006 & 6	\\
\\
$B_T$	& Tycho $B_T$ mag & 12.10 $\pm$ 0.20		& 2	\\
$V_T$	& Tycho $V_T$ mag & 10.84 $\pm$ 0.07		& 2	\\
$B$\tablenotemark{a}		
        & APASS Johnson $B$ mag	& 11.701 $\pm$	0.025& 3	\\
$V$		& APASS Johnson $V$ mag	& 10.892 $\pm$	0.016& 3	\\
$G$     & {\it Gaia} $G$ mag    & 10.5785 $\pm$ 0.0006    & 1\\
$g'$	& APASS Sloan $g'$ mag	& 11.622 $\pm$ **0.000	& 3	\\
$r'$	& APASS Sloan $r'$ mag	& 10.526 $\pm$ **0.000	& 3	\\
$i'$	& APASS Sloan $i'$ mag	& 10.182 $\pm$ 0.009 & 3	\\
\\
$J$		& 2MASS $J$ mag & 8.984  $\pm$ 0.022	& 4	\\
$H$		& 2MASS $H$ mag & 8.409 $\pm$ 0.042	    & 4	\\
$K_S$	& 2MASS $K_S$ mag & 8.313 $\pm$ 0.023 & 4	\\
\\
\textit{WISE1}		& \textit{WISE1} mag & 8.233 $\pm$ 0.023 	& 5	\\
\textit{WISE2}		& \textit{WISE2} mag & 8.293 $\pm$ 0.020 	& 5 \\
\textit{WISE3}		& \textit{WISE3} mag &8.219 $\pm$ 0.019 	& 5	\\
\textit{WISE4}		& \textit{WISE4} mag & 8.241 $\pm$ 0.127 	& 5	\\
\\
$\mu_{\alpha}$		& PM in R.A. (mas yr$^{-1}$)  & -2.929 $\pm$ 0.047	& 1 \\
$\mu_{\delta}$		& PM in decl. (mas yr$^{-1}$)   	&  31.003 $\pm$ 0.044 &  1 \\
$\pi$ & Parallax (mas)  & 17.0297 $\pm$ 0.0230 & 1 \\
$RV$ & Systemic RV (\kms)   & 3.16 $\pm$ 0.34 & 1 \\
\hline
\\[-6ex]
\end{tabular*}
\begin{flushleft} 
\footnotesize{\vspace{6pt}
    {\bf References.} (1) \citet{GaiaDR22018}; (2) \citet{Hog:2000}; (3) \citet{Henden:2016}; (4) \citet{Cutri:2003}; (5) \citet{Cutri:2014}; (6) \citet{stassun:2018b}
}
\end{flushleft}
\label{tab:stellar}
\end{table}

%% file: toi712_ef2.tex
\begin{table*}
\scriptsize
\setlength{\tabcolsep}{2pt}
\centering
\caption{\thisstar\ planetary and transit parameters: median values and 68\% confidence intervals}
\begin{tabular*}{\textwidth}{l @{\extracolsep{\fill}} lccc}
  \hline
  \hline
Parameter & Description (Units) & \multicolumn{3}{c}{Values} \\
\hline
\multicolumn{2}{l}{\textbf{Stellar Parameters}}\smallskip\\
~~~~$M_*$\dotfill &Mass (\msun)\dotfill &$0.732^{+0.027}_{-0.025}$\\
~~~~$R_*$\dotfill &Radius (\rsun)\dotfill &$0.674^{+0.017}_{-0.016}$\\
~~~~$L_*$\dotfill &Luminosity (\lsun)\dotfill &$0.1871\pm0.0061$\\
~~~~$\rho_*$\dotfill &Density (cgs)\dotfill &$3.36^{+0.24}_{-0.22}$\\
~~~~$\log{g}$\dotfill &Surface gravity (cgs)\dotfill &$4.645^{+0.021}_{-0.022}$\\
~~~~$T_{\rm eff}$\dotfill &Effective Temperature (K)\dotfill &$4622^{+61}_{-59}$\\
~~~~$[{\rm Fe/H}]$\dotfill &Metallicity (dex)\dotfill &$-0.020^{+0.12}_{-0.043}$\\
~~~~$Age$\dotfill &Age (Gyr)\dotfill &$0.83^{+0.25}_{-0.28}$\\
~~~~$EEP$\dotfill &Equal Evolutionary Phase$^{1}$ \dotfill &$271.4^{+7.8}_{-13}$\\
~~~~$A_V$\dotfill &V-band extinction (mag)\dotfill &$0.083^{+0.049}_{-0.052}$\\
~~~~$\sigma_{SED}$\dotfill &SED photometry error scaling \dotfill &$1.11^{+0.40}_{-0.26}$\\
~~~~$\varpi$\dotfill &Parallax (mas)\dotfill &$17.059\pm0.032$\\
~~~~$d$\dotfill &Distance (pc)\dotfill &$58.62\pm0.11$\\
[1ex]
\hline
\multicolumn{2}{l}{\textbf{Planetary Parameters}} & b & c & d \smallskip\\
~~~~$P$\dotfill &Period (days)\dotfill &$9.531361^{+0.000018}_{-0.000017}$&$51.69906\pm0.00017$&$84.83960^{+0.00043}_{-0.00040}$\\
~~~~$R_P$\dotfill &Radius (\re)\dotfill &$2.049^{+0.12}_{-0.080}$&$2.701^{+0.092}_{-0.082}$&$2.474^{+0.090}_{-0.082}$\\
~~~~$R_P/R_*$\dotfill &Radius of planet in stellar radii \dotfill &$0.02772^{+0.0018}_{-0.00076}$&$0.03669^{+0.00074}_{-0.00059}$&$0.03362^{+0.00079}_{-0.00074}$\\
~~~~$\delta$\dotfill &Transit depth (fraction) \dotfill &$0.000768^{+0.00010}_{-0.000042}$&$0.001346^{+0.000055}_{-0.000043}$&$0.001130^{+0.000053}_{-0.000049}$\\
~~~~$T_C$\dotfill &Time of conjunction$^{2}$ (\bjdtdb)\dotfill &$1385.5457^{+0.0012}_{-0.0010}$&$1429.4054^{+0.0019}_{-0.0017}$&$1470.8665^{+0.0023}_{-0.0025}$\\
~~~~$T_0$\dotfill &Optimal conjunction Time$^{3}$ (\bjdtdb)\dotfill &$1757.26886^{+0.00068}_{-0.00059}$&$1946.39598^{+0.00093}_{-0.00082}$&$1640.5457\pm0.0019$\\
~~~~$T_P$\dotfill &Time of Periastron (\bjdtdb)\dotfill &$1385.49^{+0.43}_{-0.35}$&$1425^{+11}_{-15}$&$1396^{+32}_{-22}$\\
~~~~$T_S$\dotfill &Time of eclipse (\bjdtdb)\dotfill &$1380.9^{+3.5}_{-3.6}$&$1403.6\pm3.2$&$1513.3^{+3.9}_{-3.8}$\\
~~~~$a$\dotfill &Semi-major axis (AU)\dotfill &$0.07928^{+0.00096}_{-0.00091}$&$0.2447^{+0.0030}_{-0.0028}$&$0.3405^{+0.0041}_{-0.0039}$\\
~~~~$i$\dotfill &Inclination (Degrees)\dotfill &$88.22^{+1.1}_{-0.53}$&$89.78^{+0.14}_{-0.11}$&$89.817^{+0.11}_{-0.066}$\\
~~~~$b$\dotfill &Transit Impact parameter \dotfill &$0.41^{+0.32}_{-0.28}$&$0.28^{+0.16}_{-0.19}$&$0.34^{+0.14}_{-0.21}$\\
~~~~$e$\dotfill &Eccentricity \dotfill &$0.54^{+0.26}_{-0.20}$&$0.089^{+0.083}_{-0.056}$&$0.073^{+0.064}_{-0.049}$\\
~~~~$\omega_*$\dotfill &Argument of Periastron (Degrees)\dotfill &$68^{+88}_{-77}$&$115^{+96}_{-95}$&$30^{+110}_{-140}$\\
~~~~$e\cos{\omega_*}$\dotfill & \dotfill &$0.02^{+0.61}_{-0.62}$&$0.001\pm0.097$&$-0.000^{+0.072}_{-0.070}$\\
~~~~$e\sin{\omega_*}$\dotfill & \dotfill &$0.21^{+0.21}_{-0.33}$&$0.029^{+0.046}_{-0.056}$&$0.011\pm0.054$\\
~~~~$\tau$\dotfill &Ingress/egress transit duration (minutes)\dotfill &$3.33^{+2.88}_{-0.56}$&$11.12^{+1.87}_{-0.86}$&$12.5^{+2.0}_{-1.3}$\\
~~~~$T_{14}$\dotfill &Total transit duration (hours)\dotfill &$1.735^{+0.046}_{-0.031}$&$4.843^{+0.058}_{-0.062}$&$5.702^{+0.098}_{-0.094}$\\
~~~~$T_{FWHM}$\dotfill &FWHM transit duration (hours)\dotfill &$1.670^{+0.026}_{-0.024}$&$4.649^{+0.060}_{-0.062}$&$5.486^{+0.101}_{-0.096}$\\
~~~~$\fave$\dotfill &Incident Flux (\fluxcgs)\dotfill &$0.0307^{+0.0056}_{-0.0074}$&$0.00419\pm0.00017$&$0.002176^{+0.000083}_{-0.000080}$\\
~~~~$T_{eq}$\dotfill &Equilibrium temperature$^{4}$ (K)\dotfill &$650.0^{+6.0}_{-5.9}$&$369.9^{+3.4}_{-3.3}$&$313.6^{+2.9}_{-2.8}$\\
~~~~$M_P$\dotfill &Predicted mass$^{5}$ (\me)\dotfill &$5.6^{+2.0}_{-1.3}$&$8.7^{+3.1}_{-1.9}$&$7.5^{+2.7}_{-1.7}$\\
~~~~$M_P/M_*$\dotfill &Predicted mass ratio$^{5}$ \dotfill &$0.0000231^{+0.0000082}_{-0.0000055}$&$0.0000357^{+0.000013}_{-0.0000080}$&$0.0000310^{+0.000011}_{-0.0000072}$\\
~~~~$K$\dotfill &Predicted RV semi-amplitude$^{5}$ (m/s)\dotfill &$2.65^{+1.3}_{-0.72}$&$1.85^{+0.65}_{-0.41}$&$1.36^{+0.48}_{-0.31}$\\
~~~~$\rho_P$\dotfill &Predicted density$^{5}$ (cgs)\dotfill &$3.53^{+1.2}_{-0.84}$&$2.42^{+0.83}_{-0.55}$&$2.73^{+0.95}_{-0.61}$\\
~~~~$logg_P$\dotfill &Predicted surface gravity$^{5}$ \dotfill &$3.11^{+0.13}_{-0.11}$&$3.07^{+0.13}_{-0.11}$&$3.08^{+0.13}_{-0.11}$\\
[1ex]
\hline
\multicolumn{2}{l}{\textbf{Wavelength Parameters}}&TESS\smallskip\\
~~~~$u_{1}$\dotfill &linear limb-darkening coeff \dotfill &$0.463^{+0.043}_{-0.047}$\\
~~~~$u_{2}$\dotfill &quadratic limb-darkening coeff \dotfill &$0.163^{+0.049}_{-0.046}$\\
[1ex]
\hline
\multicolumn{2}{l}{\textbf{Transit Parameters}}&\tess\smallskip\\ 
~~~~$\sigma^{2}$\dotfill &Added Variance \dotfill &$3.29\times10^{-7}\pm0.78\times10^{-7}$\\
~~~~$F_0$\dotfill &Baseline flux \dotfill &$0.9999793\pm0.0000062$\\
[1ex]
\hline \\[-6ex]
\end{tabular*}
\begin{flushleft} 
\footnotesize{\vspace{6pt}{\bf Note---}\\[0.25ex]
$^1$\ Corresponds to static points in a star's evolutionary history. See \S2 in \citet{Dotter:2016}.\\[0.25ex]
$^2$\ Though the time of minimum projected separation is a more correct "transit time, time of conjunction is commonly reported as the transit time, and is in practice identical in all but the most extreme combinations of eccentricity and viewing angles \citep[see][]{Eastman:2019}.\\[0.25ex]
$^3$\ Optimal time of conjunction minimizes the covariance between $T_C$ and Period.\\[0.25ex]
$^4$\ Assumes no albedo and perfect redistribution, making this identical to a surface-averaged irradiation temperature. \\[0.25ex]
$^5$\ Uses measured radius and estimated mass from \citet{Chen:2017}.
}
\end{flushleft}
\label{tab:exofast}
\end{table*}

%% file: toi712.05.tex
\begin{table}
\scriptsize
\centering
\tablewidth{\textwidth}
\caption{Parameters of the candidate TOI-712.05}
\label{tab:toi712_05tab}
\begin{tabular*}{\columnwidth}{l @{\extracolsep{\fill}} l c}
\hline
\hline
Parameter & Description (Units) & Values \\
\hline
$P$  & Period (days) &$4.321298^{+0.000093}_{-0.000069}$\\
$R_P$ & Radius ($R_\oplus$) &$0.81\pm0.11$\\
$R_P / R_*$ & Radius of planet in stellar radii &$0.0110\pm0.0015$\\
$\left(R_P/R_*\right)^2$ & Planet-to-star area ratio &$0.000120^{+0.000035}_{-0.000031}$\\
$\delta_{\rm TESS}$ &Transit depth in TESS band &$0.000122\pm0.000022$\\
$T_C$ & Time of conjunction (BJD$_{\textrm{TDB}}$) &$1388.3796^{+0.0076}_{-0.012}$\\
$a$ & Semi-major axis (AU) &$0.04679^{+0.00041}_{-0.00040}$\\
$a/R_*$ & Semi-major axis in stellar radii &$14.90^{+0.31}_{-0.29}$\\
$i$ & Inclination (Degrees) &$87.09^{+1.1}_{-0.38}$\\
$b$ & Transit impact parameter &$0.757^{+0.093}_{-0.28}$\\
$e$ & Eccentricity & 0.0 (fixed) \\ 
$\omega_*$ & Argument of Periastron (Degrees) & 90.0 (fixed)\\
$T_{14}$ & Total transit duration (hours) &$1.49^{+0.48}_{-0.26}$\\
\hline
\\[-6ex]
\end{tabular*}
\end{table}

%% file: _TOI712.bbl
\begin{thebibliography}{}
\expandafter\ifx\csname natexlab\endcsname\relax\def\natexlab#1{#1}\fi
\providecommand{\url}[1]{\href{#1}{#1}}
\providecommand{\dodoi}[1]{doi:~\href{http://doi.org/#1}{\nolinkurl{#1}}}
\providecommand{\doeprint}[1]{\href{http://ascl.net/#1}{\nolinkurl{http://ascl.net/#1}}}
\providecommand{\doarXiv}[1]{\href{https://arxiv.org/abs/#1}{\nolinkurl{https://arxiv.org/abs/#1}}}

\bibitem[{{Abe} {et~al.}(2013){Abe}, {Gon{\c{c}}alves}, {Agabi}, {Alapini},
  {Guillot}, {M{\'e}karnia}, {Rivet}, {Schmider}, {Crouzet}, {Fortney}, {Pont},
  {Barbieri}, {Daban}, {Fante{\"\i}-Caujolle}, {Gouvret}, {Bresson}, {Roussel},
  {Bonhomme}, {Robini}, {Dugu{\'e}}, {Bondoux}, {P{\'e}ron}, {Petit},
  {Szul{\'a}gyi}, {Fruth}, {Erikson}, {Rauer}, {Fressin}, {Valbousquet},
  {Blanc}, {Le van Suu}, \& {Aigrain}}]{Abe:2013}
{Abe}, L., {Gon{\c{c}}alves}, I., {Agabi}, A., {et~al.} 2013, \aap, 553, A49,
  \dodoi{10.1051/0004-6361/201220351}

\bibitem[{{Ag{\"u}eros} {et~al.}(2018){Ag{\"u}eros}, {Bowsher}, {Bochanski},
  {Cargile}, {Covey}, {Douglas}, {Kraus}, {Kundert}, {Law}, {Ahmadi}, \&
  {Arce}}]{agueros2018}
{Ag{\"u}eros}, M.~A., {Bowsher}, E.~C., {Bochanski}, J.~J., {et~al.} 2018,
  \apj, 862, 33, \dodoi{10.3847/1538-4357/aac6ed}

\bibitem[{{Astropy Collaboration} {et~al.}(2013){Astropy Collaboration},
  {Robitaille}, {Tollerud}, {Greenfield}, {Droettboom}, {Bray}, {Aldcroft},
  {Davis}, {Ginsburg}, {Price-Whelan}, {Kerzendorf}, {Conley}, {Crighton},
  {Barbary}, {Muna}, {Ferguson}, {Grollier}, {Parikh}, {Nair}, {Unther},
  {Deil}, {Woillez}, {Conseil}, {Kramer}, {Turner}, {Singer}, {Fox}, {Weaver},
  {Zabalza}, {Edwards}, {Azalee Bostroem}, {Burke}, {Casey}, {Crawford},
  {Dencheva}, {Ely}, {Jenness}, {Labrie}, {Lim}, {Pierfederici}, {Pontzen},
  {Ptak}, {Refsdal}, {Servillat}, \& {Streicher}}]{astropy}
{Astropy Collaboration}, {Robitaille}, T.~P., {Tollerud}, E.~J., {et~al.} 2013,
  \aap, 558, A33, \dodoi{10.1051/0004-6361/201322068}

\bibitem[{{Ballard} \& {Johnson}(2016)}]{Ballard2016}
{Ballard}, S., \& {Johnson}, J.~A. 2016, \apj, 816, 66,
  \dodoi{10.3847/0004-637X/816/2/66}

\bibitem[{{Barnes}(2007)}]{Barnes2007}
{Barnes}, S.~A. 2007, \apj, 669, 1167, \dodoi{10.1086/519295}

\bibitem[{{Borucki} {et~al.}(2010){Borucki}, {Koch}, {Basri}, {Batalha},
  {Brown}, {Caldwell}, {Caldwell}, {Christensen-Dalsgaard}, {Cochran},
  {DeVore}, {Dunham}, {Dupree}, {Gautier}, {Geary}, {Gilliland}, {Gould},
  {Howell}, {Jenkins}, {Kondo}, {Latham}, {Marcy}, {Meibom}, {Kjeldsen},
  {Lissauer}, {Monet}, {Morrison}, {Sasselov}, {Tarter}, {Boss}, {Brownlee},
  {Owen}, {Buzasi}, {Charbonneau}, {Doyle}, {Fortney}, {Ford}, {Holman},
  {Seager}, {Steffen}, {Welsh}, {Rowe}, {Anderson}, {Buchhave}, {Ciardi},
  {Walkowicz}, {Sherry}, {Horch}, {Isaacson}, {Everett}, {Fischer}, {Torres},
  {Johnson}, {Endl}, {MacQueen}, {Bryson}, {Dotson}, {Haas}, {Kolodziejczak},
  {Van Cleve}, {Chandrasekaran}, {Twicken}, {Quintana}, {Clarke}, {Allen},
  {Li}, {Wu}, {Tenenbaum}, {Verner}, {Bruhweiler}, {Barnes}, \&
  {Prsa}}]{Kepler}
{Borucki}, W.~J., {Koch}, D., {Basri}, G., {et~al.} 2010, Science, 327, 977,
  \dodoi{10.1126/science.1185402}

\bibitem[{{Brice{\~n}o} {et~al.}(2007){Brice{\~n}o}, {Preibisch}, {Sherry},
  {Mamajek}, {Mathieu}, {Walter}, \& {Zinnecker}}]{briceno2007}
{Brice{\~n}o}, C., {Preibisch}, T., {Sherry}, W.~H., {et~al.} 2007, in
  Protostars and Planets V, ed. B.~{Reipurth}, D.~{Jewitt}, \& K.~{Keil}, 345.
\newblock \doarXiv{astro-ph/0602446}

\bibitem[{{Brown} {et~al.}(2013){Brown}, {Baliber}, {Bianco}, {Bowman},
  {Burleson}, {Conway}, {Crellin}, {Depagne}, {De Vera}, {Dilday}, {Dragomir},
  {Dubberley}, {Eastman}, {Elphick}, {Falarski}, {Foale}, {Ford}, {Fulton},
  {Garza}, {Gomez}, {Graham}, {Greene}, {Haldeman}, {Hawkins}, {Haworth},
  {Haynes}, {Hidas}, {Hjelstrom}, {Howell}, {Hygelund}, {Lister}, {Lobdill},
  {Martinez}, {Mullins}, {Norbury}, {Parrent}, {Paulson}, {Petry}, {Pickles},
  {Posner}, {Rosing}, {Ross}, {Sand}, {Saunders}, {Shobbrook}, {Shporer},
  {Street}, {Thomas}, {Tsapras}, {Tufts}, {Valenti}, {Vander Horst}, {Walker},
  {White}, \& {Willis}}]{Brown:2013}
{Brown}, T.~M., {Baliber}, N., {Bianco}, F.~B., {et~al.} 2013, Publications of
  the Astronomical Society of the Pacific, 125, 1031, \dodoi{10.1086/673168}

\bibitem[{{Buchhave} {et~al.}(2012){Buchhave}, {Latham}, {Johansen},
  {Bizzarro}, {Torres}, {Rowe}, {Batalha}, {Borucki}, {Brugamyer}, {Caldwell},
  {Bryson}, {Ciardi}, {Cochran}, {Endl}, {Esquerdo}, {Ford}, {Geary},
  {Gilliland}, {Hansen}, {Isaacson}, {Laird}, {Lucas}, {Marcy}, {Morse},
  {Robertson}, {Shporer}, {Stefanik}, {Still}, \& {Quinn}}]{Buchhave:2012}
{Buchhave}, L.~A., {Latham}, D.~W., {Johansen}, A., {et~al.} 2012, \nat, 486,
  375, \dodoi{10.1038/nature11121}

\bibitem[{{Carleo} {et~al.}(2021){Carleo}, {Desidera}, {Nardiello},
  {Malavolta}, {Lanza}, {Livingston}, {Locci}, {Marzari}, {Messina}, {Turrini},
  {Baratella}, {Borsa}, {D'Orazi}, {Nascimbeni}, {Pinamonti}, {Rainer}, {Alei},
  {Bignamini}, {Gratton}, {Micela}, {Montalto}, {Sozzetti}, {Squicciarini},
  {Affer}, {Benatti}, {Biazzo}, {Bonomo}, {Claudi}, {Cosentino}, {Covino},
  {Damasso}, {Esposito}, {Fiorenzano}, {Frustagli}, {Giacobbe}, {Harutyunyan},
  {Leto}, {Magazz{\`u}}, {Maggio}, {Mainella}, {Maldonado}, {Mallonn},
  {Mancini}, {Molinari}, {Molinaro}, {Pagano}, {Pedani}, {Piotto}, {Poretti},
  {Redfield}, \& {Scandariato}}]{Carleo2021}
{Carleo}, I., {Desidera}, S., {Nardiello}, D., {et~al.} 2021, \aap, 645, A71,
  \dodoi{10.1051/0004-6361/202039042}

\bibitem[{{Chambers}(1999)}]{chambers1999}
{Chambers}, J.~E. 1999, \mnras, 304, 793,
  \dodoi{10.1046/j.1365-8711.1999.02379.x}

\bibitem[{{Chambers} {et~al.}(1996){Chambers}, {Wetherill}, \&
  {Boss}}]{chambers1996}
{Chambers}, J.~E., {Wetherill}, G.~W., \& {Boss}, A.~P. 1996, \icarus, 119,
  261, \dodoi{10.1006/icar.1996.0019}

\bibitem[{{Chen} \& {Kipping}(2017)}]{Chen:2017}
{Chen}, J., \& {Kipping}, D. 2017, \apj, 834, 17,
  \dodoi{10.3847/1538-4357/834/1/17}

\bibitem[{{Choi} {et~al.}(2016){Choi}, {Dotter}, {Conroy}, {Cantiello},
  {Paxton}, \& {Johnson}}]{Choi2016}
{Choi}, J., {Dotter}, A., {Conroy}, C., {et~al.} 2016, \apj, 823, 102,
  \dodoi{10.3847/0004-637X/823/2/102}

\bibitem[{{Claret}(2017)}]{Claret2017}
{Claret}, A. 2017, \aap, 600, A30, \dodoi{10.1051/0004-6361/201629705}

\bibitem[{{Collier Cameron}(2007)}]{collier2007}
{Collier Cameron}, A. 2007, Astronomische Nachrichten, 328, 1030,
  \dodoi{10.1002/asna.200710880}

\bibitem[{{Collins}(2019)}]{collins:2019}
{Collins}, K. 2019, in American Astronomical Society Meeting Abstracts, Vol.
  233, American Astronomical Society Meeting Abstracts \#233, 140.05

\bibitem[{{Collins} {et~al.}(2017){Collins}, {Kielkopf}, {Stassun}, \&
  {Hessman}}]{Collins:2017}
{Collins}, K.~A., {Kielkopf}, J.~F., {Stassun}, K.~G., \& {Hessman}, F.~V.
  2017, \aj, 153, 77, \dodoi{10.3847/1538-3881/153/2/77}

\bibitem[{{Curtis} {et~al.}(2019{\natexlab{a}}){Curtis}, {Ag{\"u}eros},
  {Douglas}, \& {Meibom}}]{Curtis2019a}
{Curtis}, J.~L., {Ag{\"u}eros}, M.~A., {Douglas}, S.~T., \& {Meibom}, S.
  2019{\natexlab{a}}, \apj, 879, 49, \dodoi{10.3847/1538-4357/ab2393}

\bibitem[{{Curtis} {et~al.}(2019{\natexlab{b}}){Curtis}, {Ag{\"u}eros},
  {Mamajek}, {Wright}, \& {Cummings}}]{Curtis2019b}
{Curtis}, J.~L., {Ag{\"u}eros}, M.~A., {Mamajek}, E.~E., {Wright}, J.~T., \&
  {Cummings}, J.~D. 2019{\natexlab{b}}, \aj, 158, 77,
  \dodoi{10.3847/1538-3881/ab2899}

\bibitem[{{Curtis} {et~al.}(2020){Curtis}, {Ag{\"u}eros}, {Matt}, {Covey},
  {Douglas}, {Angus}, {Saar}, {Cody}, {Vanderburg}, {Law}, {Kraus}, {Latham},
  {Baranec}, {Riddle}, {Ziegler}, {Lund}, {Torres}, {Meibom}, {Aguirre}, \&
  {Wright}}]{curtis2020}
{Curtis}, J.~L., {Ag{\"u}eros}, M.~A., {Matt}, S.~P., {et~al.} 2020, \apj, 904,
  140, \dodoi{10.3847/1538-4357/abbf58}

\bibitem[{{Cutri} \& {et al.}(2014)}]{Cutri:2014}
{Cutri}, R.~M., \& {et al.} 2014, VizieR Online Data Catalog, 2328, 0

\bibitem[{{Cutri} {et~al.}(2003){Cutri}, {Skrutskie}, {van Dyk}, {Beichman},
  {Carpenter}, {Chester}, {Cambresy}, {Evans}, {Fowler}, {Gizis}, {Howard},
  {Huchra}, {Jarrett}, {Kopan}, {Kirkpatrick}, {Light}, {Marsh}, {McCallon},
  {Schneider}, {Stiening}, {Sykes}, {Weinberg}, {Wheaton}, {Wheelock}, \&
  {Zacarias}}]{Cutri:2003}
{Cutri}, R.~M., {Skrutskie}, M.~F., {van Dyk}, S., {et~al.} 2003, VizieR Online
  Data Catalog, 2246, 0

\bibitem[{{Davenport}(2016)}]{davenport2016}
{Davenport}, J. R.~A. 2016, \apj, 829, 23, \dodoi{10.3847/0004-637X/829/1/23}

\bibitem[{{David} {et~al.}(2016){David}, {Hillenbrand}, {Petigura},
  {Carpenter}, {Crossfield}, {Hinkley}, {Ciardi}, {Howard}, {Isaacson}, {Cody},
  {Schlieder}, {Beichman}, \& {Barenfeld}}]{David2016}
{David}, T.~J., {Hillenbrand}, L.~A., {Petigura}, E.~A., {et~al.} 2016, \nat,
  534, 658, \dodoi{10.1038/nature18293}

\bibitem[{{David} {et~al.}(2019){David}, {Cody}, {Hedges}, {Mamajek},
  {Hillenbrand}, {Ciardi}, {Beichman}, {Petigura}, {Fulton}, {Isaacson},
  {Howard}, {Gagn{\'e}}, {Saunders}, {Rebull}, {Stauffer}, {Vasisht}, \&
  {Hinkley}}]{David2019a}
{David}, T.~J., {Cody}, A.~M., {Hedges}, C.~L., {et~al.} 2019, \aj, 158, 79,
  \dodoi{10.3847/1538-3881/ab290f}

\bibitem[{{David} {et~al.}(2021){David}, {Contardo}, {Sandoval}, {Angus}, {Lu},
  {Bedell}, {Curtis}, {Foreman-Mackey}, {Fulton}, {Grunblatt}, \&
  {Petigura}}]{david:2021}
{David}, T.~J., {Contardo}, G., {Sandoval}, A., {et~al.} 2021, \aj, 161, 265,
  \dodoi{10.3847/1538-3881/abf439}

\bibitem[{{Donati} {et~al.}(1997){Donati}, {Semel}, {Carter}, {Rees}, \&
  {Collier Cameron}}]{Donati1997}
{Donati}, J.-F., {Semel}, M., {Carter}, B.~D., {Rees}, D.~E., \& {Collier
  Cameron}, A. 1997, \mnras, 291, 658, \dodoi{10.1093/mnras/291.4.658}

\bibitem[{{Dotter}(2016{\natexlab{a}})}]{Dotter2016}
{Dotter}, A. 2016{\natexlab{a}}, \apjs, 222, 8,
  \dodoi{10.3847/0067-0049/222/1/8}

\bibitem[{{Dotter}(2016{\natexlab{b}})}]{Dotter:2016}
---. 2016{\natexlab{b}}, \apjs, 222, 8, \dodoi{10.3847/0067-0049/222/1/8}

\bibitem[{{Eastman} {et~al.}(2019){Eastman}, {Rodriguez}, {Agol}, {Stassun},
  {Beatty}, {Vanderburg}, {Gaudi}, {Collins}, \& {Luger}}]{Eastman:2019}
{Eastman}, J.~D., {Rodriguez}, J.~E., {Agol}, E., {et~al.} 2019, arXiv
  e-prints, arXiv:1907.09480.
\newblock \doarXiv{1907.09480}

\bibitem[{{Evans}(2018)}]{Evans2018}
{Evans}, D.~F. 2018, Research Notes of the American Astronomical Society, 2,
  20, \dodoi{10.3847/2515-5172/aac173}

\bibitem[{{Fabrycky} {et~al.}(2014){Fabrycky}, {Lissauer}, {Ragozzine}, {Rowe},
  {Steffen}, {Agol}, {Barclay}, {Batalha}, {Borucki}, {Ciardi}, {Ford},
  {Gautier}, {Geary}, {Holman}, {Jenkins}, {Li}, {Morehead}, {Morris},
  {Shporer}, {Smith}, {Still}, \& {Van Cleve}}]{Fabrycky2014}
{Fabrycky}, D.~C., {Lissauer}, J.~J., {Ragozzine}, D., {et~al.} 2014, \apj,
  790, 146, \dodoi{10.1088/0004-637X/790/2/146}

\bibitem[{{Fang} \& {Margot}(2012)}]{fang2012}
{Fang}, J., \& {Margot}, J.-L. 2012, \apj, 761, 92,
  \dodoi{10.1088/0004-637X/761/2/92}

\bibitem[{{Feinstein} {et~al.}(2020{\natexlab{a}}){Feinstein}, {Montet}, \&
  {Ansdell}}]{stella2020}
{Feinstein}, A., {Montet}, B., \& {Ansdell}, M. 2020{\natexlab{a}}, The Journal
  of Open Source Software, 5, 2347, \dodoi{10.21105/joss.02347}

\bibitem[{{Feinstein} {et~al.}(2020{\natexlab{b}}){Feinstein}, {Montet},
  {Ansdell}, {Nord}, {Bean}, {G{\"u}nther}, {Gully-Santiago}, \&
  {Schlieder}}]{feinstein2020}
{Feinstein}, A.~D., {Montet}, B.~T., {Ansdell}, M., {et~al.}
  2020{\natexlab{b}}, \aj, 160, 219, \dodoi{10.3847/1538-3881/abac0a}

\bibitem[{{Findeisen} \& {Hillenbrand}(2010)}]{Findeis2010}
{Findeisen}, K., \& {Hillenbrand}, L. 2010, \aj, 139, 1338,
  \dodoi{10.1088/0004-6256/139/4/1338}

\bibitem[{{Ford}(2006)}]{Ford2006}
{Ford}, E.~B. 2006, \apj, 642, 505, \dodoi{10.1086/500802}

\bibitem[{{Foreman-Mackey}(2019)}]{exoplanet}
{Foreman-Mackey}, D. 2019, {exoplanet: Probabilistic modeling of transit or
  radial velocity observations of exoplanets}.
\newblock \doeprint{1910.005}

\bibitem[{{Gagn{\'e}} {et~al.}(2021){Gagn{\'e}}, {Faherty}, {Moranta}, \&
  {Popinchalk}}]{Gagne2021}
{Gagn{\'e}}, J., {Faherty}, J.~K., {Moranta}, L., \& {Popinchalk}, M. 2021,
  \apjl, 915, L29, \dodoi{10.3847/2041-8213/ac0e9a}

\bibitem[{{Gagn{\'e}} {et~al.}(2018){Gagn{\'e}}, {Mamajek}, {Malo}, {Riedel},
  {Rodriguez}, {Lafreni{\`e}re}, {Faherty}, {Roy-Loubier}, {Pueyo}, {Robin}, \&
  {Doyon}}]{banyan}
{Gagn{\'e}}, J., {Mamajek}, E.~E., {Malo}, L., {et~al.} 2018, \apj, 856, 23,
  \dodoi{10.3847/1538-4357/aaae09}

\bibitem[{{Gaia Collaboration} {et~al.}(2016){Gaia Collaboration}, {Prusti},
  {de Bruijne}, {Brown}, {Vallenari}, {Babusiaux}, {Bailer-Jones}, {Bastian},
  {Biermann}, {Evans}, {Eyer}, {Jansen}, {Jordi}, {Klioner}, {Lammers},
  {Lindegren}, {Luri}, {Mignard}, {Milligan}, {Panem}, {Poinsignon},
  {Pourbaix}, {Randich}, {Sarri}, {Sartoretti}, {Siddiqui}, {Soubiran},
  {Valette}, {van Leeuwen}, {Walton}, {Aerts}, {Arenou}, {Cropper}, {Drimmel},
  {H{\o}g}, {Katz}, {Lattanzi}, {O'Mullane}, {Grebel}, {Holland}, {Huc},
  {Passot}, {Bramante}, {Cacciari}, {Casta{\~n}eda}, {Chaoul}, {Cheek}, {De
  Angeli}, {Fabricius}, {Guerra}, {Hern{\'a}ndez}, {Jean-Antoine-Piccolo},
  {Masana}, {Messineo}, {Mowlavi}, {Nienartowicz}, {Ord{\'o}{\~n}ez-Blanco},
  {Panuzzo}, {Portell}, {Richards}, {Riello}, {Seabroke}, {Tanga},
  {Th{\'e}venin}, {Torra}, {Els}, {Gracia-Abril}, {Comoretto},
  {Garcia-Reinaldos}, {Lock}, {Mercier}, {Altmann}, {Andrae}, {Astraatmadja},
  {Bellas-Velidis}, {Benson}, {Berthier}, {Blomme}, {Busso}, {Carry},
  {Cellino}, {Clementini}, {Cowell}, {Creevey}, {Cuypers}, {Davidson}, {De
  Ridder}, {de Torres}, {Delchambre}, {Dell'Oro}, {Ducourant}, {Fr{\'e}mat},
  {Garc{\'\i}a-Torres}, {Gosset}, {Halbwachs}, {Hambly}, {Harrison}, {Hauser},
  {Hestroffer}, {Hodgkin}, {Huckle}, {Hutton}, {Jasniewicz}, {Jordan},
  {Kontizas}, {Korn}, {Lanzafame}, {Manteiga}, {Moitinho}, {Muinonen},
  {Osinde}, {Pancino}, {Pauwels}, {Petit}, {Recio-Blanco}, {Robin}, {Sarro},
  {Siopis}, {Smith}, {Smith}, {Sozzetti}, {Thuillot}, {van Reeven}, {Viala},
  {Abbas}, {Abreu Aramburu}, {Accart}, {Aguado}, {Allan}, {Allasia},
  {Altavilla}, {{\'A}lvarez}, {Alves}, {Anderson}, {Andrei}, {Anglada Varela},
  {Antiche}, {Antoja}, {Ant{\'o}n}, {Arcay}, {Atzei}, {Ayache}, {Bach},
  {Baker}, {Balaguer-N{\'u}{\~n}ez}, {Barache}, {Barata}, {Barbier}, {Barblan},
  {Baroni}, {Barrado y Navascu{\'e}s}, {Barros}, {Barstow}, {Becciani},
  {Bellazzini}, {Bellei}, {Bello Garc{\'\i}a}, {Belokurov}, {Bendjoya},
  {Berihuete}, {Bianchi}, {Bienaym{\'e}}, {Billebaud}, {Blagorodnova},
  {Blanco-Cuaresma}, {Boch}, {Bombrun}, {Borrachero}, {Bouquillon}, {Bourda},
  {Bouy}, {Bragaglia}, {Breddels}, {Brouillet}, {Br{\"u}semeister},
  {Bucciarelli}, {Budnik}, {Burgess}, {Burgon}, {Burlacu}, {Busonero}, {Buzzi},
  {Caffau}, {Cambras}, {Campbell}, {Cancelliere}, {Cantat-Gaudin}, {Carlucci},
  {Carrasco}, {Castellani}, {Charlot}, {Charnas}, {Charvet}, {Chassat},
  {Chiavassa}, {Clotet}, {Cocozza}, {Collins}, {Collins}, {Costigan}, {Crifo},
  {Cross}, {Crosta}, {Crowley}, {Dafonte}, {Damerdji}, {Dapergolas}, {David},
  {David}, {De Cat}, {de Felice}, {de Laverny}, {De Luise}, {De March}, {de
  Martino}, {de Souza}, {Debosscher}, {del Pozo}, {Delbo}, {Delgado},
  {Delgado}, {di Marco}, {Di Matteo}, {Diakite}, {Distefano}, {Dolding}, {Dos
  Anjos}, {Drazinos}, {Dur{\'a}n}, {Dzigan}, {Ecale}, {Edvardsson}, {Enke},
  {Erdmann}, {Escolar}, {Espina}, {Evans}, {Eynard Bontemps}, {Fabre},
  {Fabrizio}, {Faigler}, {Falc{\~a}o}, {Farr{\`a}s Casas}, {Faye}, {Federici},
  {Fedorets}, {Fern{\'a}ndez-Hern{\'a}ndez}, {Fernique}, {Fienga}, {Figueras},
  {Filippi}, {Findeisen}, {Fonti}, {Fouesneau}, {Fraile}, {Fraser}, {Fuchs},
  {Furnell}, {Gai}, {Galleti}, {Galluccio}, {Garabato}, {Garc{\'\i}a-Sedano},
  {Gar{\'e}}, {Garofalo}, {Garralda}, {Gavras}, {Gerssen}, {Geyer}, {Gilmore},
  {Girona}, {Giuffrida}, {Gomes}, {Gonz{\'a}lez-Marcos},
  {Gonz{\'a}lez-N{\'u}{\~n}ez}, {Gonz{\'a}lez-Vidal}, {Granvik}, {Guerrier},
  {Guillout}, {Guiraud}, {G{\'u}rpide}, {Guti{\'e}rrez-S{\'a}nchez}, {Guy},
  {Haigron}, {Hatzidimitriou}, {Haywood}, {Heiter}, {Helmi}, {Hobbs},
  {Hofmann}, {Holl}, {Holland}, {Hunt}, {Hypki}, {Icardi}, {Irwin}, {Jevardat
  de Fombelle}, {Jofr{\'e}}, {Jonker}, {Jorissen}, {Julbe}, {Karampelas},
  {Kochoska}, {Kohley}, {Kolenberg}, {Kontizas}, {Koposov}, {Kordopatis},
  {Koubsky}, {Kowalczyk}, {Krone-Martins}, {Kudryashova}, {Kull}, {Bachchan},
  {Lacoste-Seris}, {Lanza}, {Lavigne}, {Le Poncin-Lafitte}, {Lebreton},
  {Lebzelter}, {Leccia}, {Leclerc}, {Lecoeur-Taibi}, {Lemaitre}, {Lenhardt},
  {Leroux}, {Liao}, {Licata}, {Lindstr{\o}m}, {Lister}, {Livanou}, {Lobel},
  {L{\"o}ffler}, {L{\'o}pez}, {Lopez-Lozano}, {Lorenz}, {Loureiro},
  {MacDonald}, {Magalh{\~a}es Fernandes}, {Managau}, {Mann}, {Mantelet},
  {Marchal}, {Marchant}, {Marconi}, {Marie}, {Marinoni}, {Marrese},
  {Marschalk{\'o}}, {Marshall}, {Mart{\'\i}n-Fleitas}, {Martino}, {Mary},
  {Matijevi{\v{c}}}, {Mazeh}, {McMillan}, {Messina}, {Mestre}, {Michalik},
  {Millar}, {Miranda}, {Molina}, {Molinaro}, {Molinaro}, {Moln{\'a}r},
  {Moniez}, {Montegriffo}, {Monteiro}, {Mor}, {Mora}, {Morbidelli}, {Morel},
  {Morgenthaler}, {Morley}, {Morris}, {Mulone}, {Muraveva}, {Musella},
  {Narbonne}, {Nelemans}, {Nicastro}, {Noval}, {Ord{\'e}novic},
  {Ordieres-Mer{\'e}}, {Osborne}, {Pagani}, {Pagano}, {Pailler}, {Palacin},
  {Palaversa}, {Parsons}, {Paulsen}, {Pecoraro}, {Pedrosa}, {Pentik{\"a}inen},
  {Pereira}, {Pichon}, {Piersimoni}, {Pineau}, {Plachy}, {Plum}, {Poujoulet},
  {Pr{\v{s}}a}, {Pulone}, {Ragaini}, {Rago}, {Rambaux}, {Ramos-Lerate},
  {Ranalli}, {Rauw}, {Read}, {Regibo}, {Renk}, {Reyl{\'e}}, {Ribeiro},
  {Rimoldini}, {Ripepi}, {Riva}, {Rixon}, {Roelens}, {Romero-G{\'o}mez},
  {Rowell}, {Royer}, {Rudolph}, {Ruiz-Dern}, {Sadowski}, {Sagrist{\`a}
  Sell{\'e}s}, {Sahlmann}, {Salgado}, {Salguero}, {Sarasso}, {Savietto},
  {Schnorhk}, {Schultheis}, {Sciacca}, {Segol}, {Segovia}, {Segransan},
  {Serpell}, {Shih}, {Smareglia}, {Smart}, {Smith}, {Solano}, {Solitro},
  {Sordo}, {Soria Nieto}, {Souchay}, {Spagna}, {Spoto}, {Stampa}, {Steele},
  {Steidelm{\"u}ller}, {Stephenson}, {Stoev}, {Suess}, {S{\"u}veges}, {Surdej},
  {Szabados}, {Szegedi-Elek}, {Tapiador}, {Taris}, {Tauran}, {Taylor},
  {Teixeira}, {Terrett}, {Tingley}, {Trager}, {Turon}, {Ulla}, {Utrilla},
  {Valentini}, {van Elteren}, {Van Hemelryck}, {van Leeuwen}, {Varadi},
  {Vecchiato}, {Veljanoski}, {Via}, {Vicente}, {Vogt}, {Voss}, {Votruba},
  {Voutsinas}, {Walmsley}, {Weiler}, {Weingrill}, {Werner}, {Wevers},
  {Whitehead}, {Wyrzykowski}, {Yoldas}, {{\v{Z}}erjal}, {Zucker}, {Zurbach},
  {Zwitter}, {Alecu}, {Allen}, {Allende Prieto}, {Amorim},
  {Anglada-Escud{\'e}}, {Arsenijevic}, {Azaz}, {Balm}, {Beck}, {Bernstein},
  {Bigot}, {Bijaoui}, {Blasco}, {Bonfigli}, {Bono}, {Boudreault}, {Bressan},
  {Brown}, {Brunet}, {Bunclark}, {Buonanno}, {Butkevich}, {Carret}, {Carrion},
  {Chemin}, {Ch{\'e}reau}, {Corcione}, {Darmigny}, {de Boer}, {de Teodoro}, {de
  Zeeuw}, {Delle Luche}, {Domingues}, {Dubath}, {Fodor}, {Fr{\'e}zouls},
  {Fries}, {Fustes}, {Fyfe}, {Gallardo}, {Gallegos}, {Gardiol}, {Gebran},
  {Gomboc}, {G{\'o}mez}, {Grux}, {Gueguen}, {Heyrovsky}, {Hoar}, {Iannicola},
  {Isasi Parache}, {Janotto}, {Joliet}, {Jonckheere}, {Keil}, {Kim},
  {Klagyivik}, {Klar}, {Knude}, {Kochukhov}, {Kolka}, {Kos}, {Kutka}, {Lainey},
  {LeBouquin}, {Liu}, {Loreggia}, {Makarov}, {Marseille}, {Martayan},
  {Martinez-Rubi}, {Massart}, {Meynadier}, {Mignot}, {Munari}, {Nguyen},
  {Nordlander}, {Ocvirk}, {O'Flaherty}, {Olias Sanz}, {Ortiz}, {Osorio},
  {Oszkiewicz}, {Ouzounis}, {Palmer}, {Park}, {Pasquato}, {Peltzer}, {Peralta},
  {P{\'e}turaud}, {Pieniluoma}, {Pigozzi}, {Poels}, {Prat}, {Prod'homme},
  {Raison}, {Rebordao}, {Risquez}, {Rocca-Volmerange}, {Rosen}, {Ruiz-Fuertes},
  {Russo}, {Sembay}, {Serraller Vizcaino}, {Short}, {Siebert}, {Silva},
  {Sinachopoulos}, {Slezak}, {Soffel}, {Sosnowska}, {Strai{\v{z}}ys}, {ter
  Linden}, {Terrell}, {Theil}, {Tiede}, {Troisi}, {Tsalmantza}, {Tur},
  {Vaccari}, {Vachier}, {Valles}, {Van Hamme}, {Veltz}, {Virtanen}, {Wallut},
  {Wichmann}, {Wilkinson}, {Ziaeepour}, \& {Zschocke}}]{Gaia2016}
{Gaia Collaboration}, {Prusti}, T., {de Bruijne}, J.~H.~J., {et~al.} 2016,
  \aap, 595, A1, \dodoi{10.1051/0004-6361/201629272}

\bibitem[{{Gaia Collaboration} {et~al.}(2018){Gaia Collaboration}, {Brown},
  {Vallenari}, {Prusti}, {de Bruijne}, {Babusiaux}, {Bailer-Jones}, {Biermann},
  {Evans}, {Eyer}, {Jansen}, {Jordi}, {Klioner}, {Lammers}, {Lindegren},
  {Luri}, {Mignard}, {Panem}, {Pourbaix}, {Randich}, {Sartoretti}, {Siddiqui},
  {Soubiran}, {van Leeuwen}, {Walton}, {Arenou}, {Bastian}, {Cropper},
  {Drimmel}, {Katz}, {Lattanzi}, {Bakker}, {Cacciari}, {Casta{\~n}eda},
  {Chaoul}, {Cheek}, {De Angeli}, {Fabricius}, {Guerra}, {Holl}, {Masana},
  {Messineo}, {Mowlavi}, {Nienartowicz}, {Panuzzo}, {Portell}, {Riello},
  {Seabroke}, {Tanga}, {Th{\'e}venin}, {Gracia-Abril}, {Comoretto},
  {Garcia-Reinaldos}, {Teyssier}, {Altmann}, {Andrae}, {Audard},
  {Bellas-Velidis}, {Benson}, {Berthier}, {Blomme}, {Burgess}, {Busso},
  {Carry}, {Cellino}, {Clementini}, {Clotet}, {Creevey}, {Davidson}, {De
  Ridder}, {Delchambre}, {Dell'Oro}, {Ducourant},
  {Fern{\'a}ndez-Hern{\'a}ndez}, {Fouesneau}, {Fr{\'e}mat}, {Galluccio},
  {Garc{\'\i}a-Torres}, {Gonz{\'a}lez-N{\'u}{\~n}ez}, {Gonz{\'a}lez-Vidal},
  {Gosset}, {Guy}, {Halbwachs}, {Hambly}, {Harrison}, {Hern{\'a}ndez},
  {Hestroffer}, {Hodgkin}, {Hutton}, {Jasniewicz}, {Jean-Antoine-Piccolo},
  {Jordan}, {Korn}, {Krone-Martins}, {Lanzafame}, {Lebzelter}, {L{\"o}ffler},
  {Manteiga}, {Marrese}, {Mart{\'\i}n-Fleitas}, {Moitinho}, {Mora}, {Muinonen},
  {Osinde}, {Pancino}, {Pauwels}, {Petit}, {Recio-Blanco}, {Richards},
  {Rimoldini}, {Robin}, {Sarro}, {Siopis}, {Smith}, {Sozzetti}, {S{\"u}veges},
  {Torra}, {van Reeven}, {Abbas}, {Abreu Aramburu}, {Accart}, {Aerts},
  {Altavilla}, {{\'A}lvarez}, {Alvarez}, {Alves}, {Anderson}, {Andrei},
  {Anglada Varela}, {Antiche}, {Antoja}, {Arcay}, {Astraatmadja}, {Bach},
  {Baker}, {Balaguer-N{\'u}{\~n}ez}, {Balm}, {Barache}, {Barata}, {Barbato},
  {Barblan}, {Barklem}, {Barrado}, {Barros}, {Barstow}, {Bartholom{\'e}
  Mu{\~n}oz}, {Bassilana}, {Becciani}, {Bellazzini}, {Berihuete}, {Bertone},
  {Bianchi}, {Bienaym{\'e}}, {Blanco-Cuaresma}, {Boch}, {Boeche}, {Bombrun},
  {Borrachero}, {Bossini}, {Bouquillon}, {Bourda}, {Bragaglia}, {Bramante},
  {Breddels}, {Bressan}, {Brouillet}, {Br{\"u}semeister}, {Brugaletta},
  {Bucciarelli}, {Burlacu}, {Busonero}, {Butkevich}, {Buzzi}, {Caffau},
  {Cancelliere}, {Cannizzaro}, {Cantat-Gaudin}, {Carballo}, {Carlucci},
  {Carrasco}, {Casamiquela}, {Castellani}, {Castro-Ginard}, {Charlot},
  {Chemin}, {Chiavassa}, {Cocozza}, {Costigan}, {Cowell}, {Crifo}, {Crosta},
  {Crowley}, {Cuypers}, {Dafonte}, {Damerdji}, {Dapergolas}, {David}, {David},
  {de Laverny}, {De Luise}, {De March}, {de Martino}, {de Souza}, {de Torres},
  {Debosscher}, {del Pozo}, {Delbo}, {Delgado}, {Delgado}, {Di Matteo},
  {Diakite}, {Diener}, {Distefano}, {Dolding}, {Drazinos}, {Dur{\'a}n},
  {Edvardsson}, {Enke}, {Eriksson}, {Esquej}, {Eynard Bontemps}, {Fabre},
  {Fabrizio}, {Faigler}, {Falc{\~a}o}, {Farr{\`a}s Casas}, {Federici},
  {Fedorets}, {Fernique}, {Figueras}, {Filippi}, {Findeisen}, {Fonti},
  {Fraile}, {Fraser}, {Fr{\'e}zouls}, {Gai}, {Galleti}, {Garabato},
  {Garc{\'\i}a-Sedano}, {Garofalo}, {Garralda}, {Gavel}, {Gavras}, {Gerssen},
  {Geyer}, {Giacobbe}, {Gilmore}, {Girona}, {Giuffrida}, {Glass}, {Gomes},
  {Granvik}, {Gueguen}, {Guerrier}, {Guiraud}, {Guti{\'e}rrez-S{\'a}nchez},
  {Haigron}, {Hatzidimitriou}, {Hauser}, {Haywood}, {Heiter}, {Helmi}, {Heu},
  {Hilger}, {Hobbs}, {Hofmann}, {Holland}, {Huckle}, {Hypki}, {Icardi},
  {Jan{\ss}en}, {Jevardat de Fombelle}, {Jonker}, {Juh{\'a}sz}, {Julbe},
  {Karampelas}, {Kewley}, {Klar}, {Kochoska}, {Kohley}, {Kolenberg},
  {Kontizas}, {Kontizas}, {Koposov}, {Kordopatis}, {Kostrzewa-Rutkowska},
  {Koubsky}, {Lambert}, {Lanza}, {Lasne}, {Lavigne}, {Le Fustec}, {Le
  Poncin-Lafitte}, {Lebreton}, {Leccia}, {Leclerc}, {Lecoeur-Taibi},
  {Lenhardt}, {Leroux}, {Liao}, {Licata}, {Lindstr{\o}m}, {Lister}, {Livanou},
  {Lobel}, {L{\'o}pez}, {Managau}, {Mann}, {Mantelet}, {Marchal}, {Marchant},
  {Marconi}, {Marinoni}, {Marschalk{\'o}}, {Marshall}, {Martino}, {Marton},
  {Mary}, {Massari}, {Matijevi{\v{c}}}, {Mazeh}, {McMillan}, {Messina},
  {Michalik}, {Millar}, {Molina}, {Molinaro}, {Moln{\'a}r}, {Montegriffo},
  {Mor}, {Morbidelli}, {Morel}, {Morris}, {Mulone}, {Muraveva}, {Musella},
  {Nelemans}, {Nicastro}, {Noval}, {O'Mullane}, {Ord{\'e}novic},
  {Ord{\'o}{\~n}ez-Blanco}, {Osborne}, {Pagani}, {Pagano}, {Pailler},
  {Palacin}, {Palaversa}, {Panahi}, {Pawlak}, {Piersimoni}, {Pineau}, {Plachy},
  {Plum}, {Poggio}, {Poujoulet}, {Pr{\v{s}}a}, {Pulone}, {Racero}, {Ragaini},
  {Rambaux}, {Ramos-Lerate}, {Regibo}, {Reyl{\'e}}, {Riclet}, {Ripepi}, {Riva},
  {Rivard}, {Rixon}, {Roegiers}, {Roelens}, {Romero-G{\'o}mez}, {Rowell},
  {Royer}, {Ruiz-Dern}, {Sadowski}, {Sagrist{\`a} Sell{\'e}s}, {Sahlmann},
  {Salgado}, {Salguero}, {Sanna}, {Santana-Ros}, {Sarasso}, {Savietto},
  {Schultheis}, {Sciacca}, {Segol}, {Segovia}, {S{\'e}gransan}, {Shih},
  {Siltala}, {Silva}, {Smart}, {Smith}, {Solano}, {Solitro}, {Sordo}, {Soria
  Nieto}, {Souchay}, {Spagna}, {Spoto}, {Stampa}, {Steele},
  {Steidelm{\"u}ller}, {Stephenson}, {Stoev}, {Suess}, {Surdej}, {Szabados},
  {Szegedi-Elek}, {Tapiador}, {Taris}, {Tauran}, {Taylor}, {Teixeira},
  {Terrett}, {Teyssandier}, {Thuillot}, {Titarenko}, {Torra Clotet}, {Turon},
  {Ulla}, {Utrilla}, {Uzzi}, {Vaillant}, {Valentini}, {Valette}, {van Elteren},
  {Van Hemelryck}, {van Leeuwen}, {Vaschetto}, {Vecchiato}, {Veljanoski},
  {Viala}, {Vicente}, {Vogt}, {von Essen}, {Voss}, {Votruba}, {Voutsinas},
  {Walmsley}, {Weiler}, {Wertz}, {Wevers}, {Wyrzykowski}, {Yoldas},
  {{\v{Z}}erjal}, {Ziaeepour}, {Zorec}, {Zschocke}, {Zucker}, {Zurbach}, \&
  {Zwitter}}]{GaiaDR22018}
{Gaia Collaboration}, {Brown}, A.~G.~A., {Vallenari}, A., {et~al.} 2018, \aap,
  616, A1, \dodoi{10.1051/0004-6361/201833051}

\bibitem[{{Ginzburg} {et~al.}(2018){Ginzburg}, {Schlichting}, \&
  {Sari}}]{ginzburg2018}
{Ginzburg}, S., {Schlichting}, H.~E., \& {Sari}, R. 2018, \mnras, 476, 759,
  \dodoi{10.1093/mnras/sty290}

\bibitem[{{Gladman}(1993)}]{gladman1993}
{Gladman}, B. 1993, \icarus, 106, 247, \dodoi{10.1006/icar.1993.1169}

\bibitem[{{Gray}(2005)}]{Gray:2005}
{Gray}, D.~F. 2005, {The Observation and Analysis of Stellar Photospheres}
  ({Cambridge University Press})

\bibitem[{{Guerrero} {et~al.}(2021){Guerrero}, {Seager}, {Huang}, {Vanderburg},
  {Garcia Soto}, {Mireles}, {Hesse}, {Fong}, {Glidden}, {Shporer}, {Latham},
  {Collins}, {Quinn}, {Burt}, {Dragomir}, {Crossfield}, {Vanderspek},
  {Fausnaugh}, {Burke}, {Ricker}, {Daylan}, {Essack}, {G{\"u}nther}, {Osborn},
  {Pepper}, {Rowden}, {Sha}, {Villanueva}, {Yahalomi}, {Yu}, {Ballard},
  {Batalha}, {Berardo}, {Chontos}, {Dittmann}, {Esquerdo}, {Mikal-Evans},
  {Jayaraman}, {Krishnamurthy}, {Louie}, {Mehrle}, {Niraula}, {Rackham},
  {Rodriguez}, {Rowden}, {Sousa-Silva}, {Watanabe}, {Wong}, {Zhan},
  {Zivanovic}, {Christiansen}, {Ciardi}, {Swain}, {Lund}, {Mullally},
  {Fleming}, {Rodriguez}, {Boyd}, {Quintana}, {Barclay}, {Col{\'o}n},
  {Rinehart}, {Schlieder}, {Clampin}, {Jenkins}, {Twicken}, {Caldwell},
  {Coughlin}, {Henze}, {Lissauer}, {Morris}, {Rose}, {Smith}, {Tenenbaum},
  {Ting}, {Wohler}, {Bakos}, {Bean}, {Berta-Thompson}, {Bieryla}, {Bouma},
  {Buchhave}, {Butler}, {Charbonneau}, {Doty}, {Ge}, {Holman}, {Howard},
  {Kaltenegger}, {Kane}, {Kjeldsen}, {Kreidberg}, {Lin}, {Minsky}, {Narita},
  {Paegert}, {P{\'a}l}, {Palle}, {Sasselov}, {Spencer}, {Sozzetti}, {Stassun},
  {Torres}, {Udry}, \& {Winn}}]{Guerrero2021}
{Guerrero}, N.~M., {Seager}, S., {Huang}, C.~X., {et~al.} 2021, \apjs, 254, 39,
  \dodoi{10.3847/1538-4365/abefe1}

\bibitem[{{Guillot} {et~al.}(2015){Guillot}, {Abe}, {Agabi}, {Rivet}, {Daban},
  {M{\'e}karnia}, {Aristidi}, {Schmider}, {Crouzet}, {Gon{\c{c}}alves},
  {Gouvret}, {Ottogalli}, {Faradji}, {Blanc}, {Bondoux}, \&
  {Valbousquet}}]{guillot2015}
{Guillot}, T., {Abe}, L., {Agabi}, A., {et~al.} 2015, Astronomische
  Nachrichten, 336, 638, \dodoi{10.1002/asna.201512174}

\bibitem[{{G{\"u}nther} {et~al.}(2020){G{\"u}nther}, {Zhan}, {Seager},
  {Rimmer}, {Ranjan}, {Stassun}, {Oelkers}, {Daylan}, {Newton}, {Kristiansen},
  {Olah}, {Gillen}, {Rappaport}, {Ricker}, {Vanderspek}, {Latham}, {Winn},
  {Jenkins}, {Glidden}, {Fausnaugh}, {Levine}, {Dittmann}, {Quinn},
  {Krishnamurthy}, \& {Ting}}]{gunther2020}
{G{\"u}nther}, M.~N., {Zhan}, Z., {Seager}, S., {et~al.} 2020, \aj, 159, 60,
  \dodoi{10.3847/1538-3881/ab5d3a}

\bibitem[{{Hadden} \& {Lithwick}(2014)}]{Hadden2014}
{Hadden}, S., \& {Lithwick}, Y. 2014, \apj, 787, 80,
  \dodoi{10.1088/0004-637X/787/1/80}

\bibitem[{{Hadden} \& {Lithwick}(2018)}]{hadden2018b}
---. 2018, \aj, 156, 95, \dodoi{10.3847/1538-3881/aad32c}

\bibitem[{{Hartmann} \& {Noyes}(1987)}]{hartmann1987}
{Hartmann}, L.~W., \& {Noyes}, R.~W. 1987, \araa, 25, 271,
  \dodoi{10.1146/annurev.aa.25.090187.001415}

\bibitem[{{He} {et~al.}(2020){He}, {Ford}, {Ragozzine}, \& {Carrera}}]{he2020}
{He}, M.~Y., {Ford}, E.~B., {Ragozzine}, D., \& {Carrera}, D. 2020, \aj, 160,
  276, \dodoi{10.3847/1538-3881/abba18}

\bibitem[{{Hedges} {et~al.}(2021){Hedges}, {Hughes}, {Zhou}, {David}, {Becker},
  {Giacalone}, {Vanderburg}, {Rodriguez}, {Bieryla}, {Wirth}, {Atherton},
  {Fetherolf}, {Collins}, {Price-Whelan}, {Bedell}, {Quinn}, {Gan}, {Ricker},
  {Latham}, {Vanderspek}, {Seager}, {Winn}, {Jenkins}, {Kielkopf}, {Schwarz},
  {Dressing}, {Gonzales}, {Crossfield}, {Matthews}, {Jensen}, {Furlan},
  {Gnilka}, {Howell}, {Lester}, {Scott}, {Feliz}, {Lund}, {Siverd}, {Stevens},
  {Narita}, {Fukui}, {Murgas}, {Palle}, {Sutton}, {Stassun}, {Bouma}, {Vezie},
  {Villase{\~n}or}, {Quintana}, \& {Smith}}]{hedges2021}
{Hedges}, C., {Hughes}, A., {Zhou}, G., {et~al.} 2021, \aj, 162, 54,
  \dodoi{10.3847/1538-3881/ac06cd}

\bibitem[{{Henden} {et~al.}(2016){Henden}, {Templeton}, {Terrell}, {Smith},
  {Levine}, \& {Welch}}]{Henden:2016}
{Henden}, A.~A., {Templeton}, M., {Terrell}, D., {et~al.} 2016, VizieR Online
  Data Catalog, II/336

\bibitem[{{Hippke} \& {Heller}(2019)}]{tls}
{Hippke}, M., \& {Heller}, R. 2019, \aap, 623, A39,
  \dodoi{10.1051/0004-6361/201834672}

\bibitem[{{H{\o}g} {et~al.}(2000){H{\o}g}, {Fabricius}, {Makarov}, {Urban},
  {Corbin}, {Wycoff}, {Bastian}, {Schwekendiek}, \& {Wicenec}}]{Hog:2000}
{H{\o}g}, E., {Fabricius}, C., {Makarov}, V.~V., {et~al.} 2000, \aap, 355, L27

\bibitem[{{Howell} {et~al.}(2014){Howell}, {Sobeck}, {Haas}, {Still},
  {Barclay}, {Mullally}, {Troeltzsch}, {Aigrain}, {Bryson}, {Caldwell},
  {Chaplin}, {Cochran}, {Huber}, {Marcy}, {Miglio}, {Najita}, {Smith},
  {Twicken}, \& {Fortney}}]{k2}
{Howell}, S.~B., {Sobeck}, C., {Haas}, M., {et~al.} 2014, \pasp, 126, 398,
  \dodoi{10.1086/676406}

\bibitem[{{Huang} {et~al.}(2020{\natexlab{a}}){Huang}, {Vanderburg}, {P{\'a}l},
  {Sha}, {Yu}, {Fong}, {Fausnaugh}, {Shporer}, {Guerrero}, {Vanderspek}, \&
  {Ricker}}]{qlp2020a}
{Huang}, C.~X., {Vanderburg}, A., {P{\'a}l}, A., {et~al.} 2020{\natexlab{a}},
  Research Notes of the American Astronomical Society, 4, 204,
  \dodoi{10.3847/2515-5172/abca2e}

\bibitem[{{Huang} {et~al.}(2020{\natexlab{b}}){Huang}, {Vanderburg}, {P{\'a}l},
  {Sha}, {Yu}, {Fong}, {Fausnaugh}, {Shporer}, {Guerrero}, {Vanderspek}, \&
  {Ricker}}]{qlp2020b}
---. 2020{\natexlab{b}}, Research Notes of the American Astronomical Society,
  4, 206, \dodoi{10.3847/2515-5172/abca2d}

\bibitem[{{Hunter}(2007)}]{matplotlib}
{Hunter}, J.~D. 2007, Computing in Science and Engineering, 9, 90,
  \dodoi{10.1109/MCSE.2007.55}

\bibitem[{{Jenkins}(2002)}]{jenkins2002}
{Jenkins}, J.~M. 2002, \apj, 575, 493, \dodoi{10.1086/341136}

\bibitem[{{Jenkins} {et~al.}(2020){Jenkins}, {Tenenbaum}, {Seader}, {Burke},
  {McCauliff}, {Smith}, {Twicken}, \& {Chandrasekaran}}]{jenkins2020tps}
{Jenkins}, J.~M., {Tenenbaum}, P., {Seader}, S., {et~al.} 2020, {Kepler Data
  Processing Handbook: Transiting Planet Search}, Kepler Science Document
  KSCI-19081-003

\bibitem[{{Jenkins} {et~al.}(2010{\natexlab{a}}){Jenkins}, {Caldwell},
  {Chandrasekaran}, {Twicken}, {Bryson}, {Quintana}, {Clarke}, {Li}, {Allen},
  {Tenenbaum}, {Wu}, {Klaus}, {Middour}, {Cote}, {McCauliff}, {Girouard},
  {Gunter}, {Wohler}, {Sommers}, {Hall}, {Uddin}, {Wu}, {Bhavsar}, {Van Cleve},
  {Pletcher}, {Dotson}, {Haas}, {Gilliland}, {Koch}, \&
  {Borucki}}]{Jenkins2010}
{Jenkins}, J.~M., {Caldwell}, D.~A., {Chandrasekaran}, H., {et~al.}
  2010{\natexlab{a}}, \apjl, 713, L87, \dodoi{10.1088/2041-8205/713/2/L87}

\bibitem[{{Jenkins} {et~al.}(2010{\natexlab{b}}){Jenkins}, {Chandrasekaran},
  {McCauliff}, {Caldwell}, {Tenenbaum}, {Li}, {Klaus}, {Cote}, \&
  {Middour}}]{jenkins2010tps}
{Jenkins}, J.~M., {Chandrasekaran}, H., {McCauliff}, S.~D., {et~al.}
  2010{\natexlab{b}}, in Society of Photo-Optical Instrumentation Engineers
  (SPIE) Conference Series, Vol. 7740, Software and Cyberinfrastructure for
  Astronomy, ed. N.~M. {Radziwill} \& A.~{Bridger}, 77400D,
  \dodoi{10.1117/12.856764}

\bibitem[{{Jenkins} {et~al.}(2016){Jenkins}, {Twicken}, {McCauliff},
  {Campbell}, {Sanderfer}, {Lung}, {Mansouri-Samani}, {Girouard}, {Tenenbaum},
  {Klaus}, {Smith}, {Caldwell}, {Chacon}, {Henze}, {Heiges}, {Latham},
  {Morgan}, {Swade}, {Rinehart}, \& {Vanderspek}}]{Jenkins2016}
{Jenkins}, J.~M., {Twicken}, J.~D., {McCauliff}, S., {et~al.} 2016, in
  \procspie, Vol. 9913, Software and Cyberinfrastructure for Astronomy IV,
  99133E, \dodoi{10.1117/12.2233418}

\bibitem[{{Jensen}(2013)}]{Jensen:2013}
{Jensen}, E. 2013, {Tapir: A web interface for transit/eclipse observability},
  Astrophysics Source Code Library.
\newblock \doeprint{1306.007}

\bibitem[{{Jin} \& {Mordasini}(2018)}]{jin:2018}
{Jin}, S., \& {Mordasini}, C. 2018, \apj, 853, 163,
  \dodoi{10.3847/1538-4357/aa9f1e}

\bibitem[{{Kane}(2014)}]{kane2014a}
{Kane}, S.~R. 2014, \apj, 782, 111, \dodoi{10.1088/0004-637X/782/2/111}

\bibitem[{{Kane}(2016)}]{kane2016d}
---. 2016, \apj, 830, 105, \dodoi{10.3847/0004-637X/830/2/105}

\bibitem[{{Kane}(2019)}]{kane2019c}
---. 2019, \aj, 158, 72, \dodoi{10.3847/1538-3881/ab2a09}

\bibitem[{{Kane} {et~al.}(2014){Kane}, {Kopparapu}, \&
  {Domagal-Goldman}}]{kane2014e}
{Kane}, S.~R., {Kopparapu}, R.~K., \& {Domagal-Goldman}, S.~D. 2014, \apjl,
  794, L5, \dodoi{10.1088/2041-8205/794/1/L5}

\bibitem[{{Kane} \& {Raymond}(2014)}]{kane2014b}
{Kane}, S.~R., \& {Raymond}, S.~N. 2014, \apj, 784, 104,
  \dodoi{10.1088/0004-637X/784/2/104}

\bibitem[{{Kane} {et~al.}(2016){Kane}, {Hill}, {Kasting}, {Kopparapu},
  {Quintana}, {Barclay}, {Batalha}, {Borucki}, {Ciardi}, {Haghighipour},
  {Hinkel}, {Kaltenegger}, {Selsis}, \& {Torres}}]{kane2016c}
{Kane}, S.~R., {Hill}, M.~L., {Kasting}, J.~F., {et~al.} 2016, \apj, 830, 1,
  \dodoi{10.3847/0004-637X/830/1/1}

\bibitem[{{Kasting} {et~al.}(1993){Kasting}, {Whitmire}, \&
  {Reynolds}}]{kasting1993a}
{Kasting}, J.~F., {Whitmire}, D.~P., \& {Reynolds}, R.~T. 1993, \icarus, 101,
  108, \dodoi{10.1006/icar.1993.1010}

\bibitem[{{Kempton} {et~al.}(2018){Kempton}, {Bean}, {Louie}, {Deming}, {Koll},
  {Mansfield}, {Christiansen}, {L{\'o}pez-Morales}, {Swain}, {Zellem},
  {Ballard}, {Barclay}, {Barstow}, {Batalha}, {Beatty}, {Berta-Thompson},
  {Birkby}, {Buchhave}, {Charbonneau}, {Cowan}, {Crossfield}, {de Val-Borro},
  {Doyon}, {Dragomir}, {Gaidos}, {Heng}, {Hu}, {Kane}, {Kreidberg}, {Mallonn},
  {Morley}, {Narita}, {Nascimbeni}, {Pall{\'e}}, {Quintana}, {Rauscher},
  {Seager}, {Shkolnik}, {Sing}, {Sozzetti}, {Stassun}, {Valenti}, \& {von
  Essen}}]{kempton2018}
{Kempton}, E. M.~R., {Bean}, J.~L., {Louie}, D.~R., {et~al.} 2018, \pasp, 130,
  114401, \dodoi{10.1088/1538-3873/aadf6f}

\bibitem[{{Kopparapu} {et~al.}(2014){Kopparapu}, {Ramirez}, {SchottelKotte},
  {Kasting}, {Domagal-Goldman}, \& {Eymet}}]{kopparapu2014}
{Kopparapu}, R.~K., {Ramirez}, R.~M., {SchottelKotte}, J., {et~al.} 2014, \apj,
  787, L29, \dodoi{10.1088/2041-8205/787/2/L29}

\bibitem[{{Kopparapu} {et~al.}(2013){Kopparapu}, {Ramirez}, {Kasting}, {Eymet},
  {Robinson}, {Mahadevan}, {Terrien}, {Domagal-Goldman}, {Meadows}, \&
  {Deshpande}}]{kopparapu2013a}
{Kopparapu}, R.~K., {Ramirez}, R., {Kasting}, J.~F., {et~al.} 2013, \apj, 765,
  131, \dodoi{10.1088/0004-637X/765/2/131}

\bibitem[{{Kounkel} \& {Covey}(2019)}]{Kounkel2019}
{Kounkel}, M., \& {Covey}, K. 2019, \aj, 158, 122,
  \dodoi{10.3847/1538-3881/ab339a}

\bibitem[{{Kreidberg}(2015)}]{batman}
{Kreidberg}, L. 2015, \pasp, 127, 1161, \dodoi{10.1086/683602}

\bibitem[{{K{\"u}ker} \& {R{\"u}diger}(2011)}]{kuker2011}
{K{\"u}ker}, M., \& {R{\"u}diger}, G. 2011, Astronomische Nachrichten, 332,
  933, \dodoi{10.1002/asna.201111628}

\bibitem[{{Latham} {et~al.}(2011){Latham}, {Rowe}, {Quinn}, {Batalha},
  {Borucki}, {Brown}, {Bryson}, {Buchhave}, {Caldwell}, {Carter},
  {Christiansen}, {Ciardi}, {Cochran}, {Dunham}, {Fabrycky}, {Ford}, {Gautier},
  {Gilliland}, {Holman}, {Howell}, {Ibrahim}, {Isaacson}, {Jenkins}, {Koch},
  {Lissauer}, {Marcy}, {Quintana}, {Ragozzine}, {Sasselov}, {Shporer},
  {Steffen}, {Welsh}, \& {Wohler}}]{latham2011}
{Latham}, D.~W., {Rowe}, J.~F., {Quinn}, S.~N., {et~al.} 2011, \apjl, 732, L24,
  \dodoi{10.1088/2041-8205/732/2/L24}

\bibitem[{{Li} {et~al.}(2019){Li}, {Tenenbaum}, {Twicken}, {Burke}, {Jenkins},
  {Quintana}, {Rowe}, \& {Seader}}]{li2019}
{Li}, J., {Tenenbaum}, P., {Twicken}, J.~D., {et~al.} 2019, \pasp, 131, 024506,
  \dodoi{10.1088/1538-3873/aaf44d}

\bibitem[{{Lindegren} {et~al.}(2018){Lindegren}, {Hern{\'a}ndez}, {Bombrun},
  {Klioner}, {Bastian}, {Ramos-Lerate}, {de Torres}, {Steidelm{\"u}ller},
  {Stephenson}, {Hobbs}, {Lammers}, {Biermann}, {Geyer}, {Hilger}, {Michalik},
  {Stampa}, {McMillan}, {Casta{\~n}eda}, {Clotet}, {Comoretto}, {Davidson},
  {Fabricius}, {Gracia}, {Hambly}, {Hutton}, {Mora}, {Portell}, {van Leeuwen},
  {Abbas}, {Abreu}, {Altmann}, {Andrei}, {Anglada}, {Balaguer-N{\'u}{\~n}ez},
  {Barache}, {Becciani}, {Bertone}, {Bianchi}, {Bouquillon}, {Bourda},
  {Br{\"u}semeister}, {Bucciarelli}, {Busonero}, {Buzzi}, {Cancelliere},
  {Carlucci}, {Charlot}, {Cheek}, {Crosta}, {Crowley}, {de Bruijne}, {de
  Felice}, {Drimmel}, {Esquej}, {Fienga}, {Fraile}, {Gai}, {Garralda},
  {Gonz{\'a}lez-Vidal}, {Guerra}, {Hauser}, {Hofmann}, {Holl}, {Jordan},
  {Lattanzi}, {Lenhardt}, {Liao}, {Licata}, {Lister}, {L{\"o}ffler},
  {Marchant}, {Martin-Fleitas}, {Messineo}, {Mignard}, {Morbidelli}, {Poggio},
  {Riva}, {Rowell}, {Salguero}, {Sarasso}, {Sciacca}, {Siddiqui}, {Smart},
  {Spagna}, {Steele}, {Taris}, {Torra}, {van Elteren}, {van Reeven}, \&
  {Vecchiato}}]{Lindegren2018}
{Lindegren}, L., {Hern{\'a}ndez}, J., {Bombrun}, A., {et~al.} 2018, \aap, 616,
  A2, \dodoi{10.1051/0004-6361/201832727}

\bibitem[{{Lissauer} {et~al.}(2011){Lissauer}, {Ragozzine}, {Fabrycky},
  {Steffen}, {Ford}, {Jenkins}, {Shporer}, {Holman}, {Rowe}, {Quintana},
  {Batalha}, {Borucki}, {Bryson}, {Caldwell}, {Carter}, {Ciardi}, {Dunham},
  {Fortney}, {Gautier}, {Howell}, {Koch}, {Latham}, {Marcy}, {Morehead}, \&
  {Sasselov}}]{Lissauer2011}
{Lissauer}, J.~J., {Ragozzine}, D., {Fabrycky}, D.~C., {et~al.} 2011, \apjs,
  197, 8, \dodoi{10.1088/0067-0049/197/1/8}

\bibitem[{{Lissauer} {et~al.}(2012){Lissauer}, {Marcy}, {Rowe}, {Bryson},
  {Adams}, {Buchhave}, {Ciardi}, {Cochran}, {Fabrycky}, {Ford}, {Fressin},
  {Geary}, {Gilliland}, {Holman}, {Howell}, {Jenkins}, {Kinemuchi}, {Koch},
  {Morehead}, {Ragozzine}, {Seader}, {Tanenbaum}, {Torres}, \&
  {Twicken}}]{lissauer2012}
{Lissauer}, J.~J., {Marcy}, G.~W., {Rowe}, J.~F., {et~al.} 2012, \apj, 750,
  112, \dodoi{10.1088/0004-637X/750/2/112}

\bibitem[{{Livingston} {et~al.}(2018){Livingston}, {Dai}, {Hirano}, {Gandolfi},
  {Nowak}, {Endl}, {Velasco}, {Fukui}, {Narita}, {Prieto-Arranz}, {Barragan},
  {Cusano}, {Albrecht}, {Cabrera}, {Cochran}, {Csizmadia}, {Deeg},
  {Eigm{\"u}ller}, {Erikson}, {Fridlund}, {Grziwa}, {Guenther}, {Hatzes},
  {Kawauchi}, {Korth}, {Nespral}, {Palle}, {P{\"a}tzold}, {Persson}, {Rauer},
  {Smith}, {Tamura}, {Tanaka}, {Van Eylen}, {Watanabe}, \&
  {Winn}}]{Livingston2018}
{Livingston}, J.~H., {Dai}, F., {Hirano}, T., {et~al.} 2018, \aj, 155, 115,
  \dodoi{10.3847/1538-3881/aaa841}

\bibitem[{{Livingston} {et~al.}(2019){Livingston}, {Dai}, {Hirano}, {Gandolfi},
  {Trani}, {Nowak}, {Cochran}, {Endl}, {Albrecht}, {Barragan}, {Cabrera},
  {Csizmadia}, {de Leon}, {Deeg}, {Eigm{\"u}ller}, {Erikson}, {Fridlund},
  {Fukui}, {Grziwa}, {Guenther}, {Hatzes}, {Korth}, {Kuzuhara}, {Monta{\~n}es},
  {Narita}, {Nespral}, {Palle}, {P{\"a}tzold}, {Persson}, {Prieto-Arranz},
  {Rauer}, {Tamura}, {Van Eylen}, \& {Winn}}]{Livingston2019}
---. 2019, \mnras, 484, 8, \dodoi{10.1093/mnras/sty3464}

\bibitem[{{Lomb}(1976)}]{lomb1976}
{Lomb}, N.~R. 1976, \apss, 39, 447, \dodoi{10.1007/BF00648343}

\bibitem[{{Lovis} \& {Mayor}(2007)}]{Lovis}
{Lovis}, C., \& {Mayor}, M. 2007, \aap, 472, 657,
  \dodoi{10.1051/0004-6361:20077375}

\bibitem[{{Maehara} {et~al.}(2015){Maehara}, {Shibayama}, {Notsu}, {Notsu},
  {Honda}, {Nogami}, \& {Shibata}}]{maehara2015}
{Maehara}, H., {Shibayama}, T., {Notsu}, Y., {et~al.} 2015, Earth, Planets and
  Space, 67, 59, \dodoi{10.1186/s40623-015-0217-z}

\bibitem[{{Mamajek} \& {Hillenbrand}(2008)}]{Mamajek2008}
{Mamajek}, E.~E., \& {Hillenbrand}, L.~A. 2008, \apj, 687, 1264,
  \dodoi{10.1086/591785}

\bibitem[{{Mann} {et~al.}(2016{\natexlab{a}}){Mann}, {Gaidos}, {Mace},
  {Johnson}, {Bowler}, {LaCourse}, {Jacobs}, {Vanderburg}, {Kraus}, {Kaplan},
  \& {Jaffe}}]{Mann2016}
{Mann}, A.~W., {Gaidos}, E., {Mace}, G.~N., {et~al.} 2016{\natexlab{a}}, \apj,
  818, 46, \dodoi{10.3847/0004-637X/818/1/46}

\bibitem[{{Mann} {et~al.}(2016{\natexlab{b}}){Mann}, {Newton}, {Rizzuto},
  {Irwin}, {Feiden}, {Gaidos}, {Mace}, {Kraus}, {James}, {Ansdell},
  {Charbonneau}, {Covey}, {Ireland}, {Jaffe}, {Johnson}, {Kidder}, \&
  {Vanderburg}}]{Mann2016b}
{Mann}, A.~W., {Newton}, E.~R., {Rizzuto}, A.~C., {et~al.} 2016{\natexlab{b}},
  \aj, 152, 61, \dodoi{10.3847/0004-6256/152/3/61}

\bibitem[{{Mann} {et~al.}(2017){Mann}, {Gaidos}, {Vanderburg}, {Rizzuto},
  {Ansdell}, {Medina}, {Mace}, {Kraus}, \& {Sokal}}]{Mann2017}
{Mann}, A.~W., {Gaidos}, E., {Vanderburg}, A., {et~al.} 2017, \aj, 153, 64,
  \dodoi{10.1088/1361-6528/aa5276}

\bibitem[{{Mann} {et~al.}(2018){Mann}, {Vanderburg}, {Rizzuto}, {Kraus},
  {Berlind}, {Bieryla}, {Calkins}, {Esquerdo}, {Latham}, {Mace}, {Morris},
  {Quinn}, {Sokal}, \& {Stefanik}}]{Mann2018}
{Mann}, A.~W., {Vanderburg}, A., {Rizzuto}, A.~C., {et~al.} 2018, \aj, 155, 4,
  \dodoi{10.3847/1538-3881/aa9791}

\bibitem[{{Mann} {et~al.}(2020){Mann}, {Johnson}, {Vanderburg}, {Kraus},
  {Rizzuto}, {Wood}, {Bush}, {Rockcliffe}, {Newton}, {Latham}, {Mamajek},
  {Zhou}, {Quinn}, {Thao}, {Benatti}, {Cosentino}, {Desidera}, {Harutyunyan},
  {Lovis}, {Mortier}, {Pepe}, {Poretti}, {Wilson}, {Kristiansen}, {Gagliano},
  {Jacobs}, {LaCourse}, {Omohundro}, {Schwengeler}, {Terentev}, {Kane}, {Hill},
  {Rabus}, {Esquerdo}, {Berlind}, {Collins}, {Murawski}, {Sallam}, {Aitken},
  {Massey}, {Ricker}, {Vanderspek}, {Seager}, {Winn}, {Jenkins}, {Barclay},
  {Caldwell}, {Dragomir}, {Doty}, {Glidden}, {Tenenbaum}, {Torres}, {Twicken},
  \& {Villanueva}}]{Mann2020}
{Mann}, A.~W., {Johnson}, M.~C., {Vanderburg}, A., {et~al.} 2020, \aj, 160,
  179, \dodoi{10.3847/1538-3881/abae64}

\bibitem[{{Mart{\'\i}n} {et~al.}(2018){Mart{\'\i}n}, {Lodieu}, {Pavlenko}, \&
  {B{\'e}jar}}]{martin2018}
{Mart{\'\i}n}, E.~L., {Lodieu}, N., {Pavlenko}, Y., \& {B{\'e}jar}, V. J.~S.
  2018, \apj, 856, 40, \dodoi{10.3847/1538-4357/aaaeb8}

\bibitem[{{McCully} {et~al.}(2018){McCully}, {Volgenau}, {Harbeck}, {Lister},
  {Saunders}, {Turner}, {Siiverd}, \& {Bowman}}]{McCully:2018}
{McCully}, C., {Volgenau}, N.~H., {Harbeck}, D.-R., {et~al.} 2018, in Society
  of Photo-Optical Instrumentation Engineers (SPIE) Conference Series, Vol.
  10707, \procspie, 107070K, \dodoi{10.1117/12.2314340}

\bibitem[{{Medina} {et~al.}(2020){Medina}, {Winters}, {Irwin}, \&
  {Charbonneau}}]{medina2020}
{Medina}, A.~A., {Winters}, J.~G., {Irwin}, J.~M., \& {Charbonneau}, D. 2020,
  \apj, 905, 107, \dodoi{10.3847/1538-4357/abc686}

\bibitem[{{Meingast} {et~al.}(2019){Meingast}, {Alves}, \&
  {F{\"u}rnkranz}}]{Meingast2019}
{Meingast}, S., {Alves}, J., \& {F{\"u}rnkranz}, V. 2019, \aap, 622, L13,
  \dodoi{10.1051/0004-6361/201834950}

\bibitem[{{M{\'e}karnia} {et~al.}(2016){M{\'e}karnia}, {Guillot}, {Rivet},
  {Schmider}, {Abe}, {Gon{\c{c}}alves}, {Agabi}, {Crouzet}, {Fruth},
  {Barbieri}, {Bayliss}, {Zhou}, {Aristidi}, {Szulagyi}, {Daban},
  {Fante{\"\i}-Caujolle}, {Gouvret}, {Erikson}, {Rauer}, {Bouchy}, {Gerakis},
  \& {Bouchez}}]{Mekarnia:2016}
{M{\'e}karnia}, D., {Guillot}, T., {Rivet}, J.~P., {et~al.} 2016, \mnras, 463,
  45, \dodoi{10.1093/mnras/stw1934}

\bibitem[{{Millholland} {et~al.}(2021){Millholland}, {He}, {Ford}, {Ragozzine},
  {Fabrycky}, \& {Winn}}]{millholland2021}
{Millholland}, S.~C., {He}, M.~Y., {Ford}, E.~B., {et~al.} 2021, \aj, 162, 166,
  \dodoi{10.3847/1538-3881/ac0f7a}

\bibitem[{{Moorhead} {et~al.}(2011){Moorhead}, {Ford}, {Morehead}, {Rowe},
  {Borucki}, {Batalha}, {Bryson}, {Caldwell}, {Fabrycky}, {Gautier}, {Koch},
  {Holman}, {Jenkins}, {Li}, {Lissauer}, {Lucas}, {Marcy}, {Quinn}, {Quintana},
  {Ragozzine}, {Shporer}, {Still}, \& {Torres}}]{moorhead2011}
{Moorhead}, A.~V., {Ford}, E.~B., {Morehead}, R.~C., {et~al.} 2011, \apjs, 197,
  1, \dodoi{10.1088/0067-0049/197/1/1}

\bibitem[{{Newton} {et~al.}(2019){Newton}, {Mann}, {Tofflemire}, {Pearce},
  {Rizzuto}, {Vanderburg}, {Martinez}, {Wang}, {Ruffio}, {Kraus}, {Johnson},
  {Thao}, {Wood}, {Rampalli}, {Nielsen}, {Collins}, {Dragomir}, {Hellier},
  {Anderson}, {Barclay}, {Brown}, {Feiden}, {Hart}, {Isopi}, {Kielkopf},
  {Mallia}, {Nelson}, {Rodriguez}, {Stockdale}, {Waite}, {Wright}, {Lissauer},
  {Ricker}, {Vanderspek}, {Latham}, {Seager}, {Winn}, {Jenkins}, {Bouma},
  {Burke}, {Davies}, {Fausnaugh}, {Li}, {Morris}, {Mukai}, {Villase{\~n}or},
  {Villeneuva}, {De Rosa}, {Macintosh}, {Mengel}, {Okumura}, \&
  {Wittenmyer}}]{Newton2019}
{Newton}, E.~R., {Mann}, A.~W., {Tofflemire}, B.~M., {et~al.} 2019, \apjl, 880,
  L17, \dodoi{10.3847/2041-8213/ab2988}

\bibitem[{{Obermeier} {et~al.}(2016){Obermeier}, {Henning}, {Schlieder},
  {Crossfield}, {Petigura}, {Howard}, {Sinukoff}, {Isaacson}, {Ciardi},
  {David}, {Hillenbrand}, {Beichman}, {Howell}, {Horch}, {Everett}, {Hirsch},
  {Teske}, {Christiansen}, {L{\'e}pine}, {Aller}, {Liu}, {Saglia},
  {Livingston}, \& {Kluge}}]{Obermeier2016}
{Obermeier}, C., {Henning}, T., {Schlieder}, J.~E., {et~al.} 2016, \aj, 152,
  223, \dodoi{10.3847/1538-3881/152/6/223}

\bibitem[{{Ostberg} \& {Kane}(2019)}]{ostberg2019}
{Ostberg}, C., \& {Kane}, S.~R. 2019, \aj, 158, 195,
  \dodoi{10.3847/1538-3881/ab44b0}

\bibitem[{{Owen} \& {Wu}(2013)}]{owen:2013}
{Owen}, J.~E., \& {Wu}, Y. 2013, \apj, 775, 105,
  \dodoi{10.1088/0004-637X/775/2/105}

\bibitem[{{Paulson} {et~al.}(2004){Paulson}, {Cochran}, \&
  {Hatzes}}]{Paulson2004}
{Paulson}, D.~B., {Cochran}, W.~D., \& {Hatzes}, A.~P. 2004, \aj, 127, 3579,
  \dodoi{10.1086/420710}

\bibitem[{{Paxton} {et~al.}(2011){Paxton}, {Bildsten}, {Dotter}, {Herwig},
  {Lesaffre}, \& {Timmes}}]{Paxton2011}
{Paxton}, B., {Bildsten}, L., {Dotter}, A., {et~al.} 2011, \apjs, 192, 3,
  \dodoi{10.1088/0067-0049/192/1/3}

\bibitem[{{Paxton} {et~al.}(2013){Paxton}, {Cantiello}, {Arras}, {Bildsten},
  {Brown}, {Dotter}, {Mankovich}, {Montgomery}, {Stello}, {Timmes}, \&
  {Townsend}}]{Paxton2013}
{Paxton}, B., {Cantiello}, M., {Arras}, P., {et~al.} 2013, \apjs, 208, 4,
  \dodoi{10.1088/0067-0049/208/1/4}

\bibitem[{{Paxton} {et~al.}(2015){Paxton}, {Marchant}, {Schwab}, {Bauer},
  {Bildsten}, {Cantiello}, {Dessart}, {Farmer}, {Hu}, {Langer}, {Townsend},
  {Townsley}, \& {Timmes}}]{Paxton2015}
{Paxton}, B., {Marchant}, P., {Schwab}, J., {et~al.} 2015, \apjs, 220, 15,
  \dodoi{10.1088/0067-0049/220/1/15}

\bibitem[{{Pepper} {et~al.}(2017){Pepper}, {Gillen}, {Parviainen},
  {Hillenbrand}, {Cody}, {Aigrain}, {Stauffer}, {Vrba}, {David}, {Lillo-Box},
  {Stassun}, {Conroy}, {Pope}, \& {Barrado}}]{Pepper2017}
{Pepper}, J., {Gillen}, E., {Parviainen}, H., {et~al.} 2017, \aj, 153, 177,
  \dodoi{10.3847/1538-3881/aa62ab}

\bibitem[{{Perryman} {et~al.}(1998){Perryman}, {Brown}, {Lebreton}, {Gomez},
  {Turon}, {Cayrel de Strobel}, {Mermilliod}, {Robichon}, {Kovalevsky}, \&
  {Crifo}}]{perryman1998}
{Perryman}, M.~A.~C., {Brown}, A.~G.~A., {Lebreton}, Y., {et~al.} 1998, \aap,
  331, 81.
\newblock \doarXiv{astro-ph/9707253}

\bibitem[{{Plavchan} {et~al.}(2020){Plavchan}, {Barclay}, {Gagn{\'e}}, {Gao},
  {Cale}, {Matzko}, {Dragomir}, {Quinn}, {Feliz}, {Stassun}, {Crossfield},
  {Berardo}, {Latham}, {Tieu}, {Anglada-Escud{\'e}}, {Ricker}, {Vanderspek},
  {Seager}, {Winn}, {Jenkins}, {Rinehart}, {Krishnamurthy}, {Dynes}, {Doty},
  {Adams}, {Afanasev}, {Beichman}, {Bottom}, {Bowler}, {Brinkworth}, {Brown},
  {Cancino}, {Ciardi}, {Clampin}, {Clark}, {Collins}, {Davison},
  {Foreman-Mackey}, {Furlan}, {Gaidos}, {Geneser}, {Giddens}, {Gilbert},
  {Hall}, {Hellier}, {Henry}, {Horner}, {Howard}, {Huang}, {Huber}, {Kane},
  {Kenworthy}, {Kielkopf}, {Kipping}, {Klenke}, {Kruse}, {Latouf}, {Lowrance},
  {Mennesson}, {Mengel}, {Mills}, {Morton}, {Narita}, {Newton}, {Nishimoto},
  {Okumura}, {Palle}, {Pepper}, {Quintana}, {Roberge}, {Roccatagliata},
  {Schlieder}, {Tanner}, {Teske}, {Tinney}, {Vanderburg}, {von Braun}, {Walp},
  {Wang}, {Wang}, {Weigand}, {White}, {Wittenmyer}, {Wright}, {Youngblood},
  {Zhang}, \& {Zilberman}}]{Plavchan2020}
{Plavchan}, P., {Barclay}, T., {Gagn{\'e}}, J., {et~al.} 2020, \nat, 582, 497,
  \dodoi{10.1038/s41586-020-2400-z}

\bibitem[{{Quinn} {et~al.}(2012){Quinn}, {White}, {Latham}, {Buchhave},
  {Cantrell}, {Dahm}, {F{\H{u}}r{\'e}sz}, {Szentgyorgyi}, {Geary}, {Torres},
  {Bieryla}, {Berlind}, {Calkins}, {Esquerdo}, \& {Stefanik}}]{Quinn2012}
{Quinn}, S.~N., {White}, R.~J., {Latham}, D.~W., {et~al.} 2012, \apjl, 756,
  L33, \dodoi{10.1088/2041-8205/756/2/L33}

\bibitem[{{Quinn} {et~al.}(2014){Quinn}, {White}, {Latham}, {Buchhave},
  {Torres}, {Stefanik}, {Berlind}, {Bieryla}, {Calkins}, {Esquerdo},
  {F{\H{u}}r{\'e}sz}, {Geary}, \& {Szentgyorgyi}}]{Quinn2014}
---. 2014, \apj, 787, 27, \dodoi{10.1088/0004-637X/787/1/27}

\bibitem[{{Richey-Yowell} {et~al.}(2019){Richey-Yowell}, {Shkolnik},
  {Schneider}, {Osby}, {Barman}, \& {Meadows}}]{Richey-Yowell:2019}
{Richey-Yowell}, T., {Shkolnik}, E.~L., {Schneider}, A.~C., {et~al.} 2019,
  \apj, 872, 17, \dodoi{10.3847/1538-4357/aafa74}

\bibitem[{{Ricker} {et~al.}(2016){Ricker}, {Vanderspek}, {Winn}, {Seager},
  {Berta-Thompson}, {Levine}, {Villasenor}, {Latham}, {Charbonneau}, {Holman},
  {Johnson}, {Sasselov}, {Szentgyorgyi}, {Torres}, {Bakos}, {Brown},
  {Christensen-Dalsgaard}, {Kjeldsen}, {Clampin}, {Rinehart}, {Deming}, {Doty},
  {Dunham}, {Ida}, {Kawai}, {Sato}, {Jenkins}, {Lissauer}, {Jernigan},
  {Kaltenegger}, {Laughlin}, {Lin}, {McCullough}, {Narita}, {Pepper},
  {Stassun}, \& {Udry}}]{Ricker2016}
{Ricker}, G.~R., {Vanderspek}, R., {Winn}, J., {et~al.} 2016, in Society of
  Photo-Optical Instrumentation Engineers (SPIE) Conference Series, Vol. 9904,
  Space Telescopes and Instrumentation 2016: Optical, Infrared, and Millimeter
  Wave, ed. H.~A. {MacEwen}, G.~G. {Fazio}, M.~{Lystrup}, N.~{Batalha},
  N.~{Siegler}, \& E.~C. {Tong}, 99042B, \dodoi{10.1117/12.2232071}

\bibitem[{{Rizzuto} {et~al.}(2017){Rizzuto}, {Mann}, {Vanderburg}, {Kraus}, \&
  {Covey}}]{Rizzuto2017}
{Rizzuto}, A.~C., {Mann}, A.~W., {Vanderburg}, A., {Kraus}, A.~L., \& {Covey},
  K.~R. 2017, \aj, 154, 224, \dodoi{10.3847/1538-3881/aa9070}

\bibitem[{{Rizzuto} {et~al.}(2018){Rizzuto}, {Vanderburg}, {Mann}, {Kraus},
  {Dressing}, {Ag{\"u}eros}, {Douglas}, \& {Krolikowski}}]{Rizzuto2018}
{Rizzuto}, A.~C., {Vanderburg}, A., {Mann}, A.~W., {et~al.} 2018, \aj, 156,
  195, \dodoi{10.3847/1538-3881/aadf37}

\bibitem[{{Rizzuto} {et~al.}(2020){Rizzuto}, {Newton}, {Mann}, {Tofflemire},
  {Vanderburg}, {Kraus}, {Wood}, {Quinn}, {Zhou}, {Thao}, {Law}, {Ziegler}, \&
  {Brice{\~n}o}}]{Rizzuto2020}
{Rizzuto}, A.~C., {Newton}, E.~R., {Mann}, A.~W., {et~al.} 2020, \aj, 160, 33,
  \dodoi{10.3847/1538-3881/ab94b7}

\bibitem[{{Saar} \& {Donahue}(1997)}]{Saar1997}
{Saar}, S.~H., \& {Donahue}, R.~A. 1997, \apj, 485, 319, \dodoi{10.1086/304392}

\bibitem[{{Sato} {et~al.}(2007){Sato}, {Izumiura}, {Toyota}, {Kambe}, {Takeda},
  {Masuda}, {Omiya}, {Murata}, {Itoh}, {Ando}, {Yoshida}, {Ikoma}, {Kokubo}, \&
  {Ida}}]{Sato2007}
{Sato}, B., {Izumiura}, H., {Toyota}, E., {et~al.} 2007, \apj, 661, 527,
  \dodoi{10.1086/513503}

\bibitem[{{Scargle}(1982)}]{scargle1982}
{Scargle}, J.~D. 1982, \apj, 263, 835, \dodoi{10.1086/160554}

\bibitem[{{Schlafly} \& {Finkbeiner}(2011)}]{Schlafly2011}
{Schlafly}, E.~F., \& {Finkbeiner}, D.~P. 2011, \apj, 737, 103,
  \dodoi{10.1088/0004-637X/737/2/103}

\bibitem[{{Schlegel} {et~al.}(1998){Schlegel}, {Finkbeiner}, \&
  {Davis}}]{Schlegel1998}
{Schlegel}, D.~J., {Finkbeiner}, D.~P., \& {Davis}, M. 1998, \apj, 500, 525,
  \dodoi{10.1086/305772}

\bibitem[{{Shkolnik} {et~al.}(2011){Shkolnik}, {Liu}, {Reid}, {Dupuy}, \&
  {Weinberger}}]{Shkolnik2011}
{Shkolnik}, E.~L., {Liu}, M.~C., {Reid}, I.~N., {Dupuy}, T., \& {Weinberger},
  A.~J. 2011, \apj, 727, 6, \dodoi{10.1088/0004-637X/727/1/6}

\bibitem[{{Stassun} {et~al.}(2018){Stassun}, {Oelkers}, {Pepper}, {Paegert},
  {De Lee}, {Torres}, {Latham}, {Charpinet}, {Dressing}, {Huber}, {Kane},
  {L{\'e}pine}, {Mann}, {Muirhead}, {Rojas-Ayala}, {Silvotti}, {Fleming},
  {Levine}, \& {Plavchan}}]{stassun:2018b}
{Stassun}, K.~G., {Oelkers}, R.~J., {Pepper}, J., {et~al.} 2018, \aj, 156, 102,
  \dodoi{10.3847/1538-3881/aad050}

\bibitem[{{Stumpe} {et~al.}(2012){Stumpe}, {Smith}, {Van Cleve}, {Twicken},
  {Barclay}, {Fanelli}, {Girouard}, {Jenkins}, {Kolodziejczak}, {McCauliff}, \&
  {Morris}}]{Stumpe2012}
{Stumpe}, M.~C., {Smith}, J.~C., {Van Cleve}, J.~E., {et~al.} 2012, \pasp, 124,
  985, \dodoi{10.1086/667698}

\bibitem[{{Tofflemire} {et~al.}(2021){Tofflemire}, {Rizzuto}, {Newton},
  {Kraus}, {Mann}, {Vanderburg}, {Nelson}, {Hawkins}, {Wood}, {Zhou}, {Quinn},
  {Howell}, {Collins}, {Schwarz}, {Stassun}, {Bouma}, {Essack}, {Osborn},
  {Boyd}, {F{\H{u}}r{\'e}sz}, {Glidden}, {Twicken}, {Wohler}, {McLean},
  {Ricker}, {Vanderspek}, {Latham}, {Seager}, {Winn}, \&
  {Jenkins}}]{Tofflemire2021}
{Tofflemire}, B.~M., {Rizzuto}, A.~C., {Newton}, E.~R., {et~al.} 2021, \aj,
  161, 171, \dodoi{10.3847/1538-3881/abdf53}

\bibitem[{{Tokovinin}(2018)}]{Tokovinin:2018}
{Tokovinin}, A. 2018, \pasp, 130, 035002, \dodoi{10.1088/1538-3873/aaa7d9}

\bibitem[{{Tokovinin} {et~al.}(2013){Tokovinin}, {Fischer}, {Bonati},
  {Giguere}, {Moore}, {Schwab}, {Spronck}, \& {Szymkowiak}}]{Tokovinin2013}
{Tokovinin}, A., {Fischer}, D.~A., {Bonati}, M., {et~al.} 2013, \pasp, 125,
  1336, \dodoi{10.1086/674012}

\bibitem[{{Torres} {et~al.}(2020){Torres}, {Vanderburg}, {Curtis}, {Kraus},
  {Rizzuto}, \& {Ireland}}]{torres2020}
{Torres}, G., {Vanderburg}, A., {Curtis}, J.~L., {et~al.} 2020, \apj, 896, 162,
  \dodoi{10.3847/1538-4357/ab911b}

\bibitem[{{Twicken} {et~al.}(2018){Twicken}, {Catanzarite}, {Clarke},
  {Girouard}, {Jenkins}, {Klaus}, {Li}, {McCauliff}, {Seader}, {Tenenbaum},
  {Wohler}, {Bryson}, {Burke}, {Caldwell}, {Haas}, {Henze}, \&
  {Sanderfer}}]{twicken2018}
{Twicken}, J.~D., {Catanzarite}, J.~H., {Clarke}, B.~D., {et~al.} 2018, \pasp,
  130, 064502, \dodoi{10.1088/1538-3873/aab694}

\bibitem[{{van der Walt} {et~al.}(2011){van der Walt}, {Colbert}, \&
  {Varoquaux}}]{numpy}
{van der Walt}, S., {Colbert}, S.~C., \& {Varoquaux}, G. 2011, Computing in
  Science and Engineering, 13, 22, \dodoi{10.1109/MCSE.2011.37}

\bibitem[{{Van Eylen} \& {Albrecht}(2015)}]{vaneylen2015}
{Van Eylen}, V., \& {Albrecht}, S. 2015, \apj, 808, 126,
  \dodoi{10.1088/0004-637X/808/2/126}

\bibitem[{{Vanderburg} {et~al.}(2018){Vanderburg}, {Mann}, {Rizzuto},
  {Bieryla}, {Kraus}, {Berlind}, {Calkins}, {Curtis}, {Douglas}, {Esquerdo},
  {Everett}, {Horch}, {Howell}, {Latham}, {Mayo}, {Quinn}, {Scott}, \&
  {Stefanik}}]{Vanderburg2018}
{Vanderburg}, A., {Mann}, A.~W., {Rizzuto}, A., {et~al.} 2018, \aj, 156, 46,
  \dodoi{10.3847/1538-3881/aac894}

\bibitem[{{Vanderburg} {et~al.}(2019){Vanderburg}, {Huang}, {Rodriguez},
  {Becker}, {Ricker}, {Vanderspek}, {Latham}, {Seager}, {Winn}, {Jenkins},
  {Addison}, {Bieryla}, {Brice{\~n}o}, {Bowler}, {Brown}, {Burke}, {Burt},
  {Caldwell}, {Clark}, {Crossfield}, {Dittmann}, {Dynes}, {Fulton}, {Guerrero},
  {Harbeck}, {Horner}, {Kane}, {Kielkopf}, {Kraus}, {Kreidberg}, {Law}, {Mann},
  {Mengel}, {Morton}, {Okumura}, {Pearce}, {Plavchan}, {Quinn}, {Rabus},
  {Rose}, {Rowden}, {Shporer}, {Siverd}, {Smith}, {Stassun}, {Tinney},
  {Wittenmyer}, {Wright}, {Zhang}, {Zhou}, \& {Ziegler}}]{Vanderburg:2019}
{Vanderburg}, A., {Huang}, C.~X., {Rodriguez}, J.~E., {et~al.} 2019, \apjl,
  881, L19, \dodoi{10.3847/2041-8213/ab322d}

\bibitem[{{Wisdom}(2006)}]{wisdom2006b}
{Wisdom}, J. 2006, \aj, 131, 2294, \dodoi{10.1086/500829}

\bibitem[{{Wisdom} \& {Holman}(1991)}]{wisdom1991}
{Wisdom}, J., \& {Holman}, M. 1991, \aj, 102, 1528, \dodoi{10.1086/115978}

\bibitem[{{Zhou} {et~al.}(2018){Zhou}, {Rodriguez}, {Vanderburg}, {Quinn},
  {Irwin}, {Huang}, {Latham}, {Bieryla}, {Esquerdo}, {Berlind}, \&
  {Calkins}}]{Zhou:2018}
{Zhou}, G., {Rodriguez}, J.~E., {Vanderburg}, A., {et~al.} 2018, \aj, 156, 93,
  \dodoi{10.3847/1538-3881/aad085}

\bibitem[{{Zhou} {et~al.}(2021){Zhou}, {Quinn}, {Irwin}, {Huang}, {Collins},
  {Bouma}, {Khan}, {Landrigan}, {Vanderburg}, {Rodriguez}, {Latham}, {Torres},
  {Douglas}, {Bieryla}, {Esquerdo}, {Berlind}, {Calkins}, {Buchhave},
  {Charbonneau}, {Collins}, {Kielkopf}, {Jensen}, {Tan}, {Hart}, {Carter},
  {Stockdale}, {Ziegler}, {Law}, {Mann}, {Howell}, {Matson}, {Scott}, {Furlan},
  {White}, {Hellier}, {Anderson}, {West}, {Ricker}, {Vanderspek}, {Seager},
  {Jenkins}, {Winn}, {Mireles}, {Rowden}, {Yahalomi}, {Wohler}, {Brasseur},
  {Daylan}, \& {Col{\'o}n}}]{Zhou2021}
{Zhou}, G., {Quinn}, S.~N., {Irwin}, J., {et~al.} 2021, \aj, 161, 2,
  \dodoi{10.3847/1538-3881/abba22}

\bibitem[{{Ziegler} {et~al.}(2018){Ziegler}, {Law}, {Baranec}, {Morton},
  {Riddle}, {De Lee}, {Huber}, {Mahadevan}, \& {Pepper}}]{Ziegler:2018}
{Ziegler}, C., {Law}, N.~M., {Baranec}, C., {et~al.} 2018, \aj, 156, 259,
  \dodoi{10.3847/1538-3881/aad80a}

\end{thebibliography}
